\documentclass[12pt]{article}

\usepackage{enumerate}
\usepackage{amsmath}
\usepackage{amssymb}
\usepackage{graphicx}
\usepackage{xcolor}
\usepackage{booktabs}
\usepackage{adjustbox}
\usepackage{threeparttable}
\usepackage{threeparttablex}
\usepackage{array}
\usepackage{etoolbox}
\usepackage{natbib} 
\usepackage[linesnumbered,ruled,vlined]{algorithm2e}
\usepackage{amsthm}
\usepackage{url}
\usepackage{booktabs, tabularx, threeparttablex, adjustbox, amsmath, amssymb, graphicx,xcolor,array}
\usepackage[utf8]{inputenc}
\usepackage{textcomp}
\usepackage{dsfont}
\usepackage{array}
\usepackage{newtxtext}
\usepackage{newtxmath}
\usepackage{anyfontsize}
\usepackage{makecell}
\usepackage{ragged2e}
\graphicspath{{Figure/}}
\usepackage[colorlinks=true,linkcolor=black,citecolor=blue,urlcolor=blue]{hyperref} 
\usepackage{authblk} 
\newtheorem{assumption}{Assumption}
\newtheorem{lemma}{Lemma}
\newtheorem{theorem}{Theorem}
\newtheorem{example}{Example}

\newtheorem{remark}{Remark}

\newcolumntype{Y}{>{\raggedright\arraybackslash}X}

\newcommand{\tilS}{{\tilde{S}}}
\newcommand{\tilg}{\tilde{\boldsymbol{g}}}
\newcommand{\bmug}{\boldsymbol{\mu}_g}
\newcommand{\bg}{\boldsymbol{g}}
\newcommand{\blam}{\boldsymbol{\lambda}}
\newcommand{\balp}{\boldsymbol{\alpha}}

\newcommand{\bbeta}{\boldsymbol{\beta}}
\newcommand{\independent}{\perp\!\!\!\perp}
\newcommand{\cmark}{\checkmark}
\newcommand{\xmark}{\(\times\)}

\newcommand{\bs}{\boldsymbol}

\begin{document}

\def\spacingset#1{\renewcommand{\baselinestretch}%
{#1}\small\normalsize} \spacingset{1}

\title{\bf An Estimand-Focused Approach for AUC Generalization and Cross-Study Benchmarking}

\author[1]{Jiajun Liu}
\author[2]{Guangcai Mao}
\author[1]{Xiaofei Wang\footnote{Address for correspondence: Xiaofei Wang, Department of Biostatistics and Bioinformatics, Duke University, Durham, NC 27710, U.S.A. Email: xiaofei.wang@duke.edu}}

\affil[1]{\small Department of Biostatistics and Bioinformatics, Duke University, Durham, NC 27710, U.S.A.}

\affil[2]{\small School of Mathematics and Statistics, Central China Normal University, No.152 Luoyu Road, Wuhan, Hubei 430079, China}

\date{}
\maketitle

\begin{abstract}
The area under the ROC curve (AUC) is the standard measure of a biomarker's discriminatory accuracy; however, AUC is rarely treated as a population-specific estimand. When validation cohorts differ from the intended target population in case mix, Na\"ive AUC estimates can mislead both generalization and cross-study comparison. We develop an estimand-focused framework that anchors biomarker AUC inference to a prespecified target population, aligning with the ICH E9(R1) estimand perspective adapted to discrimination rather than treatment effect. The framework supports two scientific goals: generalizing a study-specific AUC to a clinically relevant target population, and benchmarking AUCs across studies on a common population footing. Methodologically, we extend calibration weighting to the U-statistic formulation of AUC, allowing valid estimation even when the target population is characterized only by summary-level covariate information. This setting is common in biomarker validation, where individual-level target data are often unavailable and existing transportability methods may not be applicable. When patient-level real-world data are accessible, the proposed augmented variants provide double robustness and improved efficiency. We establish asymptotic properties and study their performances through comprehensive simulations. Furthermore, we demonstrate the proposed framework on the POWER trials, evaluating baseline stair-climb power (SCP) as a prognostic marker for 6-month survival in advanced non-small-cell lung cancer (NSCLC). Unlike prior work on transporting model-based predictive accuracy, our framework targets the biomarker-level estimand directly and addresses cross-study comparability — an issue not resolved by current methods.
\end{abstract}

\bigskip
\noindent%
\vfill

\newpage
\spacingset{1.9} 

\section{Introduction}\label{sec:intro}
Biomarkers play a foundational role in clinical and translational research. As they provide interpretable signals of disease presence, progression, and future risk, their applications range from cancer detection and diagnosis to prognosis, treatment planning, and clinical trial enrichment. In diagnostic settings, biomarkers support screening, early detection, and confirmatory decision making by helping distinguish patients with disease from those without disease \citep{pepe2001phases,mandrekar2010receiver,Schlattmann2022}. In prognostic settings, biomarkers, either individually or as part of clinically interpretable biomarker signatures, help identify patients at higher or lower risk of adverse outcomes, thereby informing treatment intensity, surveillance strategies, and future trial design \citep{mcshane2005reporting,Kent2018,Steyerberg2019}.

Across diagnostic and prognostic applications, the area under the receiver operating characteristic curve (AUC) is one of the most widely used measures of discriminatory accuracy. It can be interpreted as the probability that a randomly selected subject with the event or response has a more event-indicative marker value than a randomly selected subject without the event or response. However, despite this scale-free interpretation, the AUC is not a ``plug-and-play'' quantity. Its value depends on the population in which the biomarker is evaluated, because populations may differ in age, disease severity, comorbidities, and other baseline characteristics. This issue is especially important in biomarker research. As emphasized by the REMARK guidelines, markers that appear promising in one study may lead to inconsistent conclusions in others, making it essential to understand the context in which marker conclusions apply \citep{mcshane2005reporting}. It also matters when comparing the same marker across studies, since observed AUC differences may reflect case-mix differences rather than true differences in biomarker performance. Thus, these challenges motivate two scientific goals: generalizing the AUC of a biomarker to a prespecified target population, and benchmarking biomarker performance across studies after accounting for covariate shift.

Consistent with the ICH E9(R1) estimand framework \citep{ICH2019}, we treat the AUC as a population-specific estimand tied to a well-defined target population. Under this perspective, the AUC is not merely a performance metric; its interpretation depends on the population and study conditions. The five estimand components can be mapped to the biomarker AUC setting (Table~\ref{tab:auc_estimand}), clarifying whose outcomes are being predicted, which biomarker is evaluated, and under what conditions discrimination should be interpreted. Moreover, this estimand-focused view also highlights practical sources of distortion, including non-random sampling, temporal or geographic drift, incomplete verification, and covariate shift. Therefore, two problems become critical: how to generalize an observed AUC from a validation cohort to a target population of interest, and how to compare AUCs fairly across studies that represent different populations.
\begin{table*}[htbp]
\centering
\renewcommand{\arraystretch}{0.7} 
\caption{Mapping between the ICH E9(R1) estimand framework and the AUC setting}
\label{tab:auc_estimand}
\small 
\begin{tabularx}{\textwidth}{@{} l Y Y @{}} 
\toprule
\textbf{Component} & \textbf{Treatment-effect setting} & \textbf{AUC setting} \\
\midrule
Population & Target population for treatment effect & Target population for discrimination \\
\addlinespace
Treatment condition & Treatment comparison of interest & Binary outcome contrast (e.g., diseased vs non-diseased) \\
\addlinespace
Variable & Endpoint of interest & Biomarker of interest \\
\addlinespace
Intercurrent events & Post-randomization events (e.g., discontinuation, crossover) & Verification or sampling mechanisms affecting discrimination \\
\addlinespace
Summary measure & Risk difference, hazard ratio, etc. & Area under the ROC curve (AUC) \\
\bottomrule
\end{tabularx}
\end{table*}

These questions arise frequently in real applications, especially when the observed study cohort is not fully representative of the population in which the biomarker will ultimately be used. Despite its ubiquity, statistical methods for transporting and comparing AUCs remain limited. Generalizability and transportability methods extend inference from study samples to target populations \citep{Rothwell2005,Cole2010,Stuart2015,Dahabreh2019,Lee2023,Liu2024}, but have mainly been developed for causal effects rather than biomarker-level discrimination. Existing work on AUC estimation has also focused largely on bias within a single study, such as verification bias or test-result-dependent sampling \citep{Wang2012_Bios}.
More recently, target-population AUC estimation has been considered for prediction models under covariate shift \citep{Li2023}. However, many biomarker applications instead focus on the discriminatory ability of a specific biomarker rather than a multivariable prediction score. This distinction matters because biomarkers often provide more direct and interpretable clinical information \citep{taylor2008validation, ballman2015biomarker}, whereas transporting a prediction model can be constrained by whether the same covariates are measured in a new study.
It also changes the modeling strategy in estimation. Prediction-model AUCs naturally center nuisance modeling on the clinical outcome given covariates, whereas biomarker AUCs compare marker values across outcome groups, making the biomarker distribution conditional on disease status and covariates central. In addition, existing approaches typically require patient-level target covariate data, whereas many real applications provide only summary-level covariate information from external registries or surveys \citep{data_cdc_brfss,data_cdc_nhis}. Finally, less attention has been given to benchmarking biomarker performance across studies after aligning them to a common target population. Therefore, these gaps motivate the need for a framework that treats biomarker AUCs as population-specific quantities, can accommodate limited target-population information, and supports both generalization to a target population and fair cross-study benchmarking under covariate shift.

To address this need, this article develops an estimand-focused framework for AUC estimation, generalization, and comparison. To enable a wider application, the framework is designed for biomarker evaluation and applicable even when the target population is described only by summary-level information.
Our first contribution is conceptual: we formalize the AUC as a population-specific estimand and show that meaningful interpretation and comparison require anchoring discrimination performance to a prespecified target population. This perspective is important because biomarker discrimination is often reported as an observed AUC metric, without clearly specifying the population to which it pertains. Without this clarification, it is easy to overlook that distributional shifts can affect biomarker performance when people attempt to carry over the promising findings from a source trial to a target population. Consequently, such transport can mask the biomarker's true performance and clinical usefulness in the target population. Second, methodological, we address covariate shift by aligning covariate distributions and proposed a unified family of estimators. The framework is applicable under varying levels of target-population information, including settings where only summary-level covariate data are accessible. To improve efficiency and robustness, we also develop augmented, doubly robust estimators that remain valid when either the sampling mechanism or the biomarker-outcome model is correctly specified. The large-sample properties of the estimators are well established, and their performance under varying scenarios is investigated. Overall, our framework builds on familiar tools such as weighting and outcome modeling, but brings them into a U-statistic framework tailored to population-anchored AUC inference.

The remainder of the paper is organized as follows. Section~\ref{sec:BasicSetup} introduces the setup, data structure, and identification assumptions. Section~\ref{sec:method} presents six estimators for different data-availability scenarios, including calibration-weighting, outcome-modeling, sampling-weighting, and augmented estimators. Section~\ref{sec:simulation} reports simulation results under varying covariate-shift and model-specification settings. Section~\ref{sec:RealData} applies the methods to compare biomarker AUCs in a motivating NSCLC study, and Section~\ref{sec:conclusion} concludes with practical guidance.

\section{Basic Setup}\label{sec:BasicSetup}
\subsection{Problem Settings and Notations}\label{sec:BasicSetup_notation}
We conceptualize AUC analysis within an estimand-oriented framework centered on two objectives. First, AUC estimation should be anchored to a prespecified target population, which defines the clinical context and case mix for which diagnostic or prognostic accuracy is intended. Such anchoring ensures that the estimand answers a population-specific question: how well does the test discriminate within this intended population? Second, AUC comparison requires benchmarking multiple AUCs against a common target population, so that observed differences reflect genuine discrimination performance rather than shifts in prevalence, disease spectrum, or covariate distributions across studies. Framing AUC in this manner distinguishes design from analysis, yields interpretable population-specific estimates, and supports fair cross-study comparisons of biomarkers and predictive signatures.

\begin{figure}[ht]
    \centering
    \includegraphics[width=\linewidth]{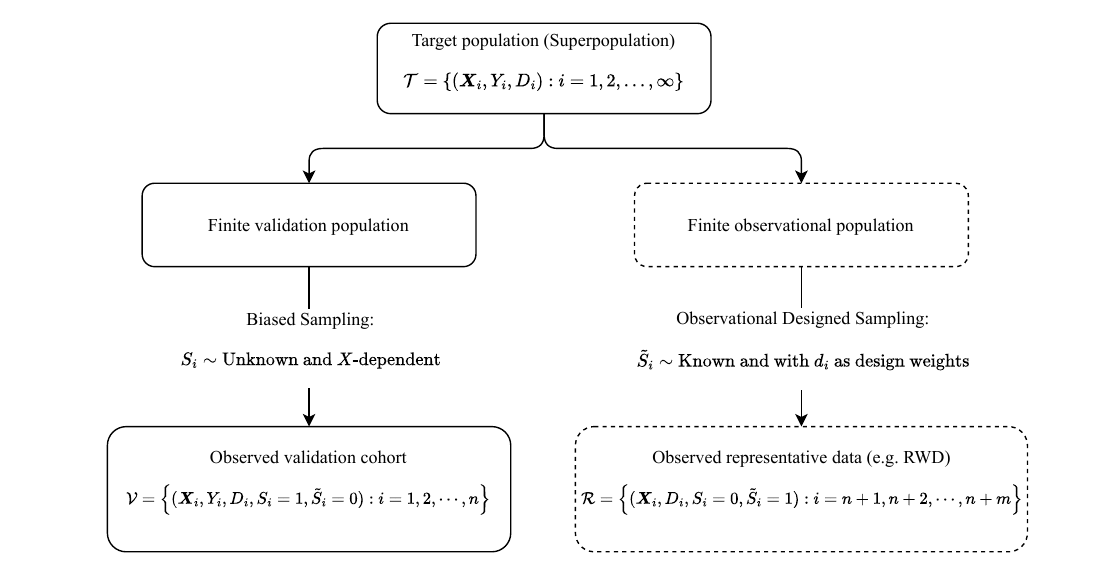}
    \caption{Sampling framework for validation cohort and observational data}
    \label{fig:data_structure1}
\end{figure}

To infer the AUC for a target population, we start with an observed study sample in which both the biomarker and outcome are available. We refer to this sample as the \emph{validation cohort}; in the generalizability literature, it is analogous to the \emph{source sample}. Our goal is to use this cohort to estimate the AUC in a prespecified target population, rather than only in the cohort itself. The main setting considered in this paper is one in which the validation cohort is not fully representative of the target population because of biased sampling, design restrictions, or other limitations in external validity. This is naturally viewed through a nested-design framework, where the validation cohort is a selected subset of a larger target population \citep{Dahabreh2021}. The inferential goal is therefore one of \emph{generalizability}: adjusting the validation-cohort AUC to approximate the target-population AUC for the same biomarker. When additional observational data that better reflect the target population are available, such as real-world data (RWD), they can assist in this adjustment. In practice, however, such external data may be available either at the individual level or only through summary-level covariate information, such as means and variances. This motivates the development of estimators that remain valid in both settings. Although we formulate the problem primarily in the nested-design setting for clarity, the same ideas extend naturally to transportability settings, where the source and target samples are distinct rather than nested. In both settings, the goal is to mitigate differences in population distributions so that the AUC is interpreted relative to the intended target population.

The data structure of the primary problem is illustrated in Figure \ref{fig:data_structure1}. Let the broader target population, or superpopulation, be denoted by $\mathcal{T}=\left\{(\boldsymbol{X}_i, Y_i, D_i): i=1,\ldots,\infty \right\}$, where $(\boldsymbol{X}_i, Y_i, D_i)$ are not directly observable. Instead, we observe a validation cohort $\mathcal{V}=\left\{(\boldsymbol{X}_i, Y_i, D_i, S_i=1, \tilS_i=0): i=1,\ldots, n\right\}$. The $Y$ represents the biomarker of interest, which is continuous and predictive of the binary response $D \in \{0, 1\}$. Here, $D = 1$ and $D = 0$ denote responders and non-responders in prognostic settings or disease-present and disease-absent subjects in diagnostic settings. Each subject has $p$-dimensional baseline covariates $\boldsymbol{X}_i=(X_{i1}, \dots, X_{ip})^\top$; and the full covariate matrix is {$\boldsymbol{\tilde X} = (\boldsymbol{X}_1, \dots, \boldsymbol{X}_n)^\top$}. 

We use two binary indicators, $S \in \{0,1\}$ and $\tilde S \in \{0,1\}$, to distinguish the two sampling processes. The indicator $S$ denotes inclusion in the validation cohort, where the biomarker and outcome are measured but the sampling mechanism may depend on covariates and is generally unknown. In contrast, $\tilde S$ denotes being selected into an auxiliary observational sample, such as RWD, whose sampling mechanism is treated as designed or known through sampling weights. By design, these two samples are disjoint, so $S\tilde S = 0$ for all units. Thus, $(S=1,\tilde S=0)$ identifies the validation cohort, $(S=0,\tilde S=1)$ denotes the observational sample, and $(S=0,\tilde S=0)$ implies the unobserved units in target population.

\begin{figure}[ht]
    \centering
    \includegraphics[width=\linewidth]{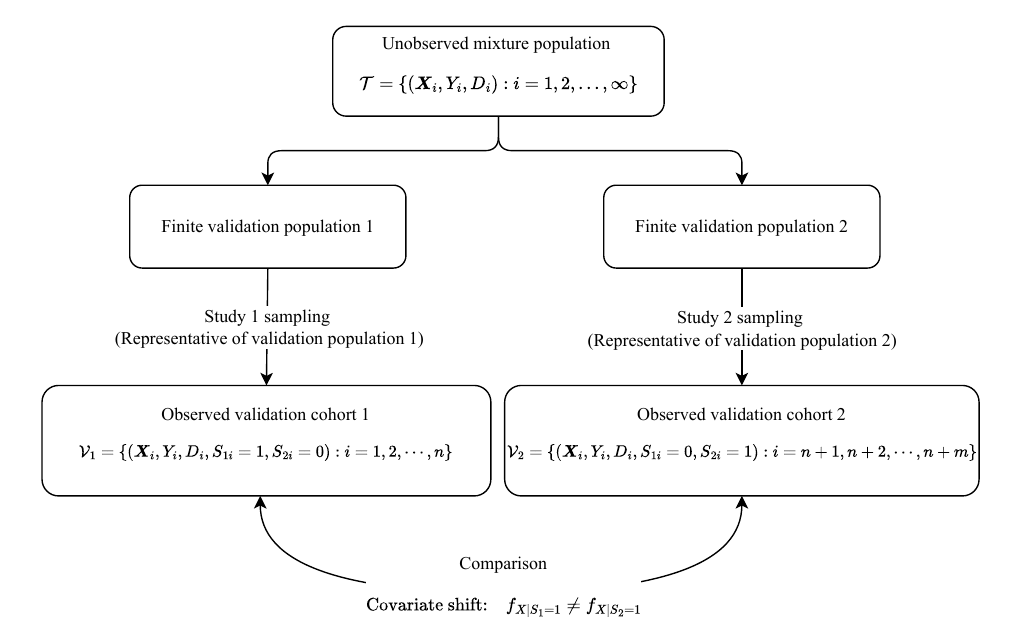}
    \caption{Sampling structure under two validation cohorts with covariate shift}
    \label{fig:data_structure2}
\end{figure}

Let $\mathcal{R}=\{(\boldsymbol{X}_i, D_i, S_i=0, \tilS_i=1): i=n+1,\ldots,n+m\}$ denote the RWD dataset of $m$ subjects, where the biomarker $Y$ is unobserved. We assume $\mathcal{R}$ helps characterize the target population through its covariate and response distributions \citep{Rudolph2023}. The combined observed data are $\mathcal{C}=\left\{(\boldsymbol{X}_i, Y_i, D_i, S_i,\tilS_i): i=1,\ldots, m+n\right\}$, where $Y_i$ is missing for individuals in $\mathcal{R}$. In the validation cohort, sampling may depend on covariates $\boldsymbol{X}$ or biomarker $Y$. Because biomarker values are typically unavailable at sampling, we assume selection depends only on $\boldsymbol{X}$. Throughout, $Y$ is treated as a biomarker related to both response $D$ and covariates $\boldsymbol{X}$.

As shown in Figure \ref{fig:data_structure2}, the benchmarking setup follows the same notation. We consider an underlying mixture population that conceptually combines the two validation cohorts, each representing its own finite validation population. When two RCTs evaluate the same biomarker, their covariate distributions may differ, creating a covariate shift that must be addressed to ensure a fair AUC comparison.

\subsection{Estimand and Assumptions}\label{sec:BasicSetup_estimand}

The target estimand is the AUC in the population of interest,
\begin{align*}
    \tau_0=\mathbb{E} \left[\mathbb{I}(Y_i>Y_j) \mid  D_i=1, D_j=0\right],\quad i\neq j,
\end{align*} 
where $i$ and $j$ are independent draws from the target population. This estimand averages pairwise comparisons of biomarker values between responders and non-responders and can be expressed as a U-statistic \citep{Bamber1975}. Identifying $\tau_0$ from the observed validation cohort $\mathcal{V}$ requires the following assumptions.

\begin{assumption}[Positivity]\label{as1}
(i) For all $\bs x$ with positive target-population density $f_{\boldsymbol{X},S}\left(\boldsymbol{X}=\boldsymbol{x}\right)\neq 0$, $\Pr \left(S=1 \mid \boldsymbol{X}=\boldsymbol{x}\right)>0$.
(ii) 
$\mathbb{E}\![\Pr(D=1 \mid \boldsymbol{X}, S=1)\,\{1-\Pr(D=1 \mid \boldsymbol{X}, S=1)\}] > 0.$
\end{assumption}

\begin{assumption}[Mean Exchangeability]\label{as2}
    $\mathbb{E}\left(D\mid \boldsymbol{X},S\right)=\mathbb{E}\left(D\mid \boldsymbol{X}\right)$.
\end{assumption}

\begin{assumption}[Conditional Independence of Biomarker and Source]\label{as3} 
    $Y\independent S\mid (\boldsymbol{X}, D)$
\end{assumption}

Assumption \ref{as1} (i) assumes that any covariate pattern represented in the target population has a nonzero probability of being included in the validation cohort. This implies that every participant in the target population has a positive probability of being selected or included in the study sample. Assumption \ref{as1} (ii) ensures that there is non-degenerate variability in response so that the biomarker AUC is estimable. 

Given Assumption \ref{as2}, the expected responses are comparable regardless of source once covariates are controlled for; specifically, for a binary outcome, 

\[\Pr(D=1 \mid \boldsymbol{X}=\boldsymbol{x}, S=1) = \Pr(D=1 \mid \boldsymbol{X}=\boldsymbol{x}).\]

Intuitively, Assumption \ref{as3} further implies that conditional on covariates $\boldsymbol{X}$ and outcome status $D$, biomarker values $Y$ do not differ systematically between the validation cohort and the broader target population. Together, these conditions enable extrapolation of biomarker distribution from the validation cohort to the target population, which is essential for identifying AUC estimators such as sampling-weighting-based and outcome-modeling-based estimators introduced in Section~\ref{sec:method}.

\section{Methods}\label{sec:method}
\subsection{Inverse Probability of Sampling Weighting}\label{sec:method_IPSW}

The inverse probability of sampling weighting (IPSW) approach is based on inverse probability weighting, which balances the covariate distribution by upweighting subjects who are less likely to be sampled \citep{Rosenbaum1983, Li2017}. Intuitively, subjects who are unlikely to be included in the validation cohort receive larger weights, thereby correcting under-representation of the covariate pattern. Following the ideas of IPSW for average treatment effect (ATE), we derive its counterpart for evaluating a biomarker AUC.

Under Assumptions~\ref{as1} - \ref{as3}, the AUC of the target population can be identified by $\mathcal{V}$ as
\begin{align}\label{eq:ipsw}
    \tau_0=\frac{\mathbb{E} \left\{w^{\rm ipsw}(\boldsymbol{X}_i,\boldsymbol{X}_j)\mathbb{I}(Y_i>Y_j, D_i=1, D_j=0) \mid S_i=1, S_j=1\right\}}{\mathbb{E} \left\{w^{\rm ipsw}(\boldsymbol{X}_i,\boldsymbol{X}_j)\mathbb{I}(D_i=1, D_j=0)\mid S_i=1, S_j=1\right\}},
\end{align}
where $i$ and $j$ are independent observations from $\mathcal{V}$. The IPSW weight is $w^{\rm ipsw}(\boldsymbol{X}_i,\boldsymbol{X}_j)=\left\{\Pr(S_i=1\mid \boldsymbol{X}_i) \Pr(S_j=1\mid \boldsymbol{X}_j)\right\}^{-1}$, with $\Pr(S_i=1\mid \boldsymbol{X}_i)=\pi(\boldsymbol{X}_i)$ given by a sampling model ($S\sim \boldsymbol{X}$) indexed by $\balp$. An unbiased estimator of $\tau_0$ is therefore
\begin{align}
    \hat{\tau}_{\text{IPSW}} = \frac{\sum_{i \neq j}^{n} w^{\rm ipsw}(\boldsymbol{X}_i, \boldsymbol{X}_j; \hat{\balp}) \mathbb{I}(Y_i > Y_j, D_i = 1, D_j = 0, S_i = 1, S_j = 1)}{\sum_{i \neq j}^{n} w^{\rm ipsw}(\boldsymbol{X}_i, \boldsymbol{X}_j; \hat{\balp}) \mathbb{I}(D_i = 1, D_j = 0, S_i = 1, S_j = 1)}.
\end{align}

Specifically, $w^{\rm ipsw}(\boldsymbol{X}_i, \boldsymbol{X}_j; \hat{\balp}) = \left\{{\pi(\boldsymbol{X}_i;\hat{\balp}) \pi(\boldsymbol{X}_j;\hat{\balp})}\right\}^{-1}$,
where $\pi(\boldsymbol{X}_i;\hat{\balp})$ is the estimated sampling probability for subject $i$, indicating the estimated probability of being in $\mathcal{V}$ given $\boldsymbol{X}_i$. In practice, this sampling score can be flexibly obtained from any suitable model of interest. The large-sample properties of $\hat{\tau}_{\text{IPSW}}$, including consistency and asymptotic normality, are established in Section \ref{app.b} and Section \ref{app.c}.

Although IPSW is theoretically identifiable and provides a principled way to handle covariate shift, its practical use is limited. When only $\mathcal{V}$ is observed, all subjects have $S=1(\tilde S=0)$, leaving no information about the $S=0$ population needed to estimate the sampling model. Real-world data $\mathcal{R}$ can partially approximate this population by supplying covariate information, but $\mathcal{V}$ and $\mathcal{R}$ are not truly complementary as some individuals may be missing from both sources. So, treating $\mathcal{V}$ as $S=1$ and $\mathcal{R}$ as $S=0$ may mis-specify the sampling model. A more serious limitation is that patient-level $\mathcal{R}$ is often unavailable, and summary-level covariate information alone cannot support the sampling model. However, IPSW remains conceptually valuable for illustrating how sampling models can be used to target the population AUC.

\subsection{Calibration Weighting}\label{sec:method_CW}
To address the practical limitations of the IPSW approach in the nested, biased sampling setting, we next propose a calibration weighting (CW) method to correct sampling bias in AUC estimation for the target population. The key idea is to calibrate the covariate distribution of the validation cohort to that of the target population by assigning individual weights so that the weighted covariates in the validation cohort match the target population's distribution. Unlike IPSW, CW does not require estimating a sampling model and can be implemented using only summary statistics of the target population, meaning that individual-level data outside the validation cohort are not necessary. These features make CW more broadly applicable. For example, when patient-level data cannot be disclosed but covariate summary statistics are available, CW remains feasible, whereas IPSW cannot be applied. In addition, IPSW depends on the correct specification of the sampling model, whereas CW's performance is not compromised by modeling assumptions. Given covariate summary statistics from the target population (the anchor population), CW can be implemented using entropy balancing to obtain the calibration weights \citep{Hainmueller2012}. We use entropy balancing because it achieves exact balance on the moments of specified covariate while keeping the weights close to uniform, which provides a stable and interpretable calibration procedure.

The property of CW is that for any vector-valued function $\bg(\boldsymbol{X})$ to be calibrated, we have 
\begin{align*}
    \mathbb{E}\left[\frac{\pi(\boldsymbol{X})^{-1} S}{\mathbb{E}\{\pi(\boldsymbol{X})^{-1} S\}} \bg(\boldsymbol{X})\right] 
    = \mathbb{E}\{\bg(\boldsymbol{X})\},
\end{align*}
where $\pi(\boldsymbol{X})^{-1} = \{ \Pr(S=1\mid \boldsymbol{X})\}^{-1}$ is the inverse sampling probability defined in Section~\ref{sec:method_IPSW}. $\bg(\cdot)$ shapes how the covariates are calibrated. Usually, first and second moments are sufficient for balance, but $\bg(\cdot)$ can include higher-order terms or transformations \citep{Hainmueller2012}.

For each subject $i$ in $\mathcal{V}$, a CW weight $q_i$ is chosen to satisfy the calibration constraint $\sum_{i=1}^{n} q_i \bg(\boldsymbol{X}_i) = \tilg$, where $\tilg$ is an estimate of $\mathbb{E}\{\bg(\boldsymbol{X})\}$ in the target population. This target moment vector may be obtained from published summary statistics or from a representative dataset $\mathcal{R}$. If $\mathcal{R}$ includes design weights (survey weights) $d_i$, then $\tilg=\sum_{i=1}^{N}\tilS_id_i\bg(\boldsymbol{X}_i)/\sum_{i=1}^{N}\tilS_id_i$ with $N$ denoting the (possibly unknown) target population size. Under equal selection probabilities ($d_i=1$), this reduces to the empirical mean of $\{\bg(\boldsymbol{X}_i):i\in\mathcal{R}\}$.
The CW weights are obtained by solving the entropy-balancing problem
\begin{align}\label{eq:cw_obj}
    \min\left\{\sum_{i=1}^n q_i\log q_i\right\},
\end{align}
subject to $q_i \geq 0$, $\sum_{i=1}^n q_i = 1$, and calibration constraint. \eqref{eq:cw_obj} minimizes the negative entropy of the weights, keeping them close to uniform and avoiding instability from extreme values \citep{Owen2001,Lee2023}. Introducing a Lagrange multiplier $\blam$, the objective becomes 
\begin{align}\label{eq:cw_lagrangeObj}
    L(\blam,q) = \sum_{i=1}^n q_i \log q_i - \blam^{T}\!\left(\sum_{i=1}^n q_i \bg(\boldsymbol{X}_i) - \tilg\right).
\end{align}

Letting $\hat{\blam}$ solve $U(\blam) = \sum_{i=1}^n e^{\blam^{T} \bg(\boldsymbol{X}_i)} \{\bg(\boldsymbol{X}_i) - \tilg\} = 0$, the resulting weights are
\begin{align*}
\hat{q}_i=q(\boldsymbol{X}_i;\hat{\blam})=\frac{\text{exp}\{{\hat{\blam}^\top \bg(\boldsymbol{X}_i)}\}}{\sum_{i=1}^n \text{exp}\{{\hat{\blam}^\top \bg(\boldsymbol{X}_i)}\}}.
\end{align*}
The CW estimator of the target population AUC is
\begin{align}
    \hat{\tau}_{\text{CW}} 
    = \frac{\sum_{i \neq j} \hat{w}^{\text{cw}}(\boldsymbol{X}_i, \boldsymbol{X}_j)\, \mathbb{I}(Y_i > Y_j, D_i = 1, D_j = 0, S_i = 1, S_j = 1)}
           {\sum_{i \neq j} \hat{w}^{\text{cw}}(\boldsymbol{X}_i, \boldsymbol{X}_j)\, \mathbb{I}(D_i = 1, D_j = 0, S_i = 1, S_j = 1)},
\end{align}
where $\hat{w}^{\text{cw}}(\boldsymbol{X}_i, \boldsymbol{X}_j) = \hat{q}_i \hat{q}_j$ are the pairwise CW weights for observations $i$ and $j$. 

\begin{assumption}\label{as:cw_loglinear}
The sampling score for inclusion in the validation cohort follows a log-linear model, i.e., $\pi(\boldsymbol{X})=\exp\{\balp_0^\top\bg(\boldsymbol{X})\}$ for some $\balp_0$ and a chosen basis function $\bg(\boldsymbol{X})$.
\end{assumption}

\begin{theorem}\label{thm:consistency_cw}
Under Assumptions~\ref{as1}-\ref{as:cw_loglinear}, the CW estimator $\hat{\tau}_{\rm{CW}}$ is consistent for $\tau_0$. 
\end{theorem}
The derivation of $\hat{\tau}_{\text{CW}}$ is provided in Section \ref{app.a2}, and the proof of Theorem~\ref{thm:consistency_cw} appears in Section \ref{app.consistency_cw}. Assumption~\ref{as:cw_loglinear} provides a sufficient condition for the consistency of $\hat{\tau}_{\text{CW}}$, but it is not necessary in general. In practice, the moment-matching construction of CW can still provide robust performance when the log-linear sampling model is only an approximation, consistent with the robustness properties of calibration weighting discussed previously \citep{Zhao2017}. Assumption~\ref{as:cw_loglinear} also establishes a direct connection between CW weights, $\hat{q}_i = q(\boldsymbol{X}_i;\hat{\blam})$, and IPSW weights, $\hat{p}_i = 1/\pi(\boldsymbol{X}_i;\hat{\balp})$. In particular, we show that $\hat{q}_i(\boldsymbol{X}_i)\xrightarrow{P} \frac{1}{N\pi(\boldsymbol{X}_i;\balp_0)}$ in Section \ref{app.consistency_cw}, implying that entropy-balancing weights in \eqref{eq:cw_obj} are asymptotically equivalent in form to IPSW weights. Similar equivalence results hold for other entropy-based objectives under logistic sampling models \citep{zhao2019covariate, Josey2021}.
Moreover, when $\mathcal{V}$ represents only a small fraction of the target population, the log-linear specification in Assumption~\ref{as:cw_loglinear} closely approximates a logistic model. Also, prior studies have shown that CW is robust to the choice of objective function \citep{Lee2023}, suggesting that its performance is not sensitive to whether a log-linear or logistic form is used in the optimization. The asymptotic normality of $\hat{\tau}_{\text{CW}}$ is shown in Section \ref{app.asyNorm_cw}.

CW addresses covariate shift by matching target moments, which remains feasible even when only summary-level covariate information is available. In practice, calibrating first and second moments is often sufficient, although richer choices of $\bg(\boldsymbol{X})$ may further improve balance \citep{Lee2023}. Unlike IPSW, CW does not rely on a fully specified likelihood-based sampling model; instead, it achieves adjustment through balancing functions of the covariates between the validation cohort and the target population. This makes CW less sensitive to misspecification of the sampling model and connects it to the broader covariate-balancing literature. Moreover, the robustness of CW can be understood in a doubly robust sense with respect to the working balance basis $\bg(\boldsymbol{X})$: consistency is retained if either the sampling model is log-linear in $\bg(\boldsymbol{X})$ or the relevant conditional outcome contrast underlying the target AUC is linear in $\bg(\boldsymbol{X})$ \citep{Zhao2017,Josey2021}. \ref{sec:CWvsIPSW} compares CW and IPSW in greater detail.

\subsection{Outcome Modeling with RWD}\label{sec:method_OMRWD}
Although CW remains feasible when only summary statistics are available, it necessarily uses less information than estimators that incorporate patient-level data. When individual-level $\mathcal{R}$ is accessible, we consider a complementary strategy based on outcome modeling. Motivated by the G-computation framework \citep{Robins1986}, we propose an Outcome Modeling with RWD (OM+RWD) estimator for evaluating the target-population AUC. This approach leverages richer information in patient-level RWD by modeling the biomarker $Y$ conditional on $\boldsymbol{X}$ and $D$, which is often more efficient and less prone to extreme variability than IPSW.

Because the estimand of interest is the biomarker AUC, which evaluates the discriminatory accuracy of $Y$, we treat biomarker $Y$ as the dependent variable. In this biomarker outcome model, the response $D$ serves as a group indicator, analogous to a ``treatment" in causal inference, while the biomarker $Y$ is the ``outcome." This perspective aligns naturally with the precision medicine context, where the goal is to characterize how well a biomarker separates responder and non-responder groups. Formally, we specify an outcome model for the biomarker as $\mathbb{E}(Y \mid \boldsymbol{X},D) = m(\boldsymbol{X},D;\bbeta)$, where $m(\cdot)$ is a flexible function indexed by a finite-dimensional parameter $\bbeta$. The model does not require a particular link or parametric structure and can be fitted separately within the $D=1$ and $D=0$ groups, yielding parameter sets $\bbeta_0$ and $\bbeta_1$, respectively. For simplicity, we use $\bbeta$ to denote the collection of group-specific parameters. These predicted outcomes are then used to estimate the target-population AUC.

Assume $Y_i \mid \boldsymbol{X}_i, D_i \sim \mathcal{G}(\boldsymbol{X}_i, D_i; \bbeta)$ independently across subjects, 
where $\mathcal{G}(\boldsymbol{X}_i, D_i; \bbeta)$ denotes any distribution with mean $M(\boldsymbol{X}_i,D_i;\bbeta)$ and variance $\sigma_1^2 D_i + \sigma_0^2 (1-D_i)$. The parameter $\bbeta$ is estimated from the outcome model fitted to $\mathcal{V} = \{(\boldsymbol{X}_i, Y_i, D_i, S_i=1, \tilde S=0): i=1,\dots,n\}$.
The form of $\mathcal{G}$ is intentionally flexible, which only requires an existing cumulative distribution function (CDF) $\mathbb{F}$ so that pairwise comparisons $\Pr(Y_i > Y_j \mid \boldsymbol{X}_i, \boldsymbol{X}_j, D_i=1, D_j=0)$ are well-defined. Under this mild requirement, OM+RWD accommodates a wide range of outcome models, provided the model predictions exhibit root-$n$ consistency.

Based on the identification result in Section \ref{app.a3}, we define OM+RWD estimator
\begin{align}
\hat{\tau}_\text{OM+RWD}=\frac{\sum_{i\neq j}^{m+n} \mathcal{P}_{ij}(\boldsymbol{C}_i,\boldsymbol{C}_j;\hat\bbeta, \hat{\sigma}_0, \hat{\sigma}_1) I_{ij,0}}{\sum_{i\neq j}^{m+n}  I_{ij,0}},
\end{align}
where $I_{ij,0}=\mathbb{I}(D_i=1,D_j=0,S_i=0,S_j=0)$, $\boldsymbol{C}_i=(\boldsymbol{X}_i, D_i, Y_i, S_i) \in \mathcal{C}$, and $\mathcal{P}_{ij}(\cdot)$ is the estimated probability $\Pr(Y_i > Y_j \mid D_i=1, D_j=0, S_i=0, S_j=0, \boldsymbol{X}_i, \boldsymbol{X}_j)$. For illustration, consider the normal outcome model

\begin{align}\label{eq:normal distribution}
    Y_i \mid \boldsymbol{X}_i, D_i \sim 
    N\bigl(M(\boldsymbol{X}_i,D_i;\bbeta), \, \sigma_1^2 D_i + \sigma_0^2 (1-D_i)\bigr).
\end{align}

Under this specification, $\Pr(Y_i > Y_j \mid \boldsymbol{X}_i, \boldsymbol{X}_j, D_i=1, D_j=0) 
= \Phi\!\left( \frac{M(\boldsymbol{X}_i,1;\bbeta) - M(\boldsymbol{X}_j,0;\bbeta)}{\sqrt{\sigma_0^2 + \sigma_1^2}} \right)$, where $\Phi(\cdot)$ is the standard normal CDF. Thus, the estimator of AUC becomes

\begin{align}\label{eq:omrwd}
\hat{\tau}_{\mathrm{OM+RWD}}=\frac{
\sum_{i\ne j}^{m+n}\Phi\!\left(\frac{M(\boldsymbol{X}_i,1,S_i=0;\hat{\bbeta})-M(\boldsymbol{X}_j,0,S_j=0;\hat{\bbeta})}{\sqrt{\hat{\sigma}_1^2+\hat{\sigma}_0^2}}\right)I_{ij,0}}{\sum_{i\ne j}^{m+n} I_{ij,0}}.
\end{align}
Here, $\hat\beta$ and the variance components $\hat\sigma_d^2$ ($d\in\{0,1\}$) are estimated entirely from $\mathcal{V}$, where $Y$ is observed, while the model is evaluated on covariates $(\boldsymbol{X}_i,D_i)$ from the representative sample $\mathcal{R}$, where $Y$ is unobserved. The variance components are estimated as $\hat{\sigma}_d^2=\frac{1}{n_d - p_d} \sum_{i : D_i = d} \{ Y_i - M(\boldsymbol{X}_i,D_i=d,S_i=1;\hat\bbeta)\}^2$, with $n_d$ subjects and $p_d$ fitted parameters in group $D=d$.

Conceptually, the OM+RWD estimator recovers the target-population AUC by splitting the roles of the two data sources: the validation cohort $\mathcal{V}$ provides the biomarker information needed to fit the outcome model, while the RWD $\mathcal{R}$ supplies the population covariate distribution over which the AUC is specified. In this way, $\mathcal{R}$ contributes the population structure and $\mathcal{V}$ contributes the biomarker signal. When patient-level RWD is available and the outcome model is correctly specified, OM+RWD provides an efficient way to estimate the biomarker AUC under covariate shift. The consistency and asymptotic normality of $\hat{\tau}_{\text{OM+RWD}}$ are shown in Section \ref{app.consistency_omrwd} and Section \ref{app.asyNorm_omrwd}. 

\subsection{Outcome Modeling without RWD}\label{sec:method_OM}
To develop an alternative outcome-modeling estimator that remains feasible when patient-level RWD are unavailable, we infer the target-population AUC by applying calibration weights that align the covariate distribution of $\mathcal{V}$ with that of the target population, which yields the Outcome Modeling (OM) estimator. Unlike OM+RWD, which uses patient-level $\mathcal{R}$ to represent the target population directly, the OM estimator anchors $\mathcal{V}$ to the target population through matching moments. The outcome model fitted in $\mathcal{V}$ supplies the biomarker information, while the CW weights enforce covariate balance. OM requires only target-population covariate summaries and $\tau_0$ remains identifiable from $\mathcal{V}$, as shown in Theorem~\ref{thm:iden_om} in Section \ref{app.a4}.

Let $\boldsymbol{V}_i=(\boldsymbol{X}_i,D_i,Y_i,S_i=1)\in \mathcal{V}$, and $\mathcal{P}_{ij}(\boldsymbol{V}_i, \boldsymbol{V}_j; \hat\bbeta, \hat{\sigma}_0,\hat{\sigma}_1)$ denotes the estimated probability of $Pr(Y_i>Y_j\mid D_i=1,D_j=0,S_i=1,S_j=1,\boldsymbol{X}_i,\boldsymbol{X}_j)$ based on the outcome model with parameters $(\hat\bbeta,\hat\sigma_0,\hat\sigma_1)$ fitted to the validation cohort. The second outcome modeling estimator, OM, is proposed as
\begin{align}
    \hat{\tau}_\text{OM}=\frac{\sum_{i\neq j}^{n}\hat{w}_{ij} \mathcal{P}_{ij}(\boldsymbol{V}_i, \boldsymbol{V}_j; \hat\bbeta, \hat{\sigma}_0,\hat{\sigma}_1) \mathbb{I}(D_i=1,D_j=0,S_i=1,S_j=1)}{\sum_{i\neq j}^{n}\hat{w}_{ij} \mathbb{I}(D_i=1,D_j=0,S_i=1,S_j=1)}.
\end{align}

Following the normal example \eqref{eq:normal distribution}, 
\begin{align*}
    \mathcal{P}_{ij}(\boldsymbol{V}_i, \boldsymbol{V}_j; \hat\bbeta, \hat{\sigma}_0,\hat{\sigma}_1)=\Phi\left(\frac{M(\boldsymbol{X}_i,D_i=1,S_i=1;\hat\bbeta)-M(\boldsymbol{X}_j,D_j=0,S_j=1;\hat\bbeta)}{\sqrt{\hat{\sigma}_1^2+\hat{\sigma}_0^2}}\right).
\end{align*}
The function $M(\boldsymbol{X}_i,D_i,S_i;\bbeta)$ is defined as in OM+RWD, but its estimation differs. In $\hat{\tau}_\text{OM+RWD}$, the model is fitted on $\mathcal{V}$ and applied to subjects in $\mathcal{R}$. In contrast, $\hat{\tau}_\text{OM}$ performs both model fitting and prediction entirely within $\mathcal{V}$. For each responder group $D=d$, we fit an outcome model using $\boldsymbol{V}_i=(\boldsymbol{X}_i, D_i=d, Y_i, S_i=1)\in\mathcal{V}\subset\mathcal{C}$ and then use these group-specific models to predict biomarker means $M(\boldsymbol{X}_i,D_i,S_i=0;\hat\bbeta)$ for all subjects in $\mathcal{V}$. The variance components $\hat\sigma_0^2$ and $\hat\sigma_1^2$ are estimated in the same way as in Section~\ref{sec:method_OMRWD}. 

For OM, the weights $\hat{w}_{ij}$ may be constructed using either IPSW or CW. Because our primary motivation is to accommodate settings where only summary-level target information is available, we adopt CW weights here, $\hat{w}_{ij}=\hat{q}_i(\boldsymbol{X}_i)\hat{q}_j(\boldsymbol{X}_j)$. As shown in Section \ref{app.consistency_cw}, $\hat{q}_i(\boldsymbol{X}_i)\xrightarrow{P} \{N\pi(\boldsymbol{X}_i;\balp_0)\}^{-1}$ as $n\rightarrow \infty$, so CW weights are asymptotically equivalent to IPSW weights. This connection to the sampling score $\pi(\boldsymbol{X}_i;\balp_0)$ ensures the consistency of $\hat{\tau}_{\text{OM}}$ and is also useful when developing the doubly robust estimator later. Consistency and asymptotic normality are established in Section \ref{app.b4} and~Section \ref{app.asyNorm_om}.

\subsection{Augmented CW}\label{sec:method_acw}
Although CW is robust and applicable when only summary-level target covariate information is available, it does not directly exploit patient-level RWD outcome information and may therefore be less efficient than estimators that combine weighting with outcome modeling. To bridge this gap, we introduce an Augmented Calibration Weighting (ACW) estimator that blends the robustness of CW with the efficiency of outcome modeling. CW provides protection against outcome model misspecification through moment-matching, whereas outcome modeling leverages patient-level RWD to improve efficiency. This integration yields a doubly robust estimator that remains valid under misspecification of outcome model and is more efficient than CW alone.

The proposed ACW estimator is constructed as a linear combination of three components,
\begin{align}\label{eq:acw_form}
\hat{\tau}_{\text{ACW}}=\hat{\tau}_{\text{CW}}-\hat{\tau}_{\text{OM}}+\hat{\tau}_{\text{OM+RWD}},
\end{align}
where 
\begin{align*}
    \hat{\tau}_{\text{CW}} = \frac{\sum_{i \neq j}^{n} \hat{w}^{\text{cw}}(\boldsymbol{X}_i, \boldsymbol{X}_j) \mathbb{I}(Y_i > Y_j, D_i = 1, D_j = 0, S_i = 1, S_j = 1)}{\sum_{i \neq j}^{n} \hat{w}^{\text{cw}}(\boldsymbol{X}_i, \boldsymbol{X}_j) \mathbb{I}(D_i = 1, D_j = 0, S_i = 1, S_j = 1)};
\end{align*}
\begin{align*}
    \hat{\tau}_\text{OM}=\frac{\sum_{i\neq j}^{n}\hat{w}_{ij} \mathcal{P}_{ij}(\boldsymbol{V}_i, \boldsymbol{V}_j; \hat\bbeta, \hat{\sigma}_0,\hat{\sigma}_1)\cdot \mathbb{I}(D_i=1,D_j=0,S_i=1,S_j=1)}{\sum_{i\neq j}^{n}\hat{w}_{ij} \mathbb{I}(D_i=1,D_j=0,S_i=1,S_j=1)}
\end{align*}
with $\boldsymbol{V}_i=(\boldsymbol{X}_i,D_i,Y_i,S_i=1)\in \mathcal{V}\subset \mathcal{C}$ and $\hat{w}_{ij}=\hat{w}^{\text{cw}}(\boldsymbol{X}_i, \boldsymbol{X}_j)=\hat{q}_i(\boldsymbol{X}_i) \hat{q}_j(\boldsymbol{X}_j)$; 
\begin{align*}
    \hat{\tau}_\text{OM+RWD}=\frac{\sum_{i\neq j}^{m+n} \mathcal{P}_{ij}(\boldsymbol{C}_i,\boldsymbol{C}_j;\hat\bbeta, \hat{\sigma}_0, \hat{\sigma}_1) \mathbb{I}(D_i = 1, D_j = 0, S_i = 0, S_j = 0)}{\sum_{i\neq j}^{m+n}  \mathbb{I}(D_i = 1, D_j = 0, S_i = 0, S_j = 0)},
\end{align*}
where $\boldsymbol{C}_i=(\boldsymbol{X}_i,D_i,Y_i,S_i)\in \mathcal{C}$, and $\mathcal{P}_{ij}(\boldsymbol{C}_i,\boldsymbol{C}_j;\hat\bbeta, \hat{\sigma}_0, \hat{\sigma}_1)$ needs both $\mathcal{V}$ and $\mathcal{R}$. Each component of $\hat{\tau}_{\text{ACW}}$ is computed following the same procedures described in previous sections.

The double robustness of $\hat{\tau}_{\text{ACW}}$ arises from two complementary cases. First, if the outcome model is correctly specified, Assumptions \ref{as1}–\ref{as3} ensure $\hat{\tau}_{\text{OM+RWD}}\xrightarrow{P}\tau_0$ (Section \ref{app.consistency_omrwd}). In this case, Assumption \ref{as:cw_loglinear} is unnecessary because $\hat{\tau}_{\text{CW}}-\hat{\tau}_{\text{OM}}\xrightarrow{P}0$, and the outcome model alone delivers consistency. Second, if the CW weights $\hat q_i(\boldsymbol{X}_i)$ are consistently estimated (guaranteed under Assumption \ref{as:cw_loglinear}), then the CW estimator is itself consistent; misspecification of the outcome model does not affect $\hat{\tau}_{\text{ACW}}$ because $\hat{\tau}_{\text{OM+RWD}}-\hat{\tau}_{\text{OM}}\xrightarrow{P}0$. Hence, $\hat{\tau}_{\text{ACW}}$ is consistent whenever either the outcome model or the CW weights are correctly specified. A full proof of consistency (double robustness) appears in Section \ref{app.consistency_acw}.

\begin{theorem}\label{thm:asymptotic_acw}
    (Asymptotic normality for ACW estimator): Assume $\{\frac{1}{\sqrt{n}}\sum_{i=1}^nq(\boldsymbol{X}_i;\blam),\blam\in \mathcal{L}\}$ is stochastically equicontinuous, and the regularity conditions in Lemma~\ref{lem3} hold. Let $\hat{\blam}\xrightarrow{P}\blam^*$ and $\hat{\bbeta}\xrightarrow{P}\bbeta^*$. If either $M(\boldsymbol{X},D;\bbeta^*)=\mathbb{E}\left[Y\mid \boldsymbol{X},D\right]$ or $q(\boldsymbol{X}_i;\blam^*)=1/\pi(\boldsymbol{X};\alpha^*)=1/Pr(S=1\mid \boldsymbol{X})$ as $n\rightarrow \infty$, then
    \begin{align*}
    \begin{split}
    \sqrt{n}\left(\hat{\tau}_{\mathrm{ACW}}-\tau_0\right)
    \rightarrow\;
    N\Bigg(
    0,\;
    &\operatorname{Var}\!\left[
    \left\{u_1(\tau_1,\blam^*,\bbeta^*)\right\}^{-1}
    \phi_1(\boldsymbol{V}_i;\tau_1,\blam^*,\bbeta^*)
    \right]
    \\
    &\quad+
    \operatorname{Var}\!\left[
    \left\{u_2(\tau_0,\bbeta^*)\right\}^{-1}
    \phi_2(\boldsymbol{C}_i;\tau_0,\bbeta^*)
    \right]
    \Bigg).
    \end{split}
    \end{align*}
    where $\tau_1$ is the probability limit of $\hat{\tau}_{\rm{ACW1}}=\hat{\tau}_{\rm{CW}}-\hat{\tau}_{\rm{OM}}$, which is typically close to zero and treated as fixed, the variance decomposition is $u_1(\tau_1,\blam,\bbeta)=\mathbb{E}\left[d^{\rm{acw1}}_0(\boldsymbol{V}_i,\boldsymbol{V}_j;{\blam})\right]=\mathbb{E}\left[q(\boldsymbol{X}_i;\blam)q(\boldsymbol{X}_j;\blam)\mathbb{I}(D_i=1,D_j=0,S_i=1,S_j=1)\right]$ and
    \begin{align*}
\phi_1(\boldsymbol{V}_i;\tau_1,\blam,\bbeta)
&= \frac{\partial U^{\rm{acw1}}(\tau_1,\blam,\bbeta)}{\partial\blam^\top}
   \left[\mathbb{E}\left\{\frac{\partial h_1(\boldsymbol{V}_i;\blam)}{\partial\blam^\top}\right\}\right]^{-1}
   h_1(\boldsymbol{V}_i;\blam) \\
&\quad + \frac{\partial U^{\rm{acw1}}(\tau_1,\blam,\bbeta)}{\partial\bbeta^\top}
   \left[\mathbb{E}\left\{\frac{\partial h_2(\boldsymbol{V}_i;\bbeta)}{\partial\bbeta^\top}\right\}\right]^{-1}
   h_2(\boldsymbol{V}_i;\bbeta)- l^*_{\rm{ACW1}}(\boldsymbol{V}_i;\tau_1,\blam,\bbeta),
\end{align*}
with $l^*_{\mathrm{ACW1}}(\boldsymbol{V}_i;\tau_1,\blam,\bbeta)=\mathbb{E}\left\{l_\mathrm{ACW1}(\boldsymbol{V}_i,\boldsymbol{V}_j;\tau_1,\blam,\bbeta)\mid\boldsymbol{V}_i\right\}+\mathbb{E}\left\{l_\mathrm{ACW1}(\boldsymbol{V}_j,\boldsymbol{V}_i;\tau_1,\blam,\bbeta)\mid\boldsymbol{V}_i\right\}$.
Additionally, $u_2(\tau,\bbeta)=\mathbb{E}\left[d^{\rm{om+rwd}}_0(\boldsymbol{C}_i,\boldsymbol{C}_j;\tau,\bbeta)\right] = \mathbb{E}\left[\mathbb{I}(D_i=1, D_j=0, S_i=0, S_j=0)\right]$, 
\begin{align*}
\phi_2(\boldsymbol{C}_i;\tau,\bbeta)=\frac{\partial U^{\rm{acw2}}(\tau,\bbeta)}{\partial\bbeta^\top}\left[\mathbb{E}\left\{\frac{\partial h_2(\boldsymbol{C}_i;\bbeta)}{\partial\bbeta^\top}\right\}\right]^{-1}h_2(\boldsymbol{C}_i;\bbeta)-l^*_{\rm{OM+RWD}}(\boldsymbol{C}_i;\tau,\bbeta),
\end{align*}
and 
$l^*_{\mathrm{OM+RWD}}(\boldsymbol{C}_i;\tau,\bbeta)=\mathbb{E}\left\{l_{\mathrm{OM+RWD}}(\boldsymbol{C}_i,\boldsymbol{C}_j;\tau,\bbeta)\mid \boldsymbol{C}_i\right\}+\mathbb{E}\left\{l_{\mathrm{OM+RWD}}(\boldsymbol{C}_j,\boldsymbol{C}_i;\tau,\bbeta)\mid \boldsymbol{C}_i\right\}.$

\end{theorem}
Theorem \ref{thm:asymptotic_acw} establishes that $\hat{\tau}_{\text{ACW}}$ is asymptotically normal under standard conditions, with proof in Section \ref{app.asyNorm_acw}. The variance naturally splits into calibration weighting and OM+RWD components, which highlights how the estimator retains validity as long as either component is correctly specified. 

The relationship between CW and ACW is therefore analogous to that between IPSW and AIPSW, but with a different weighting foundation. CW is the baseline calibration-weighting estimator and is already robust to moderate misspecification of the sampling mechanism so long as the chosen balancing functions $\bg(\boldsymbol{X})$ adequately capture the induced covariate imbalance. ACW is built on top of CW and is intended primarily to improve efficiency by incorporating the outcome model, while retaining the robustness inherited from the calibration-weighting component. In this way, ACW serves as a practical compromise between robustness and efficiency. It preserves the appeal of CW in summary-data settings while improving precision whenever patient-level RWD can be used effectively.

\subsection{Augmented IPSW}\label{sec:method_aipsw}
The validity of $\hat{\tau}_{\text{IPSW}}$ depends heavily on a correct sampling model, which is often difficult to verify and guarantee in practice. To mitigate this vulnerability, and in parallel with the construction of $\hat{\tau}_{\text{ACW}}$, we introduce an augmented IPSW (AIPSW) estimator. By incorporating outcome‐model estimators, the AIPSW achieves double robustness, providing valid inference for $\tau_0$ when either the sampling model or the outcome model is correctly specified.

The proposed AIPSW estimator is constructed as follows:
\begin{align}
\hat{\tau}_{\text{AIPSW}}=\hat{\tau}_{\text{IPSW}}-\hat{\tau}_{\text{OM}}+\hat{\tau}_{\text{OM+RWD}},
\end{align}
with
\begin{align*}
\hat{\tau}_{\text{IPSW}}= \frac{\sum_{i \neq j}^{n} w^{\text{ipsw}}(\boldsymbol{X}_i, \boldsymbol{X}_j;\hat{\balp}) \mathbb{I}(Y_i > Y_j, D_i = 1, D_j = 0, S_i = 1, S_j = 1)}{\sum_{i \neq j}^{n} w^{\text{ipsw}}(\boldsymbol{X}_i, \boldsymbol{X}_j;\hat{\balp}) \mathbb{I}(D_i = 1, D_j = 0, S_i = 1, S_j = 1)};
\end{align*}
\begin{align*}
    \hat{\tau}_\text{OM}=\frac{\sum_{i\neq j}^{n}\hat{w}_{ij} \mathcal{P}_{ij}(\boldsymbol{V}_i, \boldsymbol{V}_j; \hat\bbeta, \hat{\sigma}_0,\hat{\sigma}_1) \mathbb{I}(D_i=1,D_j=0,S_i=1,S_j=1)}{\sum_{i\neq j}^{n}\hat{w}_{ij} \mathbb{I}(D_i=1,D_j=0,S_i=1,S_j=1)},
\end{align*}
where $\boldsymbol{V}_i=(\boldsymbol{X}_i,D_i,Y_i,S_i=1)\in\mathcal{V}$ and here $\hat{w}_{ij}=w^{\text{ipsw}}(\boldsymbol{X}_i,\boldsymbol{X}_j;\hat{\balp})$; 
\begin{align*}
    \hat{\tau}_\text{OM+RWD}=\frac{\sum_{i\neq j}^{m+n} \mathcal{P}_{ij}(\boldsymbol{C}_i,\boldsymbol{C}_j;\hat\bbeta, \hat{\sigma}_0, \hat{\sigma}_1) \mathbb{I}(D_i = 1, D_j = 0, S_i = 0, S_j = 0)}{\sum_{i\neq j}^{m+n}  \mathbb{I}(D_i = 1, D_j = 0, S_i = 0, S_j = 0)},
\end{align*}
where $\boldsymbol{C}_i=(\boldsymbol{X}_i,D_i,Y_i,S_i)\in\mathcal{C}$. 
All three components are computed exactly as in the earlier estimators.
Compared to the ACW estimator, the key difference in $\hat{\tau}_{\text{AIPSW}}$ lies in the weighting. In the OM component, $\hat{w}_{ij}$ is derived from the IPSW weights rather than CW weights. This construction preserves the structure of augmentation while rooting the method in the IPSW framework. The double robustness property of $\hat{\tau}_{\text{AIPSW}}$, along with its asymptotic normality, is established in Section \ref{app.consistency_aipsw} and Section \ref{app.asyNorm_aipsw}.

\subsection{Practical guidance for proposed estimators}
The proposed estimators differ in their data requirements and modeling assumptions. While all are developed for population-anchored AUC inference under covariate shift, they apply to different levels of target-population information. The IPSW and OM+RWD estimators can be viewed as adaptations of existing weighting and outcome-modeling ideas to the biomarker AUC setting, whereas the calibration-weighting-based estimators (CW and OM) and augmented estimators (ACW and AIPSW) are newly developed within the U-statistic framework, with corresponding identification and asymptotic results.
\begin{table}[htbp]
\centering
\small
\renewcommand{\arraystretch}{0.7}
\setlength{\tabcolsep}{3pt}
\caption{Data requirements from the representative/RWD dataset $\mathcal{R}$}
\label{tab:methodsRequirements}
\begin{tabular*}{\textwidth}{@{} @{\extracolsep{\fill}} lccc @{}}
\toprule
\textbf{Estimator} 
& \textbf{Individual-level covariates} 
& \textbf{Individual-level response} 
& \textbf{Summary-level covariates} \\
& $\boldsymbol{X}$ 
& $D$ 
& $\boldsymbol{X}$ \\
\midrule
IPSW   & \cmark & \xmark & N/A \\
OM+RWD & \cmark & \cmark & N/A \\
AIPSW  & \cmark & \cmark & N/A \\
ACW    & \cmark & \cmark & \cmark \\
CW     & \xmark & \xmark & \cmark \\
OM     & \xmark & \xmark & \cmark \\
\bottomrule
\end{tabular*}
\end{table}

\begin{figure}[ht]
    \centering
    \includegraphics[width=\linewidth]{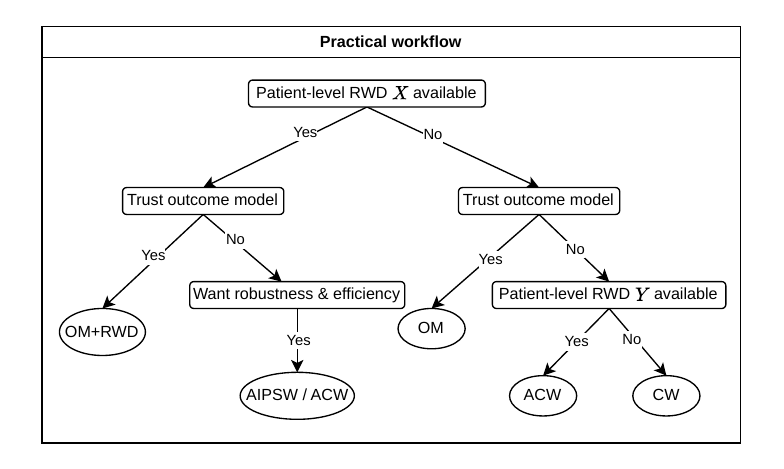}
    \caption{Decision workflow for selecting appropriate estimators based on data availability and modeling considerations}
    \label{fig:practical_workflow}
\end{figure}

Table \ref{tab:methodsRequirements} summarizes the required information from the representative or RWD dataset $\mathcal{R}$. IPSW, OM+RWD, and AIPSW require patient-level covariates, and OM+RWD and AIPSW additionally require patient-level responses. In contrast, CW and OM can be implemented using only summary-level covariate information, making them suitable when patient-level target data are unavailable. ACW integrates outcome modeling with calibration weighting: its calibration component can use summary-level covariate moments, while its augmentation term uses patient-level data to improve efficiency. Figure~\ref{fig:practical_workflow} provides a decision-oriented workflow for choosing among these estimators based on data availability, model trustworthiness, and the desired balance between robustness and efficiency.

\section{Simulation}\label{sec:simulation}

\subsection{Setup}\label{sec:simulation_setup}
To evaluate performance in estimating the target‐population AUC, we conduct simulation studies both with and without access to RWD and under varying degrees of covariate shift. In each iteration, we first generate a large finite sample $(N = 50,000)$ to approximate the finite validation population in Figure \ref{fig:data_structure1}, which in turn approximates the target population. A covariate-dependent sampling mechanism is then used to draw the observed validation cohort $\mathcal{V}$ $(n=800)$. When RWD $\mathcal{R}$ is available, an additional dataset $(m=8000)$ is independently generated from the same population and treated as representative of the target population. We generate $p=3$ baseline covariates $\boldsymbol{X}$ independently from normal and uniform distributions. The binary response $D$ depends on $\boldsymbol{X}$, and the biomarker $Y$ depends on both $\boldsymbol{X}$ and $D$. Covariate shift is induced through the sampling indicator $S_i$, drawn from the logistic model $\text{logit}\{\Pr(S_i=1 \mid \boldsymbol{X}_i)\} 
= \alpha_0 + \alpha_1 X_{i1}^2 + \alpha_2 X_{i2}^2 + \alpha_3 X_{i1}X_{i3}$, where $\balp=(\alpha_0,\alpha_1,\alpha_2,\alpha_3)$ controls covariate shift degree between $\mathcal{V}$ and target population. 

To evaluate estimator performance across a broad range of settings, we consider four model-specification scenarios involving both sampling and outcome models. As summarized in Table~\ref{tab:scena_covshift}, we consider three degrees of covariate shift: (i) no shift, where the validation cohort is a simple random sample and representative of the target population; (ii) moderate shift; and (iii) severe shift. In scenarios (ii) and (iii), the validation cohort is a biased sample and cannot recover $\tau_0$ without properly correcting for covariate shift. To examine the sensitivity of CW-based estimators to the choice of calibration basis functions, we evaluate two specifications: (i) moment balance using first and second moments, $\bg_1(\boldsymbol{X}_i)=\left[X_{i1}, X_{i2}, X_{i3}, X_{i1}^2, X_{i2}^2, X_{i3}^2\right]$, and (ii) moment + interaction balance, which augments $\bg_1(\cdot)$ with pairwise interactions, $\bg_2(\boldsymbol{X}_i)=\bg_1(\boldsymbol{X}_i)\cup\left[X_{i1}X_{i2}, X_{i1}X_{i3}, X_{i2}X_{i3}\right]$. For each scenario, we replicate the simulation for 1000 times, and perform 200 bootstrap resamples to estimate the SE of each method. Although analytic variances are derived, we use the bootstrap as a practical finite-sample SE estimator because sandwich estimation can be sensitive for U-statistics with estimated nuisance functions. Additional simulation details are provided in Section \ref{sec:SimulationDetails_setup}.

\subsection{Simulation Results}\label{sec:simulation_results}
\begin{figure*}[ht]
    \centering
    \includegraphics[width=\linewidth]{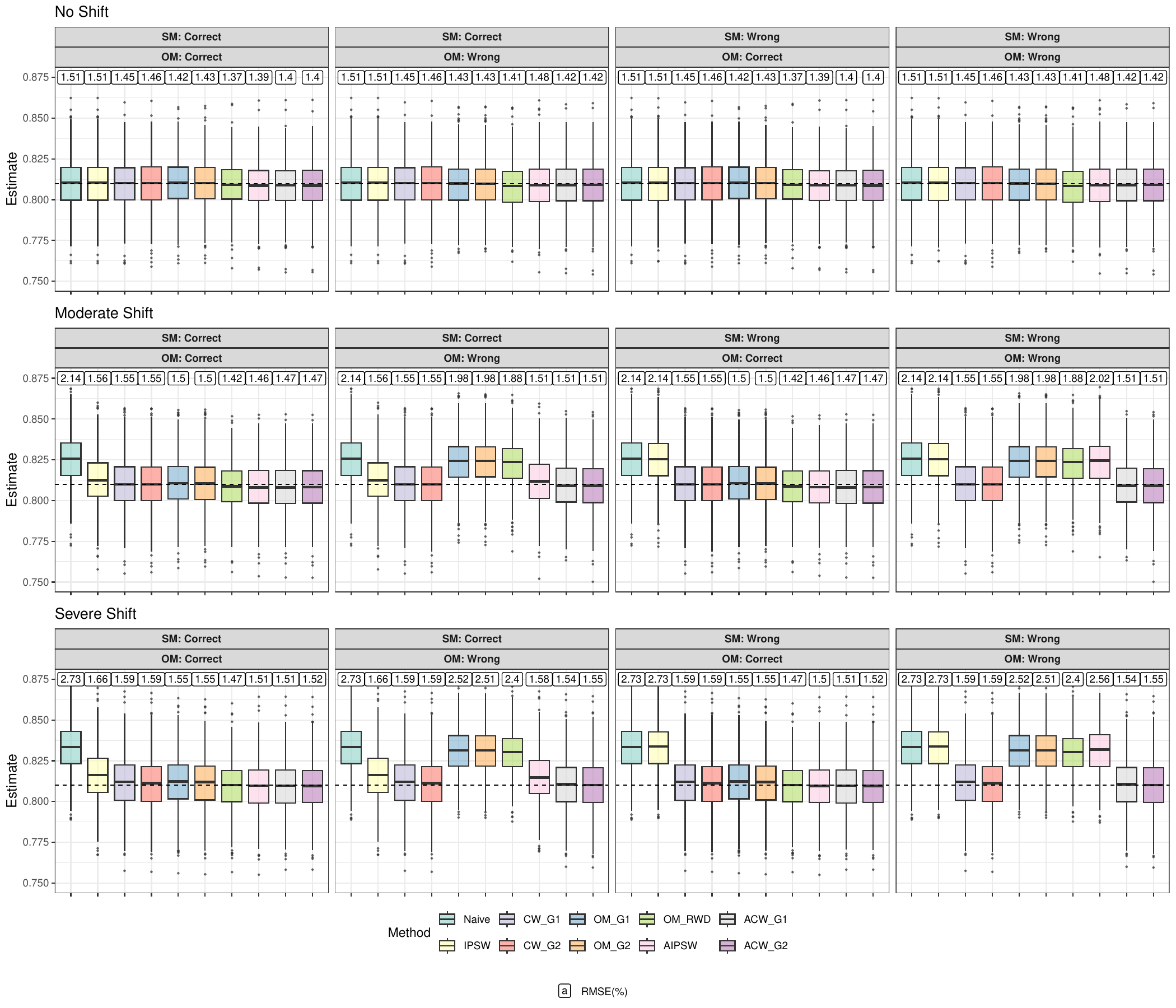}
    \caption{Simulation results across different degrees of covariate shifts and model specifications}
    \label{fig:sim_result}
\end{figure*}

Figure \ref{fig:sim_result} summarizes the performance of all estimators across degrees of covariate shift and four model-specification scenarios. The Na\"ive estimator is the empirical AUC from the validation cohort, used directly as an estimate of the target-population AUC without covariate adjustment. When the validation cohort is representative of the target population, all methods are approximately unbiased for estimating $\tau_0$. Once covariate shift is present, however, the Na\"ive estimator targets the validation-cohort AUC rather than the desired target-population AUC, leading to noticeable bias under moderate and severe shift.

The results also reflect the expected dependence on working-model specification. OM and OM+RWD are biased when the biomarker outcome model is misspecified, whereas IPSW is biased when the sampling model is misspecified. In contrast, the augmented estimators, ACW and AIPSW, remain unbiased when their double-robustness conditions hold. More specifically, AIPSW is unbiased when either the sampling model or the outcome model is correctly specified. CW and ACW do not rely on a fully specified likelihood-based sampling model; instead, their robustness comes from balancing functions of the covariates $\bg(\boldsymbol{X})$ through calibration constraints. This distinction is closely related to the covariate-balancing and entropy-balancing literature, where calibration weights can retain favorable robustness properties even when the sampling model is not correctly parameterized \citep{Zhao2017,Josey2021}. 

When both the outcome and sampling models are misspecified, IPSW, OM, OM+RWD, and AIPSW generally deviate from the true AUC ($\tau_0=0.81$), whereas CW and ACW remain close to the truth in our simulations. This does not imply universal robustness of CW-based estimators to any arbitrary misspecification. In particular, the sampling mechanism considered here is not highly complex, so balancing the first- and second-order moments is sufficient to capture the main covariate imbalance and recover the target-population AUC reasonably well. This highlights an important strength of CW: even when the sampling model is not correctly specified, calibration can still perform well as long as the chosen basis $\bg(\boldsymbol{X})$ captures the key differences between the validation and target populations. ACW further builds on this advantage by incorporating outcome-model information. Although more complex nonlinear sampling mechanisms may require richer calibration bases, CW- and ACW-based estimators remain especially useful in practice because they can be implemented with only summary-level target covariate information.

\begin{figure*}[ht]
    \centering
    \includegraphics[width=0.9\linewidth]{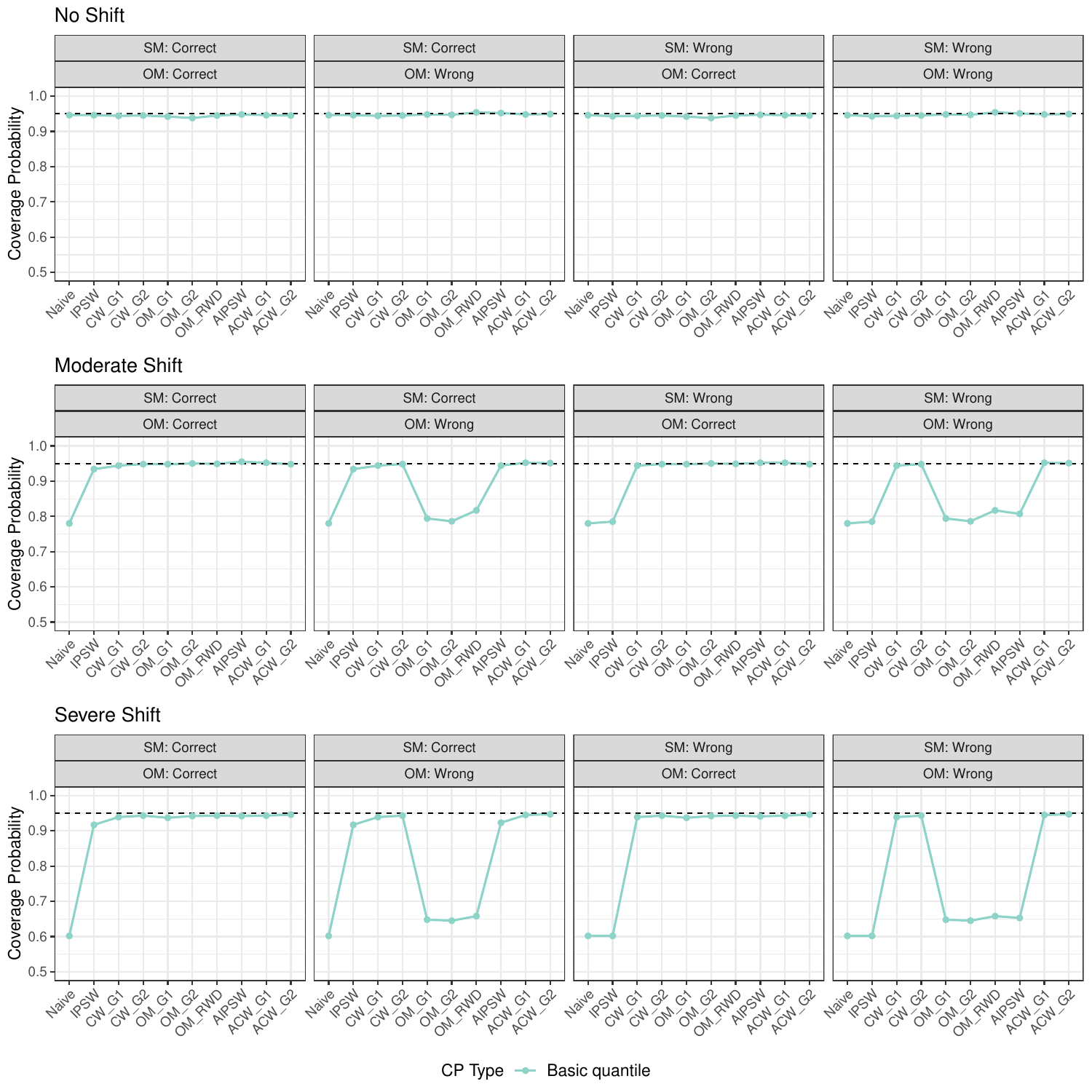}
    \caption{Coverage probability for different degrees of covariate shifts and model specifications}
    \label{fig:sim_cp}
\end{figure*}

ACW improves efficiency over CW by incorporating outcome-model information, with smaller variability and lower RMSE in several settings; similarly, AIPSW improves over IPSW. When the outcome model is correctly specified, OM+RWD achieves the lowest RMSE because it directly leverages patient-level information from both the validation cohort and RWD. CW and ACW generally have slightly lower RMSE than IPSW and AIPSW. Across scenarios, the patterns under moderate and severe shift are similar, with bias magnified as covariate shift becomes more extreme.
The choice between $g_1$ and $g_2$ has minimal impact, indicating that the CW-based estimators are not highly sensitive to modest enrichment of the calibration basis in the settings considered here, which aligns the previous conclusion \citep{Hainmueller2012}. This is practically appealing because first- and second-moment summaries (e.g. mean and variance) are among the most commonly available forms of external covariate information. In more complex settings, such as multimodal covariate distributions or high-dimensional covariates with strong nonlinear interactions, richer choices of $\bg(\boldsymbol{X})$ may be needed \citep{Lee2023}, and the performance advantage of CW/ACW may diminish toward that of IPSW/AIPSW if the relevant imbalance cannot be adequately represented by the selected calibration basis.

The coverage probability (CP) results in Figure \ref{fig:sim_cp} mirror the estimation patterns observed in Figure \ref{fig:sim_result}. With no covariate shift, all methods achieve approximately nominal 95\% coverage. Under moderate and severe shift, the Na\"ive estimator shows substantial under-coverage, whereas the calibration-weighting estimators, CW and ACW, maintain CP close to 0.95 across model-specification settings. IPSW and AIPSW show broadly similar patterns, although their CP is slightly lower than that of the corresponding CW-based estimators in some settings. The lower CP observed for OM, particularly under severe shift with a misspecified outcome model, reflects the sensitivity of OM-only estimators to outcome-model misspecification.

\section{Real Data Application}\label{sec:RealData}
In this section, we illustrate the two primary scientific goals of our framework using a common data setting. Case Study 1 focuses on generalization, where the goal is to estimate the AUC of a biomarker in a prespecified target population. Case Study 2 extends this idea to benchmarking, in which biomarker performance is compared across studies after aligning them to common target population. Together, these two case studies demonstrate how a single inferential framework supports both population-specific estimation and fair cross-study comparison.

We apply the proposed estimators to the two {Prevention and treatment Of muscle Wasting in patients with cancER} (POWER) trials to assess how well baseline stair-climb power (SCP), as a prognostic biomarker, predicts 6-month survival among patients with advanced non-small-cell lung cancer (NSCLC) receiving chemotherapy. POWER I and POWER II were identically designed, international phase III randomized trials investigating enobosarm for the prevention of cancer-related muscle loss \citep{crawford2016}. The two trials differed only in their chemotherapy backbones: \textit{platinum + taxane} in POWER I and \textit{platinum + non-taxane} in POWER II. Although both trials provide strong internal validity for their enrolled populations, it remains unclear whether the observed prognostic performance of baseline SCP would generalize to a broader target population. This motivates our \textbf{Case Study 1} in Section \ref{sec:CaseStudy_c1}, where we estimate the AUC of baseline SCP among trial-eligible NSCLC patients across the United States, a broader and clinically relevant population. By doing so, we assess whether the discrimination observed in the POWER trials is retained in the target population, which may help guide the use of baseline SCP as a prognostic biomarker in future trials.

Another motivating application arises when a biomarker is evaluated in two similar trials or laboratory settings, and researchers want to understand whether its prognostic performance is consistent across settings. For example, if baseline SCP shows different AUCs in two related trials, a natural question is whether this difference reflects true variation in biomarker performance or simply differences in the underlying patient populations. Because direct comparisons can be distorted by differences in baseline covariates, our framework uses covariate calibration to enable a fairer comparison, as demonstrated in \textbf{Case Study 2} in the following section.

\subsection{AUC Comparison through Benchmarking}\label{sec:CaseStudy_c2}
The primary analysis of these two trials found that the effect of enobosarm on SCP depends on the chemotherapy regimen \citep{crawford2016}, which motivates us to explore further whether the baseline SCP has different predictive accuracy for the survival outcome in these two trials. To compare the AUC of baseline SCP for classifying 6-month survival responders, we consider two ways to calibrate covariates: either to a single trial or to the mixture population. As long as benchmarking both trials to a common target population, the resulting AUC comparison can account for differences in baseline covariates across trials and therefore be interpreted more fairly.

\begin{figure*}[ht]
    \centering
    \includegraphics[width=\linewidth]{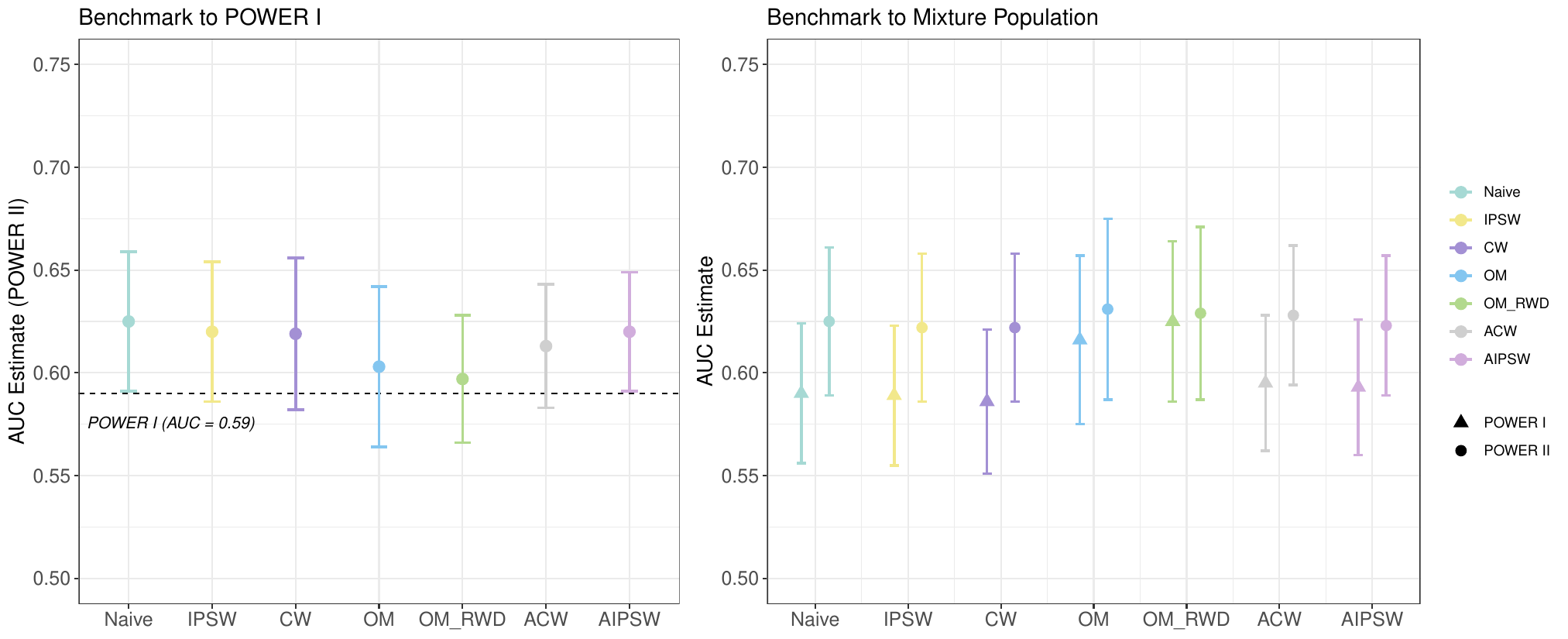}
    \caption{Comparison of AUC between POWER I and POWER II}
    \label{fig:res_Case2}
\end{figure*}

After excluding patients with missing baseline SCP values, we retain 304 patients from POWER I and 321 from POWER II. In Case Study 1, baseline weight was included as an overlapping covariate between the POWER trials and the representative dataset. However, in this comparison between POWER I and POWER II, baseline BMI is also available. Since including both BMI and weight may introduce multicollinearity, we retain baseline BMI and exclude weight from the analysis.

To compute calibration weights, we use the first and second moments of the baseline covariates in the benchmark population, i.e., either the trial population or its mixture, as the standard. For example, to benchmark POWER II to POWER I, we align the covariate distribution of POWER II to that of POWER I and compare their AUC estimates. If comparing AUCs in the mixture population, we benchmark both trials to the covariate distribution of the combined dataset. Following the setup in Case Study 1, we define the benchmark population (e.g., POWER I or the mixture population) as $S=0$, representing the target population, and the trial being weighted as $S=1$. Using the same procedure for computing calibration weights, fitting sampling score models, and estimating outcome models, we obtain the AUC results shown in Figure~\ref{fig:res_Case2}.

When calibrating POWER II to the covariate distribution of POWER I, the observed difference in the predictive accuracy of baseline SCP for 6-month survival is reduced compared to the Na\"ive approach that does not account for covariate shift. In this setting, the AUC for POWER I is fixed at 0.59 and serves as the benchmark (dashed line). Among the estimators, those that rely on outcome modeling, such as OM and OM+RWD, tend to yield smaller AUC differences between POWER I and POWER II, suggesting comparable predictive accuracy of baseline SCP across the two populations after accounting for covariate imbalance. In contrast, CW-based and IPSW-based estimators indicate an improvement in the predictive accuracy of baseline SCP in POWER II compared to POWER I, although the observed difference is diminished after covariate adjustment.

Across all benchmark choices and estimators, the POWER I and POWER II AUC estimates were generally similar, with substantial overlap in uncertainty. So, after accounting for differences in baseline covariates, the results do not suggest a large difference in the prognostic performance of baseline SCP for 6-month survival across the two chemotherapy regimens. Although the point estimates differ somewhat across estimators, these differences do not change the main conclusion. In particular, the OM-based estimators tend to give more similar AUC estimates between the two trials, whereas CW, ACW, IPSW, and AIPSW estimators generally yield slightly higher AUC estimates for POWER II. These patterns likely reflect the different ways in which the estimators rely on the outcome and sampling models. Overall, the proposed framework benchmarks the two trials on a common population footing, which avoids the potentially misleading interpretation that may arise from a Na\"ive comparison.

\section{Conclusion and Discussion}\label{sec:conclusion}
One of the central challenges in biomarker evaluation is covariate shift between a study's validation sample and the target population to which accuracy is intended to generalize. Such mismatch can distort the AUC estimation and compromise fair comparisons across studies. To overcome this, we proposed an estimand-focused framework that (i) treats the AUC as a population-specific quantity anchored to a prespecified target population and (ii) benchmarks AUCs across studies relative to that same target. This approach aligns design, analysis, and interpretation in the spirit of ICH E9\,(R1) \citep{ICH2019} and ensures that the observed differences reflect true performance rather than case-mix variation. Methodologically, we developed a unified family of calibration-based estimators that accommodate different data scenarios but share the same covariate-calibration foundation, each with established large-sample validity.

Another key contribution is extending calibration weighting, which originally developed in causal inference, to the U-statistic formulation of the AUC. This innovation enables target-anchored AUC inference even when only summary covariate information is available, a common situation in biomarker validation. When patient-level target data are available, OM+RWD and AIPSW leverage richer information and improve efficiency under standard validity conditions. Furthermore, ACW combines the strengths of both approaches: OM contributes efficiency, while CW contributes robustness by balancing covariate moments without requiring a correctly specified sampling model. Together, they yield an efficient, doubly robust structure. Our OM-based estimators follow the g-computation perspective \citep{Robins1986} and relate to recent methods for model-based AUC \citep{Li2023}, but remain distinct in focusing on biomarker-based discrimination rather than risk-prediction.

Like treatment effects in causal inference, AUC evaluation benefits from an explicit estimand framework \citep{ICH2019}. Although ICH E9(R1) was developed for treatment-effect settings, its core idea extends naturally to biomarker evaluation: the target population, outcome contrast, biomarker, relevant measurement or ascertainment conditions, and summary measure must all be clearly specified. Without these specifications, AUC estimates from different studies may be affected by spectrum, selection, and covariate shift, leading to non-comparable results and unclear interpretation of test performance.

A further practical issue is the high-dimensional setting. When the baseline covariate vector is large, CW-based methods may become computationally demanding or unstable if too many balance functions are imposed. In such cases, one possible direction is to regularize the calibration step, for example by using penalization methods such as the lasso to select a smaller set of important covariates or balance functions \citep{Tibshirani1996}. Another option is to summarize covariates through a prediction score and study how the AUC of that score transports across populations, as considered by \citet{Li2023}. This is a useful strategy when the scientific target is the predictive performance of a fitted model. However, it is not identical to the biomarker-specific problem considered here. In this setting, high-dimensional covariates are background variables to be adjusted for, not the object of evaluation. Collapsing them into a prediction score may reduce dimensionality, but it changes the target from the AUC of the original biomarker to the AUC of a model-based score. Moreover, if the prediction score is refit or recalibrated for a new population, the object being evaluated may itself change across populations. Thus, although prediction-score transport is useful when model performance is the target, it does not directly address the biomarker-specific AUC problem considered here. Extending our framework to high-dimensional settings remains an important direction for future research.

The POWER case study illustrates two main applications: generalizing a study-specific AUC to a broader clinical population and comparing AUCs across trials by benchmarking them to a common target. These analyses show how the framework works with either summary-level or patient-level information, and how to address covariate shift. In future work, we plan to extend this framework beyond covariate-based sampling to case–control and other outcome-dependent designs. For instance, modeling the sampling indicator as a function of the outcome ($S \sim D$) naturally captures covariate shift in retrospective settings. More broadly, we aim to generalize the benchmarking and transportability framework to a wider range of sampling mechanisms encountered in practice.

Looking ahead, scalable penalized or sieve-based calibration may facilitate computation in high-dimensional settings, and extending the framework from single biomarkers to multivariable signatures is a natural next step. Incorporating modern machine-learning outcome models may also improve robustness and efficiency. Although our focus is on AUC, the same principles apply to transporting clinically meaningful cut-offs, which depend on absolute risk rather than ranking. Because covariate shift alters biomarker distributions, fixed thresholds may not transport even when the AUC remains stable, making threshold recalibration across populations an important direction. Finally, since the AUC corresponds to the Mann–Whitney functional, our calibration ideas suggest a broader path toward covariate-adjusted U-statistic-based treatment effect inference within an estimand-first paradigm.

\section{Competing interests}
No competing interest is declared.


\section{Acknowledgments}
The authors thank the editors and anonymous reviewers for their constructive comments. We are grateful to Dr. Jeffrey Crawford for providing the real data and for insightful discussions. 
This work was supported in part by the U.S. Food and Drug Administration (FDA) and the National Institute on Aging of the National Institutes of Health under Grant R01AG06688.

\bibliographystyle{agsm}
\bibliography{reference}

\appendix

\setcounter{equation}{0}
\renewcommand{\theequation}{S\arabic{equation}}
\setcounter{table}{0}
\renewcommand{\thetable}{S\arabic{table}}
\setcounter{figure}{0}
\renewcommand{\thefigure}{S\arabic{figure}}
\setcounter{theorem}{0}
\renewcommand{\thetheorem}{S\arabic{theorem}}
\setcounter{lemma}{0}
\renewcommand{\thelemma}{S\arabic{lemma}}
\setcounter{remark}{0}
\renewcommand{\theremark}{S\arabic{remark}}

\newpage
\begin{center}
{\large\bf SUPPLEMENTARY MATERIAL}
\end{center}

Section~\ref{app.proof} provides the theoretical foundations of the proposed estimators, including identification results (Section~\ref{app.a}), proofs of consistency and double robustness (Section~\ref{app.b}), and asymptotic normality under large-sample theory (Section~\ref{app.c}). Section \ref{sec:OtherMethodsDiscussions} offers additional deep discussion about the methods. Section \ref{sec:app_Simulation} details the simulation setup, and Section \ref{sec:app_CaseStudy} presents additional case study results and information.

\maketitle
\section{Technical Details of Mathematical Results}\label{app.proof}
\subsection{Identifiability of Estimator}\label{app.a}
In this section, we demonstrate that the AUC in the target population, $\tau_0=\mathbb{E} \left[\mathbb{I}(Y_i>Y_j) \mid  D_i=1, D_j=0\right]$, is identifiable from the validation cohort $\mathcal{V}$.

\subsubsection{Inverse Probability of Sampling Weighting (IPSW)}\label{app.a1}
\begin{theorem}\label{thm:iden_ipsw}
    Under Assumption \ref{as1} - \ref{as3}, the AUC $\tau_0$ in the target population is identifiable by the observed validation cohort data $\mathcal{V}$ 
    \begin{align}
        \tau_0 =\frac{\mathbb{E} \left[w^{\text{ipsw}}(\boldsymbol{X}_i,\boldsymbol{X}_j) \mathbb{I}(Y_i>Y_j, D_i=1, D_j=0) \mid S_i=1, S_j=1\right]}{\mathbb{E} \left[w^{\text{ipsw}}(\boldsymbol{X}_i,\boldsymbol{X}_j) \mathbb{I}(D_i=1, D_j=0)\mid S_i=1, S_j=1\right]}
    \end{align}
\end{theorem}

Define $w^{\text{ipsw}}(\boldsymbol{X}_i,\boldsymbol{X}_j)=\left[\text{Pr}(S_i=1\mid \boldsymbol{X}_i) \text{Pr}(S_j=1\mid \boldsymbol{X}_j)\right]^{-1}=\pi(\boldsymbol{X}_i)^{-1} \pi(\boldsymbol{X}_j)^{-1}=p_i p_j$, where $p_i$ and $p_j$ are the IPSW weights for observation $i$ and $j$. $\boldsymbol{X}_i$ and $\boldsymbol{X}_j$ are the baseline covariates for a random pair of observations $i$ and $j$ ($i\neq j$). 
\begin{proof}[Proof of Theorem~\upshape\ref{thm:iden_ipsw}]
\allowdisplaybreaks
    \begin{align*}
    \tau_0 &= \mathbb{E}\left[\mathbb{I}(Y_i>Y_j) \mid D_i=1, D_j=0\right] \\
    &=\frac{\mathbb{E}\left[\mathbb{I}(Y_i>Y_j, D_i=1, D_j=0)\right]}{\text{Pr}(D_i=1, D_j=0)}\\
    &=\frac{\mathbb{E} \Bigl[\mathbb{E} \bigl[\mathbb{I}(Y_i>Y_j, D_i=1, D_j=0) \mid \boldsymbol{X}_i,\boldsymbol{X}_j \bigr]\Bigr]}{\mathbb{E}\left[\text{Pr}(D_i=1, D_j=0\mid \boldsymbol{X}_i,\boldsymbol{X}_j)\right]} \\
    &=\frac{\mathbb{E} \Bigl[ \mathbb{I}(Y_i>Y_j) \mathbb{E} \bigl[\mathbb{I}(D_i=1, D_j=0) \mid \boldsymbol{X}_i,\boldsymbol{X}_j \bigr]\Bigr]}{\mathbb{E}\left[\text{Pr}(D_i=1, D_j=0 \mid \boldsymbol{X}_i,\boldsymbol{X}_j)\right]} \\
    &=\frac{\mathbb{E} \Bigl[ \mathbb{I}(Y_i>Y_j) \text{Pr} (D_i=1, D_j=0 \mid \boldsymbol{X}_i,\boldsymbol{X}_j) \Bigr]}{\mathbb{E}\left[\text{Pr}(D_i=1, D_j=0\mid \boldsymbol{X}_i,\boldsymbol{X}_j)\right]}\\
    &=\frac{\mathbb{E} \Bigl[ \mathbb{I}(Y_i>Y_j) \text{Pr} (D_i=1, D_j=0 \mid \boldsymbol{X}_i,\boldsymbol{X}_j, S_i=1, S_j=1) \Bigr]}{\mathbb{E}\left[\text{Pr}(D_i=1, D_j=0\mid \boldsymbol{X}_i,\boldsymbol{X}_j, S_i=1, S_j=1)\right]} \qquad\text{(by Assumption \ref{as2})}\\
    &=\frac{\mathbb{E} \Bigl[  \frac{\mathbb{I}(Y_i>Y_j) \text{Pr}(D_i=1, D_j=0, S_i=1, S_j=1 \mid \boldsymbol{X}_i,\boldsymbol{X}_j)}{\text{Pr}(S_i=1,S_j=1\mid \boldsymbol{X}_i,\boldsymbol{X}_j)} \Bigr]}{\mathbb{E}\left[\frac{\text{Pr}(D_i=1, D_j=0, S_i=1, S_j=1 \mid \boldsymbol{X}_i,\boldsymbol{X}_j)}{\text{Pr}(S_i=1,S_j=1\mid \boldsymbol{X}_i,\boldsymbol{X}_j)}\right]}\qquad \text{(by $\Pr(D\mid \boldsymbol{X},S) \Pr(S\mid \boldsymbol{X})=\text{Pr}(D,S\mid \boldsymbol{X})$)} \\
    &=\frac{\mathbb{E} \Bigl[ \mathbb{I}(Y_i>Y_j) \frac{\text{Pr}(D_i=1, D_j=0, S_i=1, S_j=1 \mid \boldsymbol{X}_i,\boldsymbol{X}_j)}{\text{Pr}(S_i=1\mid \boldsymbol{X}_i) \text{Pr}(S_j=1\mid \boldsymbol{X}_j)} \Bigr]}{\mathbb{E}\left[\frac{\text{Pr}(D_i=1, D_j=0, S_i=1, S_j=1 \mid \boldsymbol{X}_i,\boldsymbol{X}_j)}{\text{Pr}(S_i=1\mid \boldsymbol{X}_i) \text{Pr}(S_j=1\mid \boldsymbol{X}_j)}\right]} \qquad\text{(by $S_i \perp\!\!\!\perp S_j$ and Assumption \ref{as1} (i))} \\
    &=\frac{\mathbb{E} \Bigl[ \mathbb{I}(Y_i>Y_j) \frac{\mathbb{E}\left[\mathbb{I}(D_i=1, D_j=0, S_i=1, S_j=1) \mid \boldsymbol{X}_i,\boldsymbol{X}_j\right]}{\text{Pr}(S_i=1\mid \boldsymbol{X}_i) \text{Pr}(S_j=1\mid \boldsymbol{X}_j)} \Bigr]}{\mathbb{E}\left[\frac{\mathbb{E}\left[\mathbb{I}(D_i=1, D_j=0, S_i=1, S_j=1) \mid \boldsymbol{X}_i,\boldsymbol{X}_j\right]}{\text{Pr}(S_i=1\mid \boldsymbol{X}_i) \text{Pr}(S_j=1\mid \boldsymbol{X}_j)}\right]} \qquad \text{(by definition of conditional expectation)}\\
    &=\frac{\mathbb{E} \Bigl[ \frac{\mathbb{E}\left[\mathbb{I}(Y_i>Y_j, D_i=1, D_j=0, S_i=1, S_j=1) \mid \boldsymbol{X}_i,\boldsymbol{X}_j\right]}{\text{Pr}(S_i=1\mid \boldsymbol{X}_i) \text{Pr}(S_j=1\mid \boldsymbol{X}_j)} \Bigr]}{\mathbb{E}\left[\frac{\mathbb{E}\left[\mathbb{I}(D_i=1, D_j=0, S_i=1, S_j=1) \mid \boldsymbol{X}_i,\boldsymbol{X}_j\right]}{\text{Pr}(S_i=1\mid \boldsymbol{X}_i) \text{Pr}(S_j=1\mid \boldsymbol{X}_j)}\right]}\\
    &=\frac{\frac{1}{\text{Pr}(S_i=1\mid \boldsymbol{X}_i) \text{Pr}(S_j=1\mid \boldsymbol{X}_j)}\mathbb{E} \left[ \mathbb{I}(Y_i>Y_j, D_i=1, D_j=0, S_i=1, S_j=1) \right]}{\frac{1}{\text{Pr}(S_i=1\mid \boldsymbol{X}_i) \text{Pr}(S_j=1\mid \boldsymbol{X}_j)}\mathbb{E} \left[\mathbb{I}(D_i=1, D_j=0, S_i=1, S_j=1) \right]}\\
    &=\frac{\mathbb{E} \left[\frac{1}{\text{Pr}(S_i=1\mid \boldsymbol{X}_i) \text{Pr}(S_j=1\mid \boldsymbol{X}_j)}\mathbb{I}(Y_i>Y_j, D_i=1, D_j=0, S_i=1, S_j=1)\right]}{\mathbb{E} \left[\frac{1}{\text{Pr}(S_i=1\mid \boldsymbol{X}_i) \text{Pr}(S_j=1\mid \boldsymbol{X}_j)}\mathbb{I}(D_i=1, D_j=0, S_i=1, S_j=1)\right]}\\
    &=\frac{\mathbb{E} \left[\frac{\mathbb{I}(Y_i>Y_j, D_i=1, D_j=0)}{\text{Pr}(S_i=1\mid \boldsymbol{X}_i) \text{Pr}(S_j=1\mid \boldsymbol{X}_j)} \mid S_i=1, S_j=1\right]}{\mathbb{E} \left[\frac{\mathbb{I}(D_i=1, D_j=0)}{\text{Pr}(S_i=1\mid \boldsymbol{X}_i) \text{Pr}(S_j=1\mid \boldsymbol{X}_j)}\mid S_i=1, S_j=1\right]}
    \end{align*}
\end{proof}
Based on Theorem \ref{thm:iden_ipsw}, we can provide an IPSW estimator as:
\begin{align}
    \hat{\tau}_{\text{IPSW}} = \frac{\sum_{i\neq j}^{n} w^{\text{ipsw}}(\boldsymbol{X}_i, \boldsymbol{X}_j; \hat{\balp})  \mathbb{I}(Y_i > Y_j, D_i = 1, D_j = 0, S_i = 1, S_j = 1)}{\sum_{i\neq j}^{n} w^{\text{ipsw}}(\boldsymbol{X}_i, \boldsymbol{X}_j; \hat{\balp})  \mathbb{I}(D_i = 1, D_j = 0, S_i = 1, S_j = 1)}.
\end{align}
The IPSW weights $w^{\text{ipsw}}(\boldsymbol{X}_i, \boldsymbol{X}_j; \hat{\balp})$ is defined as:
\begin{align*}
    w^{\text{ipsw}}(\boldsymbol{X}_i, \boldsymbol{X}_j; \hat{\balp})=\frac{1}{\pi(\boldsymbol{X}_i;\hat{\balp}) \pi(\boldsymbol{X}_j;\hat{\balp})}=p_i(\boldsymbol{X}_i;\hat{\balp}) p_j(\boldsymbol{X}_j;\hat{\balp}),
\end{align*}
where $\pi(\boldsymbol{X};\hat{\balp})$ is the estimated sampling score by fitting the sampling model on $\text{Pr}(S=1|\boldsymbol{X})$ parameterized by $\balp$. In practice, the representative dataset and the validation cohort are used jointly to model the sampling indicator $S$. Let $\hat{p}_i=p_i(\boldsymbol{X}_i;\hat{\balp})$ and $\hat{p}_j=p_j(\boldsymbol{X}_j;\hat{\balp})$ be the IPSW weights for $i$ and $j$ in the validation cohort. 

\subsubsection{Calibration Weighting (CW)}\label{app.a2}
The proof of Theorem \ref{thm:iden_cw} is similar to the proof of Theorem \ref{thm:iden_ipsw}, while the suggested estimator is different from IPSW. 
\begin{theorem}\label{thm:iden_cw}
    Under Assumption \ref{as1} - \ref{as3}, the AUC $\tau_0$ in the target population is identifiable by the observed validation cohort data $\mathcal{V}$ 
    \begin{align}
        \tau_0 =\frac{\mathbb{E} \left[w^{\text{cw}}(\boldsymbol{X}_i,\boldsymbol{X}_j) \mathbb{I}(Y_i>Y_j, D_i=1, D_j=0) \mid S_i=1, S_j=1\right]}{\mathbb{E} \left[w^{\text{cw}}(\boldsymbol{X}_i,\boldsymbol{X}_j) \mathbb{I}(D_i=1, D_j=0)\mid S_i=1, S_j=1\right]}.
    \end{align}
\end{theorem}

Define $w^{\text{cw}}(\boldsymbol{X}_i, \boldsymbol{X}_j)={q}_i(\boldsymbol{X}_i){q}_j(\boldsymbol{X}_j)$, where ${q}_i(\boldsymbol{X}_i)$ and ${q}_j(\boldsymbol{X}_j)$ are calibration weights for subject $i$ and subject $j$. The CW estimator is given by
\begin{align}
    \hat{\tau}_{\text{CW}} = \frac{\sum_{i\neq j}^{n} \hat{w}^{cw}(\boldsymbol{X}_i, \boldsymbol{X}_j)  \mathbb{I}(Y_i > Y_j, D_i = 1, D_j = 0, S_i = 1, S_j = 1)}{\sum_{i\neq j}^{n} \hat{w}^{\text{cw}}(\boldsymbol{X}_i, \boldsymbol{X}_j)  \mathbb{I}(D_i = 1, D_j = 0, S_i = 1, S_j = 1)}.
\end{align}
The CW weights $w^{\text{cw}}(\boldsymbol{X}_i, \boldsymbol{X}_j)$ are defined as:
\begin{align*}
    \hat{w}^{\text{cw}}(\boldsymbol{X}_i, \boldsymbol{X}_j)=\hat{q}_i(\boldsymbol{X}_i)\hat{q}_j(\boldsymbol{X}_j),
\end{align*}
where $\hat{q}_i(\boldsymbol{X}_i)$ and $\hat{q}_j(\boldsymbol{X}_j)$ are parameterized by $\blam$ and depend on the $\boldsymbol{X}_i$ and $\boldsymbol{X}_j$ of subject $i$ and $j$, i.e., $\hat{q}_i(\boldsymbol{X}_i)=q(\boldsymbol{X}_i;\hat\blam)$. The estimation of $q$ is discussed in detail in the main text. In  Section~\ref{app.consistency_cw}, we show that there is a direct relationship between the estimated sampling score ${\pi}_s(X;\hat\balp)$ and the calibration weight $q(X;\hat\blam)$.

\subsubsection{Outcome Modeling with RWD (OM+RWD)}\label{app.a3}
\begin{theorem}\label{thm:iden_omRWD}
Under Assumptions~\ref{as1}-\ref{as3}, the AUC $\tau_0$ in the target population is identifiable from the observed validation cohort $\mathcal{V}$ and RWD $\mathcal{R}$ via the outcome model:
\begin{align*}
\tau_0&=\frac{\mathbb{E}\!\left[\Pr(Y_i > Y_j \mid D_i=1, D_j=0, S_i=0, S_j=0, \boldsymbol{X}_i, \boldsymbol{X}_j)\, \mathbb{I}(D_i=1, D_j=0)\mid S_i=0, S_j=0\right]}{\mathbb{E}\!\left[\mathbb{I}(D_i=1, D_j=0)\mid S_i=0, S_j=0\right]}.
\end{align*}

\end{theorem}
\begin{proof}
    \begin{align*}
    \tau_0 
    &= \mathbb{E}\left[\mathbb{I}(Y_i>Y_j) \mid D_i=1, D_j=0\right] \\
    &=\mathbb{E}\left[\mathbb{I}(Y_i>Y_j) \mid D_i=1, D_j=0, S_i=0, S_j=0\right]\\
    &=\frac{\mathbb{E}\left[\mathbb{I}(Y_i>Y_j, D_i=1, D_j=0, S_i=0, S_j=0)\right]}{\mathbb{E}\left[\mathbb{I}(D_i=1, D_j=0, S_i=0, S_j=0)\right]}\\
    &=\frac{\mathbb{E} \Bigl[\mathbb{E} \bigl[\mathbb{I}(Y_i>Y_j, D_i=1, D_j=0, S_i=0, S_j=0) \mid \boldsymbol{X}_i,\boldsymbol{X}_j \bigr]\Bigr]}{\mathbb{E}\left[\mathbb{I}(D_i=1, D_j=0, S_i=0, S_j=0)\right]}\\
    &=\frac{\mathbb{E} \Bigl[\text{Pr} (Y_i>Y_j\mid D_i=1, D_j=0, S_i=0, S_j=0, \boldsymbol{X}_i,\boldsymbol{X}_j) \text{Pr}(D_i=1, D_j=0, S_i=0, S_j=0\mid \boldsymbol{X}_i,\boldsymbol{X}_j)\Bigr]}{\mathbb{E}\left[\mathbb{I}(D_i=1, D_j=0, S_i=0, S_j=0)\right]}\\
    &=\frac{\mathbb{E} \Bigl[\text{Pr} (Y_i>Y_j\mid D_i=1, D_j=0, S_i=0, S_j=0, \boldsymbol{X}_i,\boldsymbol{X}_j) \mathbb{I}(D_i=1, D_j=0, S_i=0, S_j=0)\Bigr]}{\mathbb{E}\left[\mathbb{I}(D_i=1, D_j=0, S_i=0, S_j=0)\right]}\\
    &=\frac{\mathbb{E} \Bigl[\text{Pr} (Y_i>Y_j\mid D_i=1, D_j=0, S_i=0, S_j=0, \boldsymbol{X}_i,\boldsymbol{X}_j) \mathbb{I}(D_i=1, D_j=0)\mid S_i=0, S_j=0\Bigr]}{\mathbb{E}\left[\mathbb{I}(D_i=1, D_j=0)\mid S_i=0, S_j=0\right]}
\end{align*}
\end{proof}
Based on Theorem \ref{thm:iden_omRWD}, we propose an RWD-based OM estimator under the assumption of a bi-normal distribution for $Y$ given $\boldsymbol{X}$ for $D=1$ and $D=0$.
\begin{align}
    \hat{\tau}_\text{OM+RWD}=\frac{\sum_{i\neq j}^{m+n} \Phi\left(\frac{M(\boldsymbol{X}_i,D_i=1,S_i=0;\hat\bbeta)-M(\boldsymbol{X}_j,D_j=0,S_j=0;\hat\bbeta)}{\sqrt{\hat{\sigma}_1^2+\hat{\sigma}_0^2}}\right) \mathbb{I}(D_i = 1, D_j = 0, S_i = 0, S_j = 0)}{\sum_{i\neq j}^{m+n}  \mathbb{I}(D_i = 1, D_j = 0, S_i = 0, S_j = 0)},
\end{align}
where $M(\boldsymbol{X}_i,D_i,S_i=0;\hat\bbeta)$ is the estimated outcome mean by plugging $(\boldsymbol{X}_i,D_i,S_i=0)$ ($i\in \mathcal{R}$) in the outcome model parameterized by $\bbeta$. The $\Phi(\cdot)$ is the standard normal CDF, which is bounded between 0 and 1. Since biomarker $Y$ is unobserved in representative data $\mathcal{R}$ (e.g. RWD), the $\hat\bbeta$ is estimated using $\{(\boldsymbol{X}_i,Y_i,D_i,S_i=1):i=1,\dots,n\}$ in the validation cohort $\mathcal{V}$. The outcome model, defined in Section \ref{sec:method_OMRWD}, uses biomarker $Y$ as the dependent variable and posits a relationship between $Y$, covariates $\boldsymbol{X}$, and the outcome indicator $D$. The $\hat{\sigma}_0^2$ and $\hat{\sigma}_1^2$ are the sample variances of $Y$ for $D=0$ and $D=1$ groups, which are solely based on the observed validation cohort dataset $\mathcal{V}$ and is estimated by 
\begin{align*}
    \hat{\sigma}_d^2=\frac{1}{n_d - p_d} \sum_{i : D_i = d} \left( Y_i - M(\boldsymbol{X}_i,S_i=1;\hat\bbeta) \right)^2,
\end{align*}
where $d\in\{0,1\}$, $p_d$ is the number of parameters when fixing $D=d$, $n_d$ is sample size of group $D=d$. 

\subsubsection{Outcome Modeling without RWD (OM)}\label{app.a4}
\begin{theorem}\label{thm:iden_om}
Under Assumptions~\ref{as1}-\ref{as:cw_loglinear}, the AUC for the target population is identifiable from the observed validation cohort $\mathcal{V}$ as
\begin{align*}
    \tau_{0} =\frac{\mathbb{E}\!\left[\,w_{ij}\,
        \Pr(Y_i > Y_j \mid D_i=1, D_j=0, S_i=1, S_j=1, \boldsymbol{X}_i, \boldsymbol{X}_j)\,
        \mathbb{I}(D_i=1, D_j=0) \;\middle|\; S_i=1, S_j=1\right]}
         {\mathbb{E}\!\left[\,w_{ij}\, \mathbb{I}(D_i=1, D_j=0) \;\middle|\; S_i=1, S_j=1\right]},
\end{align*}
where $w_{ij} = \bigl(\Pr(S_i=1 \mid \boldsymbol{X}_i) \Pr(S_j=1 \mid \boldsymbol{X}_j)\bigr)^{-1} = \pi(\boldsymbol{X}_i;\balp)^{-1}\,\pi(\boldsymbol{X}_j;\balp)^{-1}$.

\end{theorem}
\begin{proof}
\allowdisplaybreaks
    \begin{align*}
        \tau_0 &= \mathbb{E}\left[\mathbb{I}(Y_i>Y_j) \mid D_i=1, D_j=0\right] \\
        &=\frac{\mathbb{E}\left[\mathbb{I}(Y_i>Y_j, D_i=1, D_j=0)\right]}{\mathbb{E}\left[\mathbb{I}(D_i=1,D_j=0)\right]}\\
        &=\frac{\mathbb{E} \Bigl[\mathbb{E} \bigl[\mathbb{I}(Y_i>Y_j, D_i=1, D_j=0) \mid \boldsymbol{X}_i,\boldsymbol{X}_j \bigr]\Bigr]}{\mathbb{E}\Bigl[\mathbb{E}\left[\mathbb{I}(D_i=1, D_j=0)\mid \boldsymbol{X}_i,\boldsymbol{X}_j\right]\Bigr]}\\
        &=\frac{\mathbb{E} \Bigl[\text{Pr}(Y_i>Y_j \mid D_i=1, D_j=0, \boldsymbol{X}_i,\boldsymbol{X}_j)\text{Pr}(D_i=1, D_j=0\mid \boldsymbol{X}_i,\boldsymbol{X}_j)\Bigr]}{\mathbb{E}\Bigl[\mathbb{E}\left[\mathbb{I}(D_i=1, D_j=0)\mid \boldsymbol{X}_i,\boldsymbol{X}_j\right]\Bigr]}\\
        &=\frac{\mathbb{E} \Bigl[\text{Pr}(Y_i>Y_j \mid D_i=1, D_j=0, S_i=1, S_j=1, \boldsymbol{X}_i,\boldsymbol{X}_j)\text{Pr}(D_i=1, D_j=0\mid S_i=1, S_j=1, \boldsymbol{X}_i,\boldsymbol{X}_j)\Bigr]}{\mathbb{E}\Bigl[\mathbb{E}\left[\mathbb{I}(D_i=1, D_j=0)\mid \boldsymbol{X}_i,\boldsymbol{X}_j\right]\Bigr]}\\& \qquad\qquad\qquad\qquad\qquad\qquad\qquad\qquad\qquad\qquad\qquad\qquad\qquad\qquad\qquad (\text{by Assumptions \ref{as2} and \ref{as3}})\\
        &=\frac{\mathbb{E} \Bigl[\text{Pr}(Y_i>Y_j \mid D_i=1, D_j=0, S_i=1, S_j=1, \boldsymbol{X}_i,\boldsymbol{X}_j)\frac{\text{Pr}(D_i=1, D_j=0, S_i=1, S_j=1\mid \boldsymbol{X}_i,\boldsymbol{X}_j)}{\text{Pr}(S_i=1, S_j=1\mid \boldsymbol{X}_i, \boldsymbol{X}_j)}\Bigr]}{\mathbb{E}\Bigl[\mathbb{E}\left[\mathbb{I}(D_i=1, D_j=0)\mid \boldsymbol{X}_i,\boldsymbol{X}_j\right]\Bigr]}\\
        &=\frac{\mathbb{E} \Bigl[\text{Pr}\left(Y_i>Y_j\mid D_i=1, D_j=0, S_i=1, S_j=1, \boldsymbol{X}_i, \boldsymbol{X}_j\right)\frac{ \mathbb{E} \left[\mathbb{I}(D_i=1, D_j=0, S_i=1, S_j=1)\mid \boldsymbol{X}_i,\boldsymbol{X}_j\right]}{\text{Pr}\left(S_i=1, S_j=1\mid \boldsymbol{X}_i, \boldsymbol{X}_j\right)}\Bigr]}{\mathbb{E}\Bigl[\frac{\text{Pr}(D_i=1, D_j=0, S_i=1, S_j=1\mid \boldsymbol{X}_i,\boldsymbol{X}_j)}{\text{Pr}(S_i=1, S_j=1\mid \boldsymbol{X}_i,\boldsymbol{X}_j)}\Bigr]}\\
        &=\frac{\mathbb{E} \Bigl[\frac{ \mathbb{E} \left[\text{Pr}(Y_i>Y_j\mid D_i=1, D_j=0, S_i=1, S_j=1, \boldsymbol{X}_i, \boldsymbol{X}_j) \mathbb{I}(D_i=1, D_j=0, S_i=1, S_j=1)\mid \boldsymbol{X}_i,\boldsymbol{X}_j\right]}{\text{Pr}\left(S_i=1, S_j=1\mid \boldsymbol{X}_i, \boldsymbol{X}_j\right)}\Bigr]}{\mathbb{E}\Bigl[\frac{\text{Pr}(D_i=1, D_j=0, S_i=1, S_j=1\mid \boldsymbol{X}_i,\boldsymbol{X}_j)}{\text{Pr}(S_i=1, S_j=1\mid \boldsymbol{X}_i,\boldsymbol{X}_j)}\Bigr]}\\
        &=\frac{\mathbb{E} \left[\frac{\text{Pr}(Y_i>Y_j\mid D_i=1, D_j=0, S_i=1, S_j=1, \boldsymbol{X}_i, \boldsymbol{X}_j)\cdot \mathbb{I}(D_i=1, D_j=0, S_i=1, S_j=1)}{\text{Pr}(S_i=1, S_j=1\mid \boldsymbol{X}_i, \boldsymbol{X}_j)}\right]}{\mathbb{E}\Bigl[\frac{\mathbb{I}(D_i=1, D_j=0, S_i=1, S_j=1)}{\text{Pr}(S_i=1, S_j=1\mid \boldsymbol{X}_i,\boldsymbol{X}_j)}\Bigr]}\\
        &=\frac{\mathbb{E} \left[\frac{\text{Pr}(Y_i>Y_j\mid D_i=1, D_j=0, S_i=1, S_j=1, \boldsymbol{X}_i, \boldsymbol{X}_j)\cdot \mathbb{I}(D_i=1, D_j=0, S_i=1, S_j=1)}{\text{Pr}(S_i=1\mid \boldsymbol{X}_i)\text{Pr}(S_j=1\mid \boldsymbol{X}_j)}\right]}{\mathbb{E}\Bigl[\frac{\mathbb{I}(D_i=1, D_j=0, S_i=1, S_j=1)}{\text{Pr}(S_i=1\mid \boldsymbol{X}_i)\text{Pr}(S_j=1\mid \boldsymbol{X}_j)}\Bigr]} \qquad(\text{ by } i \independent j )\\
        &=\frac{\mathbb{E} \left[\frac{\text{Pr}(Y_i>Y_j\mid D_i=1, D_j=0, S_i=1, S_j=1, \boldsymbol{X}_i, \boldsymbol{X}_j)\cdot \mathbb{I}(D_i=1, D_j=0)}{\text{Pr}(S_i=1\mid \boldsymbol{X}_i)\text{Pr}(S_j=1\mid \boldsymbol{X}_j)}\mid S_i=1, S_j=1\right]}{\mathbb{E}\Bigl[\frac{\mathbb{I}(D_i=1, D_j=0)}{\text{Pr}(S_i=1\mid \boldsymbol{X}_i)\text{Pr}(S_j=1\mid \boldsymbol{X}_j)}\mid S_i=1, S_j=1\Bigr]}
    \end{align*}
\end{proof}

Therefore, for any distribution of $Y\mid (\boldsymbol{X},D)$, we can identify $\tau_0$ using the outcome modeling method only if its CDF $\mathbb{F}$ exists and allows the estimation of $\text{Pr}(Y_i>Y_j\mid D_i=1, D_j=0, \boldsymbol{X}_i, \boldsymbol{X}_j)$. Given Assumption \ref{as3}, we have $\text{Pr}(Y_i>Y_j\mid D_i=1, D_j=0, S_i=1, S_j=1, \boldsymbol{X}_i, \boldsymbol{X}_j)=\text{Pr}(Y_i>Y_j\mid D_i=1, D_j=0, S_i=0, S_j=0, \boldsymbol{X}_i, \boldsymbol{X}_j)$, where the finite population ($S=0$) can be representative of the target population. For example, assume that $Y\mid (\boldsymbol{X},D)$ follows a normal distribution as defined in~\eqref{eq:normal distribution}. Based on Theorem \ref{thm:iden_om}, we propose an OM estimator
\begin{align}
    \hat{\tau}_\text{OM}=\frac{\sum_{i\neq j}^{n}\hat{w}_{ij} \mathcal{P}_{ij}(\boldsymbol{V}_i, \boldsymbol{V}_j; \hat\bbeta, \hat{\sigma}_0,\hat{\sigma}_1) \mathbb{I}(D_i=1,D_j=0,S_i=1,S_j=1)}{\sum_{i\neq j}^{n}\hat{w}_{ij} \mathbb{I}(D_i=1,D_j=0,S_i=1,S_j=1)},
\end{align}
where 
\begin{align*}
    \mathcal{P}_{ij}(\boldsymbol{V}_i, \boldsymbol{V}_j; \hat\bbeta, \hat{\sigma}_0,\hat{\sigma}_1)=\Phi\left(\frac{M(\boldsymbol{X}_i,D_i=1,S_i=1;\hat\bbeta)-M(\boldsymbol{X}_j,D_j=0,S_j=1;\hat\bbeta)}{\sqrt{\hat{\sigma}_1^2+\hat{\sigma}_0^2}}\right),
\end{align*}
and $\hat{w}_{ij}=\hat{q}_i(\boldsymbol{X}_i) \hat{q}_j(\boldsymbol{X}_j)$ is derived from CW weights. More generally, $\mathcal{P}_{ij}(\boldsymbol{V}_i, \boldsymbol{V}_j; \hat\bbeta, \hat{\sigma}_0,\hat{\sigma}_1)$ can be defined as a different form, which is based on the distribution of $Y\mid (\boldsymbol{X},D)$.

\subsection{Proof of Consistency and Double Robustness}\label{app.b}
Before providing the proof of consistency for all estimators, we first introduce two key lemmas, and related details can be found in \cite{Li2023}.
\begin{lemma}\label{lem1}
Let $\mathcal{C} = \{ \boldsymbol{C}_i : i = 1, \dots, n \}$ be a set of independent and identically distributed random variables, and let $l(\boldsymbol{C}_i, \boldsymbol{C}_j; \boldsymbol{\theta}) $ be a real-valued function that depends on a pair of random variables, parameterized by $\boldsymbol{\theta}$. The parameter $\boldsymbol{\theta} \in \Theta$, which is a bounded Euclidean space. Define the U-process as
\begin{align*}
    U_n(\boldsymbol{\theta}) = \frac{1}{n(n-1)} \sum_{i \neq j} l(\boldsymbol{C}_i, \boldsymbol{C}_j; \boldsymbol{\theta}).
\end{align*}
The expectation is defined as
\begin{align*}
    U(\boldsymbol{\theta}) = \mathbb{E}\left[l(\boldsymbol{C}_i, \boldsymbol{C}_j; \boldsymbol{\theta})\right]
\end{align*}
Assume $\hat{\boldsymbol{\theta}}$ is a consistent estimator of $\boldsymbol{\theta}_0$, meaning that $\hat{\boldsymbol{\theta}}\xrightarrow{P}\boldsymbol{\theta}_0$. Given the following conditions:
\begin{enumerate}
    \item $\{U_n(\boldsymbol{\theta}),\boldsymbol{\theta}\in \Theta\}$ is stochastically equicontinuous;
    \item $\mathbb{E}\left|l(\boldsymbol{C}_i, \boldsymbol{C}_j; \boldsymbol{\theta})\right|<\infty$, $\forall $ $\boldsymbol{\theta}\in \Theta$,
\end{enumerate}
we have
\begin{align*}
    U_n(\hat{\boldsymbol{\theta}})\xrightarrow{P}U(\boldsymbol{\theta}_0).
\end{align*}
\end{lemma}

\begin{lemma}\label{lem2}
    Assume that $\{V_n(\boldsymbol{\theta})=\frac{1}{n}\sum_{i=1}^nl_1(\boldsymbol{C}_i;\boldsymbol{\theta}),\boldsymbol{\theta}\in \Theta\}$ and $\{W_n(\boldsymbol{\theta})=\frac{1}{n}\sum_{i=1}^nl_2(\boldsymbol{C}_i;\boldsymbol{\theta}),\boldsymbol{\theta}\in \Theta\}$ are both stochastically equicontinuous. If both kernel $l_1(\boldsymbol{C}_i;\boldsymbol{\theta})$ and $l_2(\boldsymbol{C}_i;\boldsymbol{\theta})$ are uniformly bounded, then the class
    \begin{align*}
        \left\{U_n(\boldsymbol{\theta})=\frac{1}{n(n-1)}\sum_{i\neq j}l_1(\boldsymbol{C}_i;\boldsymbol{\theta})l_2(\boldsymbol{C}_j;\boldsymbol{\theta}),\boldsymbol{\theta}\in \Theta \right\}
    \end{align*}
    is also stochastically equicontinuous.
\end{lemma}

\subsubsection{IPSW Estimator $\hat{\tau}_\text{IPSW}$}\label{app_con}
\begin{theorem}\label{thm:consistency_ipsw}
    (Consistency of IPSW estimator): Assume $\{\frac{1}{n}\sum_{i=1}^n\pi(\boldsymbol{X}_i;\balp),\balp\in \mathcal{A}\}$ is stochastically equicontinuous. When $\pi(\boldsymbol{X};\balp^*)=\text{Pr}(S=1\mid \boldsymbol{X})$ with $\hat{\balp}\rightarrow \balp^*$, then
    \begin{align*}
        \hat{\tau}_\text{IPSW}\xrightarrow{P}\tau_0.
    \end{align*}
\end{theorem}
\begin{proof}
    First, by Lemma \ref{lem2}, when assuming $1/\pi(\boldsymbol{X};\balp)$ is uniformly bounded and $\{\frac{1}{n}\sum_{i=1}^n\pi(\boldsymbol{X}_i;\balp),\balp\in \mathcal{A}\}$ is stochastically equicontinuous, we have $\{\frac{1}{n(n-1)}\sum_{i\neq j}w^{\text{ipsw}}(\boldsymbol{X}_i,\boldsymbol{X}_j;\balp),\balp\in \mathcal{A}\}$ to be also stochastically equicontinuous, where $w^{\text{ipsw}}(\boldsymbol{X}_i,\boldsymbol{X}_j;\balp)=\pi(\boldsymbol{X}_i;{\balp})^{-1} \pi(\boldsymbol{X}_j;{\balp})^{-1}$. Both functions $\mathbb{I}(Y_i > Y_j, D_i = 1, D_j = 0, S_i = 1, S_j = 1)$ and $\mathbb{I}(D_i = 1, D_j = 0, S_i = 1, S_j = 1)$ are bounded, and thus 
    \begin{align*}
        \left\{\frac{1}{n(n-1)}\sum_{i\neq j}d^{\text{ipsw}}_k(\boldsymbol{C}_i,\boldsymbol{C}_j;\hat{\balp})\right\}
    \end{align*}
    are stochastically equicontinuous for $k\in\{0,1\}$, where $d^{\text{ipsw}}_1(\boldsymbol{C}_i,\boldsymbol{C}_j;\hat{\balp})=w^{\text{ipsw}}(\boldsymbol{X}_i,\boldsymbol{X}_j;\hat{\balp}) \mathbb{I}(Y_i > Y_j, D_i = 1, D_j = 0, S_i = 1, S_j = 1)$ and $d^{\text{ipsw}}_0(\boldsymbol{C}_i,\boldsymbol{C}_j;\hat{\balp})=w^{\text{ipsw}}(\boldsymbol{X}_i,\boldsymbol{X}_j;\hat{\balp}) \mathbb{I}(D_i = 1, D_j = 0, S_i = 1, S_j = 1)$ are the main parts for the numerator and denominator of the $\hat{\tau}_\text{IPSW}$, respectively. In addition, $d^{\text{ipsw}}_k(\boldsymbol{C}_i,\boldsymbol{C}_j;\hat{\balp})$ ($k\in\{0,1\}$) is bounded; therefore, both conditions hold for applying Lemma \ref{lem1}. For $k\in\{0,1\}$, we have
    \begin{align*}
        \frac{1}{n(n-1)}\sum_{i\neq j}d^{\text{ipsw}}_k(\boldsymbol{C}_i,\boldsymbol{C}_j;\hat{\balp})\xrightarrow{P} \mathbb{E}\left[d^{\text{ipsw}}_k(\boldsymbol{C}_i,\boldsymbol{C}_j;\balp_0)\right],
    \end{align*}
    where $\mathbb{E}\left[d^{\text{ipsw}}_1(\boldsymbol{C}_i,\boldsymbol{C}_j;\balp_0)\right]=\mathbb{E}\left[w^{\text{ipsw}}(\boldsymbol{X}_i,\boldsymbol{X}_j) \mathbb{I}(Y_i > Y_j, D_i = 1, D_j = 0, S_i = 1, S_j = 1)\right]$ and $\mathbb{E}\left[d^{\text{ipsw}}_0(\boldsymbol{C}_i,\boldsymbol{C}_j;\balp_0)\right]=\mathbb{E}\left[w^{\text{ipsw}}(\boldsymbol{X}_i,\boldsymbol{X}_j) \mathbb{I}(D_i = 1, D_j = 0, S_i = 1, S_j = 1)\right]$. By the continuous mapping theorem, the consistency of $\hat{\tau}_{\text{IPSW}}$ is proved as
    \begin{align*}
        \hat{\tau}_{\text{IPSW}}&\xrightarrow{P} \tau_0=\frac{\mathbb{E}\left[w^{\text{ipsw}}(\boldsymbol{X}_i,\boldsymbol{X}_j) \mathbb{I}(Y_i > Y_j, D_i = 1, D_j = 0, S_i = 1, S_j = 1)\right]}{\mathbb{E}\left[w^{\text{ipsw}}(\boldsymbol{X}_i,\boldsymbol{X}_j) \mathbb{I}(D_i = 1, D_j = 0, S_i = 1, S_j = 1)\right]}.
    \end{align*}
\end{proof}

\begin{remark}
    Suppose the sampling model for $\Pr(S=1\mid \boldsymbol{X})$ is correctly specified as a logistic regression model. If both $1/\pi(\boldsymbol{X};\balp)$ and $\left\|\partial\pi(\boldsymbol{X};\balp)/\partial\balp\right\|$ are uniformly bounded, then $\hat{\tau}_{\text{IPSW}}$ has consistency. 
\end{remark}
\subsubsection{CW Estimator $\hat{\tau}_\text{CW}$}\label{app.consistency_cw}
\begin{proof}[Proof of Theorem \upshape\ref{thm:consistency_cw}]
As stated in Theorem \ref{thm:consistency_cw}, the CW estimator is consistent for $\tau_0$ when Assumption \ref{as:cw_loglinear} is held. 
Let $\boldsymbol{\mu}_{\bg0}=\mathbb{E}\{\bg(\boldsymbol{X})\}$ be the true value of $\bg(\boldsymbol{X})$ and $\bar{\bg}_0=\bg(\boldsymbol{X})-\boldsymbol{\mu}_{\bg0}$. First, based on M-estimation \citep{BoosStefanski2013}, we rewrite \eqref{eq:cw_lagrangeObj} as two estimating equations below:
\begin{align}\label{eq:app_dr1}
    \frac{1}{N} \sum_{i=1}^{N} \mathcal{E}(\boldsymbol{X}_i, \tilS_i; \bmug) 
= \frac{1}{N} \sum_{i=1}^{N} \tilS_i d_i \left\{ \bg(\boldsymbol{X}_i) - \bmug \right\} = 0,
\end{align}
\begin{align}\label{eq:app_dr2}
    \frac{1}{N} \sum_{i=1}^{N} \mathcal{D}(\boldsymbol{X}_i, S_i; \blam,\bmug) 
= \frac{1}{N} \sum_{i=1}^{N} S_i \text{exp}\{\blam^\top\bg(\boldsymbol{X}_i)\}\{\bg(\boldsymbol{X}_i)-\bmug\} = 0.
\end{align}

First, given the basic setup above, $\boldsymbol{\mu}_{\bg0}$ is the solution to $\mathbb{E}\{\mathcal{E}(\boldsymbol{X}; \bmug)\}=0$. When $\bmug=\boldsymbol{\mu}_{\bg0}$, we have
\begin{align*}
    \mathbb{E}\{\mathcal{D}(\boldsymbol{X},S;\blam,\boldsymbol{\mu}_{\bg0})\}&=\mathbb{E}\left\{\mathbb{E}\{\mathcal{D}(\boldsymbol{X},S;\blam,\boldsymbol{\mu}_{\bg0})\mid \boldsymbol{X}\}\right\}\\
    &=\mathbb{E}\{\text{Pr}(S=1|\boldsymbol{X})\ \text{exp}\{\blam^\top\bg(\boldsymbol{X})\}\{\bg(\boldsymbol{X})-\mathbb{E}\{\bg(\boldsymbol{X})\}\}\}\\
    &=\mathbb{E}\{\pi(\boldsymbol{X})\ \text{exp}\{\blam^\top\bg(\boldsymbol{X})\}\{\bg(\boldsymbol{X})-\mathbb{E}\{\bg(\boldsymbol{X})\}\}\}.
\end{align*}
To construct unbiased estimating equations, we hope $\mathbb{E}\{\mathcal{D}(\boldsymbol{X},S;\blam,\boldsymbol{\mu}_{\bg0})\}=0$, which means we hope $\pi(\boldsymbol{X})\ \text{exp}\{\blam^\top\bg(\boldsymbol{X})\}$ to be constant. Based on Assumption \ref{as:cw_loglinear}, we have $\pi(\boldsymbol{X})=\text{exp}\{\balp_0^\top\bg(\boldsymbol{X})\}$. Therefore, we have $$\pi(\boldsymbol{X})\ \text{exp}\{\blam^\top\bg(\boldsymbol{X})\}=\text{exp}\{\balp_0^\top\bg(\boldsymbol{X})\} \text{exp}\{\blam^\top\bg(\boldsymbol{X})\}=\text{exp}\{(\balp_0+\blam)^\top\bg(\boldsymbol{X})\}.$$ To make $\pi(\boldsymbol{X})\ \text{exp}\{\blam^\top\bg(\boldsymbol{X})\}$ constant, we can let $\balp_0+\blam=0$. Then, we look at the denominator of $\hat{q}_i$, i.e., ${\sum_{i=1}^n \text{exp}\{{{\blam}^\top \bg(\boldsymbol{X}_i)}\}}$.
\begin{align*}
    \frac{1}{N}{\sum_{i=1}^n \text{exp}\{{{\blam}^\top \bg(\boldsymbol{X}_i)}\}}&=\frac{1}{N}\sum_{i=1}^N S_i\ \text{exp}\{{{\blam}^\top \bg(\boldsymbol{X}_i)}\}\\
    &=\frac{1}{N}\sum_{i=1}^N S_i\ \text{exp}\{{{-\balp_0}^\top \bg(\boldsymbol{X}_i)}\}+O_p(n^{-1/2}/N)\\
    &=1+O_p(N^{-1/2})+O_p(n^{-1/2}/N)\\
    &=1+o_p(1).
\end{align*}
Thus, 
\begin{align*}
    \hat{q}_i(\boldsymbol{X}_i)&=q(\boldsymbol{X}_i;\hat{\blam})=\frac{\text{exp}\{{\hat{\blam}^\top \bg(\boldsymbol{X}_i)}\}}{\sum_{i=1}^n \text{exp}\{{\hat{\blam}^\top \bg(\boldsymbol{X}_i)}\}}\\
    &=\frac{\text{exp}\{{{-\balp_0}^\top \bg(\boldsymbol{X}_i)}\}}{N}+O_p(n^{-1/2}/N)\\
    &=\frac{\pi^{-1}_S(\boldsymbol{X}_i;\balp_0)}{N}+O_p(n^{-1/2}/N).
\end{align*}
Based on this, as $n\rightarrow\infty$, we have $\hat{q}_i(\boldsymbol{X}_i)\xrightarrow{P} (N\pi(\boldsymbol{X}_i;\balp_0))^{-1}$.
\begin{align*}
    \hat{\tau}_{\text{CW}} &= \frac{\sum_{i\neq j}^{n} \hat{q}_i(\boldsymbol{X}_i)\hat{q}_j(\boldsymbol{X}_j)  \mathbb{I}(Y_i > Y_j, D_i = 1, D_j = 0, S_i = 1, S_j = 1)}{\sum_{i\neq j}^{n} \hat{q}_i(\boldsymbol{X}_i)\hat{q}_j(\boldsymbol{X}_j)  \mathbb{I}(D_i = 1, D_j = 0, S_i = 1, S_j = 1)}\\
    &=\frac{\sum_{i\neq j}^{n} \frac{1}{\pi(\boldsymbol{X}_i;\balp_0)\pi(\boldsymbol{X}_j;\balp_0)}  \mathbb{I}(Y_i > Y_j, D_i = 1, D_j = 0, S_i = 1, S_j = 1)}{\sum_{i\neq j}^{n} \frac{1}{\pi(\boldsymbol{X}_i;\balp_0)\pi(\boldsymbol{X}_j;\balp_0)}  \mathbb{I}(D_i = 1, D_j = 0, S_i = 1, S_j = 1)}\\
    &=\tau_0+o_p(1).
\end{align*}
\end{proof}

\subsubsection{OM+RWD Estimator $\hat{\tau}_\text{OM+RWD}$}\label{app.consistency_omrwd}
\begin{theorem}\label{thm:consistency_omrwd}
    (Consistency of OM+RWD estimator): Under Assumptions \ref{as1} - \ref{as3}, assume $\hat\bbeta\xrightarrow{P}\bbeta^*$ and $Y_i\mid \boldsymbol{X}_i,D_i \sim \mathcal{G}(\boldsymbol{X}_i, D_i; \bbeta^*)$ with CDF being differentiable, then
    $$\hat{\tau}_\text{OM+RWD}\xrightarrow{P}\tau_0.$$
\end{theorem}

\begin{remark}
    For a class of smooth functions $\{l(\boldsymbol{X}_i,\boldsymbol{X}_j;\bbeta)\mid \bbeta\in \mathbb{B}\}$, the Lipschitz condition validates stochastic equicontinuity.\\
    \textit{(Lipschitz condition)} When there exists a function $b(\boldsymbol{X}_i,\boldsymbol{X}_j)$ such that $\mathbb{E}\{b(\boldsymbol{X}_i,\boldsymbol{X}_j)\}<\infty$, for all $\bbeta,\bbeta^*\in \mathbb{B}$, 
    $$\mid l(\boldsymbol{X}_i,\boldsymbol{X}_j;\bbeta)-l(\boldsymbol{X}_i,\boldsymbol{X}_j;\bbeta^*)\mid\leq b(\boldsymbol{X}_i,\boldsymbol{X}_j)\parallel \bbeta-\bbeta^*\parallel.$$
    Two substitute conditions for the above assumption are
\begin{enumerate}
    \item $l(\boldsymbol{X}_i,\boldsymbol{X}_j;\bbeta)$ is differentiable with a derivative of $\partial l(\boldsymbol{X}_i,\boldsymbol{X}_j;\bbeta)/\partial\bbeta$;
    \item $\mathbb{E}\left\{\text{sup}_{\bbeta\in\mathbb{B}}\parallel \partial l(\boldsymbol{X}_i,\boldsymbol{X}_j;\bbeta)/\partial\bbeta\parallel \right\}<\infty$. 
\end{enumerate}
\end{remark}

\begin{proof}[Proof of Theorem \upshape\ref{thm:consistency_omrwd}]
Assume $Y_i\mid \boldsymbol{X}_i,D_i \sim \mathcal{G}(\boldsymbol{X}_i, D_i; \bbeta)$. For simplicity, we assume $\mathcal{G}(\boldsymbol{X}_i, D_i; \bbeta)$ to be $N(M(\boldsymbol{X}_i,D_i;\bbeta), \sigma_1^2  D_i + \sigma_0^2  (1-D_i))$ when $Y_i\mid \boldsymbol{X}_i,D_i$ follows Normal distribution, while $\mathcal{G}(\boldsymbol{X}_i, D_i; \bbeta)$ can be generalized to other distributions. The mean is $M(\boldsymbol{X}_i,D_i=d;\bbeta)=\mathbb{E}\{Y_i\mid \boldsymbol{X}_i, D_i=d\}$($d\in\{0,1\}$) and the variances are different by $D$ as a group indicator. We give an example of the normal distribution to show the consistency of the OM estimator with RWD, while the same proof holds for any other distribution with a CDF. Given RWD is representative of the target population,
\begin{align*}
    \text{Pr}(Y_i>Y_j\mid \boldsymbol{X}_i,\boldsymbol{X}_j,D_i=1,D_j=0)&=\text{Pr}(Y_i-Y_j>0\mid \boldsymbol{X}_i,\boldsymbol{X}_j,D_i=1,D_j=0)\\
    &=\Phi\left(\frac{M(\boldsymbol{X}_i,D_i=1;\bbeta)-M(\boldsymbol{X}_j,D_j=0;\bbeta)}{\sqrt{\sigma_0^2+\sigma_1^2}}\right)
\end{align*}
Assume $\hat\bbeta\xrightarrow{P}\bbeta^*$ and $\hat{\sigma}_d^2\xrightarrow{P}\sigma_d^2$, by the continuous mapping theorem, we have
\begin{align*}
    \Phi\left(\frac{M(\boldsymbol{X}_i,D_i=1;\hat\bbeta)-M(\boldsymbol{X}_j,D_j=0;\hat\bbeta)}{\sqrt{\hat{\sigma}_0^2+\hat{\sigma}_1^2}}\right)
    \xrightarrow{P} 
    \text{Pr}(Y_i>Y_j\mid \boldsymbol{X}_i,\boldsymbol{X}_j,D_i=1,D_j=0).
\end{align*}

Based on the remark above, we then show $$\left\{\Phi\left(\frac{M(\boldsymbol{X}_i,D_i=1;\bbeta)-M(\boldsymbol{X}_j,D_j=0;\bbeta)}{\sqrt{\sigma_0^2+\sigma_1^2}}\right):\bbeta\in \mathbb{B}\right\}$$ is stochastically equicontinuous by using the Lipschitz condition. Define $$l(\boldsymbol{X}_i,\boldsymbol{X}_j;\bbeta)=\Phi\left(\frac{M(\boldsymbol{X}_i,D_i=1;\bbeta)-M(\boldsymbol{X}_j,D_j=0;\bbeta)}{\sqrt{\sigma_0^2+\sigma_1^2}}\right),$$
where $\Phi(\cdot)$ is smooth. Let
\[\Delta_{ij}(\bbeta)=\frac{M(\boldsymbol{X}_i,D_i=1;\bbeta)-M(\boldsymbol{X}_j,D_j=0;\bbeta)}{\sqrt{\sigma_0^2+\sigma_1^2}},\]
and thus 
\begin{align*}
    \frac{\partial l(\boldsymbol{X}_i,\boldsymbol{X}_j;\bbeta)}{\partial\bbeta}&=\phi\left(\Delta_{ij}(\bbeta)\right) \frac{dM(\boldsymbol{X}_i,D_i=1;\bbeta)/d\bbeta-dM(\boldsymbol{X}_j,D_j=0;\bbeta)/d\bbeta}{\sqrt{\sigma_0^2+\sigma_1^2}}\\
    &\leq \frac{1}{\sqrt{2\pi}} \frac{dM(\boldsymbol{X}_i,D_i=1;\bbeta)/d\bbeta-dM(\boldsymbol{X}_j,D_j=0;\bbeta)/d\bbeta}{\sqrt{\sigma_0^2+\sigma_1^2}},
\end{align*}
as for any $Z\sim N(0,1)$, we have $\phi(Z=z)=\frac{1}{\sqrt{2\pi}}e^{-z^2/2}\leq \frac{1}{\sqrt{2\pi}}$. Let $dM(\boldsymbol{X}_i,D_i=1;\bbeta)/d\bbeta=f(\boldsymbol{X}_i;\bbeta)$. 
Therefore,
\begin{align*}
    \parallel \frac{\partial l(\boldsymbol{X}_i,\boldsymbol{X}_j;\bbeta)}{\partial\bbeta}\parallel &\leq\frac{1}{\sqrt{2\pi(\sigma_0^2+\sigma_1^2)}}\cdot\parallel f(\boldsymbol{X}_i;\bbeta)-f(\boldsymbol{X}_j;\bbeta)\parallel\\
    &\leq\frac{1}{\sqrt{2\pi(\sigma_0^2+\sigma_1^2)}}\cdot \parallel \frac{\partial f(\tilde{X};\bbeta)}{\partial \tilde{X}}\parallel\cdot \parallel \boldsymbol{X}_i-\boldsymbol{X}_j \parallel,
\end{align*}
where $\tilde{X}=t\boldsymbol{X}_i+(1-t)\boldsymbol{X}_j$ for some $t\in (0,1)$. Assume $f(X)$ is differentiable in $X$, and its derivative is bounded uniformly over $\bbeta \in \mathbb{B}$, i.e., $\sup_{\bbeta\in \mathbb{B}}\parallel \frac{\partial f(X;\bbeta)}{\partial X}\parallel\leq L(X)$ with $\mathbb{E}\{L(X)\}<\infty$. Therefore, the Lipschitz condition is satisfied by
\begin{align*}
    \mathbb{E}\left\{\text{sup}_{\bbeta\in\mathbb{B}}\parallel \frac{\partial l(\boldsymbol{X}_i,\boldsymbol{X}_j;\bbeta)}{\partial\bbeta}\parallel \right\}\leq\frac{1}{\sqrt{2\pi(\sigma_0^2+\sigma_1^2)}} \mathbb{E}\left\{L(\tilde{X})\parallel \boldsymbol{X}_i-\boldsymbol{X}_j \parallel\right\}<\infty.
\end{align*}
Therefore, $\left\{\Phi\left(\frac{M(\boldsymbol{X}_i,D_i=1;\bbeta)-M(\boldsymbol{X}_j,D_j=0;\bbeta)}{\sqrt{\sigma_0^2+\sigma_1^2}}\right):\bbeta\in \mathbb{B}\right\}$ is stochastically equicontinuous. Because $\mathbb{I}(D_i=1,D_j=0,S_i=0, S_j=0)$ is always bounded, we have 
\begin{align*}
    \left\{\frac{1}{m(m-1)}\sum_{i\neq j}d^{\text{om+rwd}}_k(\boldsymbol{C}_i,\boldsymbol{C}_j;\bbeta)\right\}    
\end{align*}
are stochastically equicontinuous, where 
\begin{align*}
    &d^{\text{om+rwd}}_{k=1}(\boldsymbol{C}_i,\boldsymbol{C}_j;\hat\bbeta)\\
    &\qquad=\Phi\left(\frac{M(\boldsymbol{X}_i,D_i=1,S_i=0;\hat\bbeta)-M(\boldsymbol{X}_j,D_j=0,S_j=0;\hat\bbeta)}{\sqrt{\hat{\sigma}_1^2+\hat{\sigma}_0^2}}\right) \mathbb{I}(D_i = 1, D_j = 0, S_i = 0, S_j = 0),
\end{align*} 
where $d^{\text{om+rwd}}_{k=0}(\boldsymbol{C}_i,\boldsymbol{C}_j;\hat\bbeta)= \mathbb{I}(D_i = 1, D_j = 0, S_i = 0, S_j = 0).$
Therefore, given Lemma \ref{lem1}, when assuming $\hat\bbeta\xrightarrow{P}\bbeta^*$, we have 
$$\frac{1}{m(m-1)}\sum_{i\neq j}d^{\text{om+rwd}}_{1}(\boldsymbol{C}_i,\boldsymbol{C}_j;\hat\bbeta)\xrightarrow{P}\mathbb{E}\{d^{\text{om+rwd}}_{1}(\boldsymbol{C}_i,\boldsymbol{C}_j;\bbeta^*)\},$$
and
\begin{align*}
    \mathbb{E}\{d^{\text{om+rwd}}_{1}(\boldsymbol{C}_i,\boldsymbol{C}_j;\bbeta^*)\}&=\mathbb{E}\left\{\Phi\left(\Delta_{ij}(\bbeta^*)\right)\mathbb{I}(D_i=1, D_j=0, S_i=0, S_j=0)\right\}\\
    &=\mathbb{E}\left\{\text{Pr}(Y_i>Y_j\mid D_i=1, D_j=0, S_i=0, S_j=0) \mathbb{I}(D_i = 1, D_j = 0, S_i = 0, S_j = 0)\right\}.
\end{align*}
Similarly, 
$$\frac{1}{m(m-1)}\sum_{i\neq j}d^{\text{om+rwd}}_{0}(\boldsymbol{C}_i,\boldsymbol{C}_j;\hat\bbeta)\xrightarrow{P}\mathbb{E}\{d^{\text{om+rwd}}_{0}(\boldsymbol{C}_i,\boldsymbol{C}_j;\bbeta^*)\},$$
and 
$$\mathbb{E}\{d^{\text{om+rwd}}_{0}(\boldsymbol{C}_i,\boldsymbol{C}_j;\bbeta^*)\}=\mathbb{E}\left\{ \mathbb{I}(D_i = 1, D_j = 0, S_i = 0, S_j = 0)\right\}.$$
By the continuous mapping theorem, we complete the proof of Theorem \ref{thm:consistency_omrwd} by
\begin{align*}
\hat{\tau}_{\text{OM+RWD}}\xrightarrow{P}\tau_0=\frac{\mathbb{E} \left[\text{Pr} (Y_i>Y_j\mid D_i = 1, D_j = 0, S_i = 0, S_j = 0, \boldsymbol{X}_i,\boldsymbol{X}_j) \mathbb{I}(D_i = 1, D_j = 0, S_i = 0, S_j = 0)\right]}{\mathbb{E}\left[\mathbb{I}(D_i=1, D_j=0, S_i=0, S_j=0)\right]}.
\end{align*}
Therefore, $\hat{\tau}_{\text{OM+RWD}}$ is consistent for $\tau_0$ if the outcome model is correctly specified.
\end{proof}

\subsubsection{OM Estimator $\hat{\tau}_\text{OM}$}\label{app.b4}
\begin{theorem}\label{thm:consistency_om}
    (Consistency of OM estimator): Assume $\{\frac{1}{n}\sum_{i=1}^nq(\boldsymbol{X}_i;\blam),\blam\in \mathcal{L}\}$ is stochastically equicontinuous. Under Assumption \ref{as1} - \ref{as:cw_loglinear}, if $\mathbb{E}\left[Y\mid \boldsymbol{X},D,S=1\right]=M(\boldsymbol{X},D;\boldsymbol{\bbeta}^*)$ with  $\hat{\boldsymbol{\bbeta}}\rightarrow\boldsymbol{\bbeta}^*$, then 
    \begin{align*}
        \hat{\tau}_\text{OM}\xrightarrow{P}\tau_0.
    \end{align*}
\end{theorem}
\begin{proof}
    Similar to the proof of consistency of the IPSW estimator, by $q(\boldsymbol{X}_i;\blam)$ being a calibration weight is uniformly bounded and applying Lemma \ref{lem2}, we have$$ \left\{\frac{1}{n(n-1)}\sum_{i\neq j}q(\boldsymbol{X}_i;\blam)q(\boldsymbol{X}_j;\blam), \blam\in \mathcal{L}\right\}$$ to be stochastically equicontinuous, where $q(\boldsymbol{X}_i;\blam)q(\boldsymbol{X}_j;\blam)$ is based on the calibration weights for subject $i$ and $j$. As $\mathcal{P}_{ij}(\boldsymbol{V}_i,\boldsymbol{V}_j;\boldsymbol{\bbeta},\sigma_1,\sigma_0)=\text{Pr}(Y_i>Y_j\mid D_i=1, D_j=0, S_i=1,S_j=1,\boldsymbol{X}_i,\boldsymbol{X}_j)$ is bounded between 0 and 1, we can apply Lemma \ref{lem1}. Assume $\bbeta_0$ and $\blam_0$ are the true parameters of $\bbeta$ and $\blam$. For $k\in \{0,1\}$ we have 
    \begin{align*}
        \frac{1}{n(n-1)}\sum_{i\neq j}d_k^{\text{om}}(\boldsymbol{V}_i,\boldsymbol{V}_j;\hat\blam,\hat{\bbeta})\xrightarrow{P}\mathbb{E}\left\{d_k^{\text{om}}(\boldsymbol{V}_i,\boldsymbol{V}_j;\blam_0,\boldsymbol{\bbeta}_0)\right\},
    \end{align*}
    where 
    \begin{align*}
        &\mathbb{E}\{d^{\text{om}}_1(\boldsymbol{V}_i,\boldsymbol{V}_j;\bbeta_0, \blam_0)\}\\
        &=\mathbb{E}\left\{w(\boldsymbol{X}_i,\boldsymbol{X}_j) \text{Pr}(Y_i>Y_j\mid D_i=1, D_j=0, S_i=1, S_j=1, \boldsymbol{X}_i, \boldsymbol{X}_j) \mathbb{I}(D_i=1,D_j=0,S_i=1,S_j=1)\right\}
    \end{align*} 
    and 
    \begin{align*}
        \mathbb{E}\{d^{\text{om}}_0(\boldsymbol{V}_i,\boldsymbol{V}_j;\blam_0)\}&=\mathbb{E}\left\{w(\boldsymbol{X}_i,\boldsymbol{X}_j) \mathbb{I}(D_i = 1, D_j = 0, S_i = 1, S_j = 1)\right\}\\
        &=\mathbb{E}\left\{w(\boldsymbol{X}_i,\boldsymbol{X}_j) \mathbb{I}(D_i=1,D_j=0,S_i=1,S_j=1)\right\}.
    \end{align*}

    As demonstrated in Section \ref{sec:method_CW}, under Assumption \ref{as:cw_loglinear}, we have $\hat{q}_i(\boldsymbol{X}_i)\xrightarrow{P} (N\pi(\boldsymbol{X}_i;\balp_0))^{-1}$ as $n\rightarrow\infty$. Therefore, as the sample size of the validation cohort increases, the CW weights $$\hat{w}_{ij}=\hat{q}_i(\boldsymbol{X}_i)\hat{q}_j(\boldsymbol{X}_j)=\hat{q}(\boldsymbol{X}_i;\blam)\hat{q}(\boldsymbol{X}_j;\blam)\xrightarrow{P}\frac{1}{\text{Pr}(S_i\mid \boldsymbol{X}_i)\text{Pr}(S_j\mid \boldsymbol{X}_j)}.$$ Assume $\boldsymbol{\bbeta}\rightarrow \boldsymbol{\bbeta}_0$, $\hat{\sigma}_1\xrightarrow{P}\sigma_{01}$, and $\hat{\sigma}_0\xrightarrow{P}\sigma_{00}$, where $\boldsymbol{\bbeta}_0$ is the true parameter for the outcome model. Then, under Assumption \ref{as3}, we have
    \begin{align*}
        \mathcal{P}_{ij}(\boldsymbol{V}_i, \boldsymbol{V}_j; \hat\bbeta, \hat{\sigma}_0,\hat{\sigma}_1)\xrightarrow{P}\text{Pr}(Y_i>Y_j\mid D_i=1, D_j=0, \boldsymbol{X}_i, \boldsymbol{X}_j).
    \end{align*}
    Therefore, by the continuous mapping theorem, the proof is complete by
    \begin{align*}
        \hat{\tau}_{\text{OM}}\xrightarrow{P}\tau_0=\frac{\mathbb{E} \left[\frac{\text{Pr}(Y_i>Y_j\mid D_i=1, D_j=0, S_i=1, S_j=1, \boldsymbol{X}_i, \boldsymbol{X}_j) \mathbb{I}(D_i=1, D_j=0, S_i=1, S_j=1)}{\text{Pr}(S_i=1\mid \boldsymbol{X}_i)\text{Pr}(S_j=1\mid \boldsymbol{X}_j)}\right]}{\mathbb{E}\Bigl[\frac{\mathbb{I}(D_i=1, D_j=0, S_i=1, S_j=1)}{\text{Pr}(S_i=1\mid \boldsymbol{X}_i)\text{Pr}(S_j=1\mid \boldsymbol{X}_j)}\Bigr]}
    \end{align*}
\end{proof}
\subsubsection{ACW Estimator $\hat{\tau}_\text{ACW}$}\label{app.consistency_acw}
\begin{theorem}\label{thm:consistency_acw}
    Under Assumptions \ref{as1}-\ref{as:cw_loglinear},
    $\hat{\tau}_{\text{ACW}}$ is a consistent, doubly robust estimator for $\tau_0$.
\end{theorem}
\begin{proof}
    In addition to establishing the consistency of $\hat{\tau}_{\text{ACW}}$, we also demonstrate the double robustness of this augmented estimator.\\
    \textbf{Situation 1:} Under Assumption \ref{as:cw_loglinear}, $\hat{q}_i(\boldsymbol{X}_i)$ is correctly estimated with $\hat{q}_i(\boldsymbol{X}_i)\xrightarrow{P}q_0(\boldsymbol{X}_i)$, where $q_0(\boldsymbol{X}_i)$ is the true CW weights given baseline covariates $\boldsymbol{X}_i$ and implies $\hat{\blam}\xrightarrow{P}\blam_0$. However, $\hat\bbeta\xrightarrow{P}\bbeta^*$, where $\bbeta^*$ is not necessarily $\bbeta_0$. 

    Under this situation, as proved in Section~\ref{app.consistency_cw}, we have $\hat{\tau}_{\text{CW}}\xrightarrow{P}\tau_0$. Therefore, we aim to show $\hat{\tau}_{\text{OM+RWD}}-\hat{\tau}_{\text{OM}}\xrightarrow{P}0$. Define $\hat{\tau}_{\text{OM+RWD}}=U_1$ as
    \begin{align*}
        U_1=\frac{U_{11}}{U_{12}}=\frac{\frac{1}{m(m-1)}\sum_{i\neq j}^{m+n} \mathcal{P}_{ij}(\boldsymbol{C}_i,\boldsymbol{C}_j;\hat\bbeta, \hat{\sigma}_0, \hat{\sigma}_1) \mathbb{I}(D_i = 1, D_j = 0, S_i = 0, S_j = 0)}{\frac{1}{m(m-1)}\sum_{i\neq j}^{m+n}  \mathbb{I}(D_i = 1, D_j = 0, S_i = 0, S_j = 0)}.
    \end{align*}
    
    Similar to the proof for Theorem \ref{thm:consistency_omrwd}, we have $U_{11}\xrightarrow{P}\mathbb{E}\{l_{11}\}$ and $U_{12}\xrightarrow{P}\mathbb{E}\{l_{12}\}$.
    Define the kernels as
    $$l_{11}(\boldsymbol{C}_i, \boldsymbol{C}_j;\bbeta, \sigma_0,\sigma_1)=\mathcal{P}_{ij}(\boldsymbol{C}_i, \boldsymbol{C}_j; \bbeta, \sigma_0,\sigma_1) \mathbb{I}(D_i=1,D_j=0,S_i=1,S_j=1)$$
    and
    $$l_{12}(\boldsymbol{C}_i, \boldsymbol{C}_j)= \mathbb{I}(D_i=1,D_j=0,S_i=1,S_j=1).$$
    Given RWD $\mathcal{R}$ is representative of the target population, we have the outcome(biomarker) distribution for $S=0$ that can represent the target population, and thus
    
    \begin{align*}
        \mathbb{E}\{l_{11}\}&=\mathbb{E}\{(1-S_i)(1-S_j)\mathcal{P}_{ij}(\boldsymbol{C}_i, \boldsymbol{C}_j; \bbeta^*, \sigma_0^*,\sigma_1^*)\mathbb{I}(D_i=1,D_j=0)\}\\
        &=\mathbb{E}_0\{\mathcal{P}_{ij}(\boldsymbol{C}_i, \boldsymbol{C}_j; \bbeta^*, \sigma_0^*,\sigma_1^*)\mathbb{I}(D_i=1,D_j=0)\}\\
        &=\mathbb{E}_0\{\mathcal{P}_{ij}(\boldsymbol{C}_i, \boldsymbol{C}_j; \bbeta^*, \sigma_0^*,\sigma_1^*)\}\text{Pr}(D_i=1,D_j=0)
    \end{align*}
    and 
    \begin{align*}
        \mathbb{E}\{l_{12}\}&=\mathbb{E}\{(1-S_i)(1-S_j)\mathbb{I}(D_i=1,D_j=0)\}=\mathbb{E}_0\{\mathbb{I}(D_i=1,D_j=0)\}=\text{Pr}(D_i=1,D_j=0),
    \end{align*}
    where $\mathbb{E}_0$ is the target population-level expectation. So,
    \begin{align*}
        U_1\xrightarrow{P}\mathbb{E}_0\left\{\mathcal{P}_{ij}(\boldsymbol{C}_i, \boldsymbol{C}_j; \bbeta^*, \sigma_0^*,\sigma_1^*)\right\}.
    \end{align*}
    
    Then, for the second term, we define $\hat{\tau}_{\text{OM}}=U_2$ as the ratio of two $U$-processes
    \begin{align*}
        U_2=\frac{U_{21}}{U_{22}}=\frac{\frac{1}{n(n-1)}\sum_{i\neq j}^{n}\hat{q}_{i}(\boldsymbol{X}_i)\hat{q}_{j}(\boldsymbol{X}_j) \mathcal{P}_{ij}(\boldsymbol{V}_i, \boldsymbol{V}_j; \hat\bbeta, \hat{\sigma}_0,\hat{\sigma}_1) \mathbb{I}(D_i=1,D_j=0,S_i=1,S_j=1)}{\frac{1}{n(n-1)}\sum_{i\neq j}^{n}\hat{q}_{i}(\boldsymbol{X}_i)\hat{q}_{j}(\boldsymbol{X}_j) \mathbb{I}(D_i=1,D_j=0,S_i=1,S_j=1)}.
    \end{align*}
    The kernels are defined as
    $$l_{21}(\boldsymbol{V}_i, \boldsymbol{V}_j;\blam,\bbeta, \sigma_0,\sigma_1)=q_{i}(\boldsymbol{X}_i;\blam)q_{j}(\boldsymbol{X}_j;\blam) \mathcal{P}_{ij}(\boldsymbol{V}_i, \boldsymbol{V}_j; \bbeta, \sigma_0,\sigma_1) \mathbb{I}(D_i=1,D_j=0,S_i=1,S_j=1)$$
    and
    $$l_{22}(\boldsymbol{V}_i, \boldsymbol{V}_j;\blam)=q_{i}(\boldsymbol{X}_i;\blam)q_{j}(\boldsymbol{X}_j;\blam) \mathbb{I}(D_i=1,D_j=0,S_i=1,S_j=1).$$
    Similar to the proof of consistency of $\hat{\tau}_{\text{OM+RWD}}$, given equicontinuous and uniformly bounded, Lemma \ref{lem1} and Lemma \ref{lem2} are applied to validate $U_{21}\xrightarrow{P}\mathbb{E}\{l_{21}\}$ and $U_{22}\xrightarrow{P}\mathbb{E}\{l_{22}\}$. Therefore, by continuous mapping theorem, we have $U_2\xrightarrow{P}\mathbb{E}\{l_{21}\}/\mathbb{E}\{l_{22}\}$, where
    \begin{align*}
    \frac{\mathbb{E}\{l_{21}\}}{\mathbb{E}\{l_{22}\}}=\frac{\mathbb{E}\{q_0(\boldsymbol{X}_i)q_0(\boldsymbol{X}_j)\mathcal{P}_{ij}(\boldsymbol{V}_i, \boldsymbol{V}_j; \bbeta^*, \sigma_0^*,\sigma_1^*)\mathbb{I}(D_i=1,D_j=0,S_i=1,S_j=1)\}}{\mathbb{E}\{q_0(\boldsymbol{X}_i)q_0(\boldsymbol{X}_j)\mathbb{I}(D_i=1,D_j=0,S_i=1,S_j=1)\}}.
    \end{align*}
    By the definition of CW weights, for any $i\neq j$ and $h(X)$ as any function of $X$,
    \begin{align*}
        \mathbb{E}\{q_0(\boldsymbol{X}_i)q_0(\boldsymbol{X}_j)\mathbb{I}(S_i=1, S_j=1)h(\boldsymbol{X}_i,\boldsymbol{X}_j)\}=\mathbb{E}_0\{h(\boldsymbol{X}_i,\boldsymbol{X}_j)\},
    \end{align*}
    where $\mathbb{E}_0$ is the true population-level expectation. So,
    \begin{align*}
        \mathbb{E}\{l_{21}\}&=\mathbb{E}\{q_0(\boldsymbol{X}_i)q_0(\boldsymbol{X}_j)S_iS_j\mathcal{P}_{ij}(\boldsymbol{V}_i, \boldsymbol{V}_j; \bbeta^*, \sigma_0^*,\sigma_1^*)\mathbb{I}(D_i=1,D_j=0)\}\\
        &=\mathbb{E}_0\{\mathcal{P}_{ij}(\boldsymbol{V}_i, \boldsymbol{V}_j; \bbeta^*, \sigma_0^*,\sigma_1^*)\mathbb{I}(D_i=1,D_j=0)\}\\
        &=\mathbb{E}_0\left\{\mathbb{E}_0\{\mathcal{P}_{ij}(\boldsymbol{V}_i, \boldsymbol{V}_j; \bbeta^*, \sigma_0^*,\sigma_1^*)\mathbb{I}(D_i=1,D_j=0)\mid D_i,D_j\}\right\}\\
        &=\mathbb{E}_0\left\{\mathcal{P}_{ij}(\boldsymbol{V}_i, \boldsymbol{V}_j; \bbeta^*, \sigma_0^*,\sigma_1^*)\right\}\text{Pr}(D_i=1,D_j=0),
    \end{align*}
    and 
    \begin{align*}
        \mathbb{E}\{l_{22}\}&=\mathbb{E}\{q_0(\boldsymbol{X}_i)q_0(\boldsymbol{X}_j)S_iS_jI(D_i=1,D_j=0)\}=\mathbb{E}_0\{\mathbb{I}(D_i=1,D_j=0)\}\\
        &=\text{Pr}(D_i=1,D_j=0).
    \end{align*}
    Therefore,
    $$U_2\xrightarrow{P}\mathbb{E}_0\left\{\mathcal{P}_{ij}(\boldsymbol{V}_i, \boldsymbol{V}_j; \bbeta^*, \sigma_0^*,\sigma_1^*)\right\}.$$
    Since $\mathcal{R}=\mathcal{C}\backslash\mathcal{V}$ can be representative of the target population, we have $\mathbb{E}_0\left\{\mathcal{P}_{ij}(\boldsymbol{C}_i, \boldsymbol{C}_j; \bbeta^*, \sigma_0^*,\sigma_1^*)\right\}=\mathbb{E}_0\left\{\mathcal{P}_{ij}(\boldsymbol{V}_i, \boldsymbol{V}_j; \bbeta^*, \sigma_0^*,\sigma_1^*)\right\}$. Therefore, 
    \begin{align*}
        U_1-U_2=\hat{\tau}_{\text{OM+RWD}}-\hat{\tau}_{\text{OM}}\xrightarrow{P}0.
    \end{align*}
    Based on this result, given $\hat{q}_i(\boldsymbol{X}_i)\xrightarrow{P}q_0(\boldsymbol{X}_i)$, we have $\hat{\tau}_{\text{ACW}}\xrightarrow{P}\tau_0$ regardless of outcome model specification. \\
    \textbf{Situation 2:} Given outcome model is correctly specified, which means $\hat\bbeta\xrightarrow{P}\bbeta_0$, while $\hat{q}_i(\boldsymbol{X}_i)\xrightarrow{P}q^*(\boldsymbol{X}_i)$ and $q^*(\boldsymbol{X}_i)$ is not necessarily the same as $q_0(\boldsymbol{X}_i)$.

    Under this situation, as shown in Section~\ref{app.consistency_omrwd}, $\hat{\tau}_{\text{OM+RWD}}\xrightarrow{P}\tau_0$. Then, we aim to show $\hat{\tau}_{\text{CW}}-\hat{\tau}_{\text{OM}}\xrightarrow{P}0$. Similarly, we define $\hat{\tau}_{\text{CW}}=U_1$ as
    \begin{align*}
        U_1 = \frac{\frac{1}{n(n-1)}\sum_{i \neq j}^{n} \hat{w}^{\text{cw}}(\boldsymbol{X}_i, \boldsymbol{X}_j) \mathbb{I}(Y_i > Y_j, D_i = 1, D_j = 0, S_i = 1, S_j = 1)}{\frac{1}{n(n-1)}\sum_{i \neq j}^{n} \hat{w}^{\text{cw}}(\boldsymbol{X}_i, \boldsymbol{X}_j) \mathbb{I}(D_i = 1, D_j = 0, S_i = 1, S_j = 1)}.
    \end{align*}
    Define
    $$l_{11}(\boldsymbol{V}_i,\boldsymbol{V}_j;\blam)=q(\boldsymbol{X}_i;\blam)q(\boldsymbol{X}_j;\blam)\mathbb{I}(Y_i > Y_j, D_i = 1, D_j = 0, S_i = 1, S_j = 1)$$
    and 
    $$l_{12}(\boldsymbol{V}_i,\boldsymbol{V}_j;\blam)=q(\boldsymbol{X}_i;\blam)q(\boldsymbol{X}_j;\blam)\mathbb{I}(D_i = 1, D_j = 0, S_i = 1, S_j = 1).$$
    Similar to the proof in Section~\ref{app.consistency_cw}, we have $U_1\xrightarrow{P}\mathbb{E}\{l_{11}\}/\mathbb{E}\{l_{12}\}$. Then we derive the $\mathbb{E}\{l_{11}\}$ and $\mathbb{E}\{l_{12}\}$ as
    \begin{align*}
        &\mathbb{E}\{l_{11}\}=\mathbb{E}\left\{q_i^*(\boldsymbol{X}_i)q_j^*(\boldsymbol{X}_j)\mathbb{I}(Y_i > Y_j, D_i = 1, D_j = 0, S_i = 1, S_j = 1)\right\}\\
        &=\mathbb{E}\left\{\mathbb{E}\{q_i^*(\boldsymbol{X}_i)q_j^*(\boldsymbol{X}_j)\mathbb{I}(Y_i > Y_j, D_i = 1, D_j = 0, S_i = 1, S_j = 1)\mid \boldsymbol{X}_i,\boldsymbol{X}_j\}\right\}\\
        &=\mathbb{E}\left\{q_i^*(\boldsymbol{X}_i)q_j^*(\boldsymbol{X}_j)\mathbb{E}\{\mathbb{I}(Y_i > Y_j, D_i = 1, D_j = 0, S_i = 1, S_j = 1)\mid \boldsymbol{X}_i,\boldsymbol{X}_j\}\right\}\\
        &= \mathbb{E}\!\left\{q_i^*(\boldsymbol{X}_i) q_j^*(\boldsymbol{X}_j)\,\Pr\!\left(Y_i>Y_j \mid D_i{=}1, D_j{=}0, S_i{=}1, S_j{=}1, \boldsymbol{X}_i, \boldsymbol{X}_j\right)\right.\left.\,\Pr\!\left(D_i{=}1, D_j{=}0, S_i{=}1, S_j{=}1 \mid \boldsymbol{X}_i, \boldsymbol{X}_j\right)\right\}
    \end{align*}
    and 
   \begin{align*}
        &\mathbb{E}\{l_{12}\}=\mathbb{E}\left\{q_i^*(\boldsymbol{X}_i)q_j^*(\boldsymbol{X}_j)\mathbb{I}(D_i = 1, D_j = 0, S_i = 1, S_j = 1)\right\}\\
        &=\mathbb{E}\left\{\mathbb{E}\{q_i^*(\boldsymbol{X}_i)q_j^*(\boldsymbol{X}_j)\mathbb{I}(D_i = 1, D_j = 0, S_i = 1, S_j = 1)\mid \boldsymbol{X}_i,\boldsymbol{X}_j\}\right\}\\
        &=\mathbb{E}\left\{q_i^*(\boldsymbol{X}_i)q_j^*(\boldsymbol{X}_j)\mathbb{E}\{\mathbb{I}(D_i = 1, D_j = 0, S_i = 1, S_j = 1)\mid \boldsymbol{X}_i,\boldsymbol{X}_j\}\right\}\\
        &=\mathbb{E}\left\{q_i^*(\boldsymbol{X}_i)q_j^*(\boldsymbol{X}_j)\text{Pr}(D_i = 1, D_j = 0, S_i = 1, S_j = 1\mid \boldsymbol{X}_i,\boldsymbol{X}_j)\right\}.
    \end{align*}
    Therefore, 
    \begin{align*}
        U_1\xrightarrow{P}\frac{\mathbb{E}\left\{q_i^*(\boldsymbol{X}_i)q_j^*(\boldsymbol{X}_j)\text{Pr}(Y_i > Y_j\mid \mathcal{Z}^1_{ij}, \boldsymbol{X}_i, \boldsymbol{X}_j)\text{Pr}(\mathcal{Z}^1_{ij}\mid \boldsymbol{X}_i,\boldsymbol{X}_j)\right\}}{\mathbb{E}\left\{q_i^*(\boldsymbol{X}_i)q_j^*(\boldsymbol{X}_j)\text{Pr}(\mathcal{Z}^1_{ij}\mid \boldsymbol{X}_i,\boldsymbol{X}_j)\right\}},
    \end{align*}
    where $\mathcal{Z}^1_{ij}=\{D_i=1,D_j=0,S_i=1,S_j=1\}$.
    Then, we define $U_2=\hat{\tau}_{\text{OM}}$ as
    \begin{align*}
        U_2=\frac{U_{21}}{U_{22}}=\frac{\frac{1}{n(n-1)}\sum_{i\neq j}^{n}\hat{q}_{i}(\boldsymbol{X}_i)\hat{q}_{j}(\boldsymbol{X}_j) \mathcal{P}_{ij}(\boldsymbol{V}_i, \boldsymbol{V}_j; \hat\bbeta, \hat{\sigma}_0,\hat{\sigma}_1) \mathbb{I}(D_i=1,D_j=0,S_i=1,S_j=1)}{\frac{1}{n(n-1)}\sum_{i\neq j}^{n}\hat{q}_{i}(\boldsymbol{X}_i)\hat{q}_{j}(\boldsymbol{X}_j) \mathbb{I}(D_i=1,D_j=0,S_i=1,S_j=1)}.
    \end{align*}
    The kernels are defined as
    $$l_{21}(\boldsymbol{V}_i, \boldsymbol{V}_j;\blam,\bbeta, \sigma_0,\sigma_1)=q_{i}(\boldsymbol{X}_i;\blam)q_{j}(\boldsymbol{X}_j;\blam) \mathcal{P}_{ij}(\boldsymbol{V}_i, \boldsymbol{V}_j; \bbeta, \sigma_0,\sigma_1) \mathbb{I}(D_i=1,D_j=0,S_i=1,S_j=1)$$
    and
    $$l_{22}(\boldsymbol{V}_i, \boldsymbol{V}_j;\blam)=q_{i}(\boldsymbol{X}_i;\blam)q_{j}(\boldsymbol{X}_j;\blam) \mathbb{I}(D_i=1,D_j=0,S_i=1,S_j=1)$$.
    
    Therefore, we have $$U_2\xrightarrow{P}\mathbb{E}\{l_{21}\}/\mathbb{E}\{l_{22}\},$$
    \begin{align*}
        \mathbb{E}\{l_{21}\}&=\mathbb{E}\{q_i^*(\boldsymbol{X}_i)q_j^*(\boldsymbol{X}_j) \mathcal{P}_{ij}(\boldsymbol{V}_i, \boldsymbol{V}_j; \bbeta_0, \sigma_{00},\sigma_{10}) \mathbb{I}(D_i=1,D_j=0,S_i=1,S_j=1)\}\\
        &=\mathbb{E}\left\{\mathbb{E}\{q_i^*(\boldsymbol{X}_i)q_j^*(\boldsymbol{X}_j) \mathcal{P}_{ij}(\boldsymbol{V}_i, \boldsymbol{V}_j; \bbeta_0, \sigma_{00},\sigma_{10}) \mathbb{I}(D_i=1,D_j=0,S_i=1,S_j=1)\mid \boldsymbol{X}_i,\boldsymbol{X}_j\}\right\}\\
        &=\mathbb{E}\left\{q_i^*(\boldsymbol{X}_i)q_j^*(\boldsymbol{X}_j) \mathcal{P}_{ij}(\boldsymbol{V}_i, \boldsymbol{V}_j; \bbeta_0, \sigma_{00},\sigma_{10})\text{Pr}(D_i=1,D_j=0,S_i=1,S_j=1\mid \boldsymbol{X}_i,\boldsymbol{X}_j)\right\}
    \end{align*}
    and $\mathbb{E}\{l_{22}\}=\mathbb{E}\{l_{12}\}$. 
    
    By definition, we have $$\mathcal{P}_{ij}(\boldsymbol{V}_i, \boldsymbol{V}_j; \bbeta_0, \sigma_{00},\sigma_{10})=\text{Pr}(Y_i > Y_j\mid D_i=1,D_j=0,S_i=1,S_j=1, \boldsymbol{X}_i, \boldsymbol{X}_j).$$ Therefore,
    \begin{align*}
        U_2\xrightarrow{P}\frac{\mathbb{E}\left\{q_i^*(\boldsymbol{X}_i)q_j^*(\boldsymbol{X}_j)\text{Pr}(Y_i > Y_j\mid \mathcal{Z}^1_{ij}, \boldsymbol{X}_i, \boldsymbol{X}_j)\text{Pr}(\mathcal{Z}^1_{ij}\mid \boldsymbol{X}_i,\boldsymbol{X}_j)\right\}}{\mathbb{E}\left\{q_i^*(\boldsymbol{X}_i)q_j^*(\boldsymbol{X}_j)\text{Pr}(\mathcal{Z}^1_{ij}\mid \boldsymbol{X}_i,\boldsymbol{X}_j)\right\}},
    \end{align*}
    where $\mathcal{Z}^1_{ij}=\{D_i=1,D_j=0,S_i=1,S_j=1\}$, and thus
    $$U_1-U_2=\hat{\tau}_{\text{CW}}-\hat{\tau}_{\text{OM}}\xrightarrow{P}0.$$
    So, if the outcome model is correctly specified, then $\hat{\tau}_{\text{ACW}}$ is also consistent for $\tau_0$ regardless of how the CW weights are estimated.   
\end{proof}
\begin{remark}
    Under Assumption \ref{as:cw_loglinear}, $\hat{q}_i(\boldsymbol{X}_i)\xrightarrow{P}q_0(\boldsymbol{X}_i)$ is always true. Therefore, $\hat{\tau}_{\text{ACW}}$ is a double robust estimator against the misspecification of OM. 
\end{remark}

\subsubsection{AIPSW Estimator $\hat{\tau}_\text{AIPSW}$}\label{app.consistency_aipsw}
\begin{theorem}\label{thm:consistency_aipsw}
    Under Assumption \ref{as1}-\ref{as3}, $\hat{\tau}_{\text{AIPSW}}$ is consistent and a double robust estimator for $\tau_0$.
\end{theorem}
The proof for Theorem \ref{thm:consistency_aipsw} is omitted here since the proof here is very similar to the proof of Theorem \ref{thm:consistency_acw}. The difference is that the $q(\boldsymbol{X}_i;\blam)$ is replaced with $p(\boldsymbol{X}_i;\balp)=1/\pi(\boldsymbol{X}_i;\balp)$. Therefore, Situation 1 is corresponding to when the sampling model is correct and we aim to conclude $\hat{\tau}_{\text{OM+RWD}}-\hat{\tau}_{\text{OM}}\xrightarrow{P}0$, where the weights in $\hat{\tau}_{\text{OM}}$ are also replaced with $p_i(\boldsymbol{X}_i;\hat{\balp})$. Similar to the proof of Theorem \ref{thm:consistency_acw}, the Situation 2 is defined as outcome model is correct but the sampling model is not necessarily correct; we aim to show $\hat{\tau}_{\text{IPSW}}-\hat{\tau}_{\text{OM}}\xrightarrow{P}0$. 
\begin{remark}
    $\hat{\tau}_{\text{AIPSW}}$ is consistent for $\tau_0$ when either the outcome model or the sampling model is correctly specified.
\end{remark}

\subsection{Proof of Asymptotic Normality}\label{app.c}%
Before providing the proofs of asymptotic normality for all the proposed estimators, we introduce Lemma $\ref{lem3}$.
\begin{lemma}\label{lem3}
    Let $\mathcal{C} = \{ \boldsymbol{C}_i : i = 1, \dots, n \}$ be a set of independent and identically distributed random variables, and let $l(\boldsymbol{C}_i, \boldsymbol{C}_j; \tau,\boldsymbol{\theta}) $ be a real-valued function that depends on a pair of random variables, parameterized by $(\tau,\boldsymbol{\theta})$. The parameter $(\tau,\boldsymbol{\theta}) \in \mathcal{B}$, which is a bounded Euclidean space. Define the U-process as
\begin{align*}
    U_n(\tau,\boldsymbol{\theta}) = \frac{1}{n(n-1)} \sum_{i \neq j} l(\boldsymbol{C}_i, \boldsymbol{C}_j; \tau,\boldsymbol{\theta}).
\end{align*}
The expectation is defined as
\begin{align*}
    U(\tau,\boldsymbol{\theta}) = \mathbb{E}\left\{l(\boldsymbol{C}_i, \boldsymbol{C}_j; \tau,\boldsymbol{\theta})\right\}
\end{align*}
Assume $\hat{\boldsymbol{\theta}}$ is a consistent estimator of $\boldsymbol{\theta}^*$, i.e., $\hat{\boldsymbol{\theta}}\xrightarrow{P}\boldsymbol{\theta}^*$. And $\hat{\tau}$ is assumed to be the unique solution to $U_n(\tau,\hat{\boldsymbol{\theta}})=0$. Let the $\tau^*$ be the limit of $\hat{\tau}$, which is not necessarily to be $\tau_0$. List the following assumptions:
\begin{enumerate}
    \item $\left\{\sqrt{n}\{U_n(\tau,\boldsymbol{\theta})-U(\tau,\boldsymbol{\theta})\},(\tau,\boldsymbol{\theta})\in \mathcal{B}\right\}$ is stochastically equicontinuous.
    \item $U(\tau,\boldsymbol{\theta})$ is continuously differentiable in $\tau$ and $\boldsymbol{\theta}$.
    \item (Standard regularity condition)\\ The estimator $\hat{\boldsymbol{\theta}}$ is the solution of the estimating equation $$\sum_{i=1}^nh(\boldsymbol{C}_i;\hat{\boldsymbol{\theta}})=0,$$ where $h=(h_1,\cdots,h_p)^\top$ and $p=\text{dim}(\boldsymbol{\theta})$. Define $\boldsymbol{\theta}^*$ as the solution of 
    \begin{align*}
        \mathbb{E}\left\{h(\boldsymbol{C}; \boldsymbol{\theta})\right\} = 0
    \end{align*}
    and assume that this solution is well-defined and unique. Additionally, we impose several conditions. First, the expectation of the squared norm of the estimating function is finite, i.e., $\mathbb{E} \|h(\boldsymbol{C}; \boldsymbol{\theta}^*)\|^2 < \infty$. Second, $\hat{\boldsymbol{\theta}} \xrightarrow{P} \boldsymbol{\theta}^*$. Third, $\forall$ $\boldsymbol{\theta}\in \Theta$, the first derivative of $h(\boldsymbol{C}; \boldsymbol{\theta})$, i.e.,$\frac{\partial h(\boldsymbol{C}; \boldsymbol{\theta})}{\partial \boldsymbol{\theta}}$, exists, and $\mathbb{E}  \left\| \frac{\partial h(\mathbf{O}; \boldsymbol{\theta})}{\partial \boldsymbol{\theta}} \right\|  < \infty$. Furthermore, 
    \begin{align*}
        \mathbb{E} \left\{ \frac{\partial h(\boldsymbol{C}; \boldsymbol{\theta})}{\partial \boldsymbol{\theta}} \Bigg|_{\boldsymbol{\theta} = \boldsymbol{\theta}^*} \right\}
    \end{align*}
    is nonsingular. Lastly, for \( \epsilon_n \xrightarrow{P} 0 \), the uniform convergence condition holds:  
   \begin{align*}
       \sup_x \left\{ \Big| \frac{1}{n} \sum_{i=1}^{n} \left\{ \frac{\partial h(\boldsymbol{C}; x)}{\partial x} - \frac{\partial h(\boldsymbol{C}; \boldsymbol{\theta}^*)}{\partial \boldsymbol{\theta}^*} \right\} \Big| : \big| x - \boldsymbol{\theta}^*\big| < \epsilon_n \right\} \xrightarrow{P} 0.
   \end{align*}
\end{enumerate}
When all these conditions hold, we have 
\begin{align*}
    \sqrt{n}(\hat{\tau}-\tau^*)\xrightarrow{}N(0,\operatorname{Var}\left[\left(u(\tau^*,\boldsymbol{\theta}^*)\right)^{-1}\phi(\boldsymbol{C}_i;\tau^*,\boldsymbol{\theta}^*)\right]),
\end{align*}
where 
$$u(\tau,\boldsymbol{\theta})=\frac{\partial U(\tau,\boldsymbol{\theta})}{\partial\tau},$$
$$\phi(\boldsymbol{C}_i;\tau,\boldsymbol{\theta})=\frac{\partial U(\tau,\boldsymbol{\theta})}{\partial\boldsymbol{\theta}^\top}\left[\mathbb{E}\left\{\frac{\partial h(\boldsymbol{C}_i;\boldsymbol{\theta})}{\partial\boldsymbol{\theta}^\top}\right\}\right]^{-1}h(\boldsymbol{C}_i;\boldsymbol{\theta})-l^*(\boldsymbol{C}_i;\tau,\boldsymbol{\theta}),$$
and 
$$l^*(\boldsymbol{C}_i;\tau,\boldsymbol{\theta})=\mathbb{E}\left\{l(\boldsymbol{C}_i,\boldsymbol{C}_j;\tau,\boldsymbol{\theta})\mid \boldsymbol{C}_i\right\}+\mathbb{E}\left\{l(\boldsymbol{C}_j,\boldsymbol{C}_i;\tau,\boldsymbol{\theta})\mid \boldsymbol{C}_i\right\}.$$
\end{lemma}

\begin{proof}
    Following the approach of \cite{Rotnitzky2006}, we provide the proof below. 
    
    Since $\hat{\tau}$ is assumed to be the solution to $U_n(\tau,\hat{\boldsymbol{\theta}})=0$, we have
    \begin{align*}
        0 &= \sqrt{n}U_n(\hat{\tau},\hat{\boldsymbol{\theta}})\\
        &= \sqrt{n}\left[U_n(\hat{\tau},\hat{\boldsymbol{\theta}}) -U(\hat{\tau},\hat{\boldsymbol{\theta}})-(U_n({\tau}^*,\boldsymbol{{\theta}}^*)-U({\tau}^*,\boldsymbol{{\theta}}^*))\right]+\sqrt{n}[U(\hat{\tau},\hat{\boldsymbol{\theta}})-U({\tau}^*,\boldsymbol{{\theta}}^*)]+\sqrt{n}U_n({\tau}^*,\boldsymbol{{\theta}}^*)\\
        &=op(1)+\sqrt{n}\left[U(\hat{\tau},\hat{\boldsymbol{\theta}})-U({\tau}^*,\boldsymbol{{\theta}}^*)\right]+\sqrt{n}U_n({\tau}^*,\boldsymbol{{\theta}}^*) \qquad\text{(by 1st assumption in Lemma \ref{lem3})}\\
        &=op(1)+\left\{ \frac{\partial U (\tau', \boldsymbol{\theta}')}{\partial \tau} + o_p(1) \right\} \sqrt{n} (\hat{\tau} - \tau^*) + \left\{ \frac{\partial U (\tau', \boldsymbol{\theta}')}{\partial \boldsymbol{\theta}^\top} + o_p(1) \right\} \sqrt{n} (\hat{\boldsymbol{\theta}} - \boldsymbol{\theta}^*)+ \sqrt{n} U_n (\tau^*, \boldsymbol{\theta}^*)\\&\qquad\qquad\qquad\qquad\qquad\qquad\qquad\quad\qquad\qquad\qquad\qquad\text{(by 2nd assumption in Lemma \ref{lem3})},
    \end{align*}
    where $(\tau', \boldsymbol{\theta}')$ is between $(\tau^*, \boldsymbol{\theta}^*)$ and $(\hat{\tau}, \hat{\boldsymbol{\theta}})$. This means that $\mid \tau'-\tau^*\mid\leq \mid\hat{\tau}-\tau^*\mid$ and $\| \boldsymbol{\theta}'-\boldsymbol{\theta}^*\|\leq \|\hat{\boldsymbol{\theta}}-\boldsymbol{\theta}^*\|$. 
    Given the 3rd assumption in Lemma \ref{lem3}, we have 
    \begin{align*}
        0&=\sum_{i=1}^nh(\boldsymbol{C}_i;\hat{\boldsymbol{\theta}})\\
        &=\sum_{i=1}^nh(\boldsymbol{C}_i;{\boldsymbol{\theta}}^*)+\sum_{i=1}^n\frac{\partial h(\boldsymbol{C}_i;{\boldsymbol{\theta}}^*)}{\partial \boldsymbol{\theta}^\top}(\hat{\boldsymbol{\theta}}-\boldsymbol{\theta}^*)+op(n\|\hat{\boldsymbol{\theta}}-\boldsymbol{\theta}^*\|)\qquad\text{ (by Taylpr expansion)}
    \end{align*}
    Multiply both sides by $1/n$, and we have
    \begin{align*}
         \frac{1}{n}\sum_{i=1}^n\frac{\partial h(\boldsymbol{C}_i;{\boldsymbol{\theta}}^*)}{\partial \boldsymbol{\theta}^\top}(\hat{\boldsymbol{\theta}}-\boldsymbol{\theta}^*)=-\frac{1}{n}\sum_{i=1}^nh(\boldsymbol{C}_i;{\boldsymbol{\theta}}^*)+op(\|\hat{\boldsymbol{\theta}}-\boldsymbol{\theta}^*\|)
    \end{align*}
    Then, we rearrange the equation and get
    \begin{align*}
        \hat{\boldsymbol{\theta}}-\boldsymbol{\theta}^*=-\left[\frac{1}{n}\sum_{i=1}^n\frac{\partial h(\boldsymbol{C}_i;{\boldsymbol{\theta}}^*)}{\partial \boldsymbol{\theta}^\top}\right]^{-1}\frac{1}{n}\sum_{i=1}^nh(\boldsymbol{C}_i;{\boldsymbol{\theta}}^*)+op(\|\hat{\boldsymbol{\theta}}-\boldsymbol{\theta}^*\|).
    \end{align*}
    Therefore, we have
    \begin{align*}
        \sqrt{n}(\hat{\boldsymbol{\theta}}-\boldsymbol{\theta}^*)
        &=-\left[\frac{1}{n}\sum_{i=1}^n\frac{\partial h(\boldsymbol{C}_i;{\boldsymbol{\theta}}^*)}{\partial \boldsymbol{\theta}^\top}\right]^{-1}\frac{1}{\sqrt{n}}\sum_{i=1}^nh(\boldsymbol{C}_i;{\boldsymbol{\theta}}^*)+op(1)\\
        &=-\left[\mathbb{E}\left\{\frac{\partial h(\boldsymbol{C}_i;{\boldsymbol{\theta}}^*)}{\partial \boldsymbol{\theta}^\top}\right\}+op(1)\right]^{-1}\frac{1}{\sqrt{n}}\sum_{i=1}^nh(\boldsymbol{C}_i;{\boldsymbol{\theta}}^*)+op(1)\quad \text{(by Law of Large Numbers (LLN))}\\
        &=\frac{1}{\sqrt{n}}\sum_{i=1}^n\left[\mathbb{E}\left\{\frac{\partial h(\boldsymbol{C}_i;{\boldsymbol{\theta}}^*)}{\partial \boldsymbol{\theta}^\top}\right\}\right]^{-1}h(\boldsymbol{C}_i;{\boldsymbol{\theta}}^*)+op(1).
    \end{align*}
    Thus, we plug in this formula to get $\sqrt{n}(\hat{{\tau}}-\tau^*)$ as
    \begin{align*}
&\sqrt{n}\,(\hat{\tau}-\tau^*)\\
&\quad=\left[\frac{\partial U(\tau',\boldsymbol{\theta}')}{\partial \tau}\right]^{-1}\Biggl\{\frac{\partial U(\tau',\boldsymbol{\theta}')}{\partial \boldsymbol{\theta}^\top}\,\frac{1}{\sqrt{n}}\sum_{i=1}^n\left[\mathbb{E}\!\left\{\frac{\partial h(\boldsymbol{C}_i;\boldsymbol{\theta}^*)}{\partial \boldsymbol{\theta}^\top}\right\}\right]^{-1}h(\boldsymbol{C}_i;\boldsymbol{\theta}^*)\\
&\qquad\qquad-\sqrt{n}\,U_n(\tau^*,\boldsymbol{\theta}^*)\Biggr\}+ o_p(1).
\end{align*}

    Given $\hat{\tau}\xrightarrow{P}\tau^*$,$\hat{\boldsymbol{\theta}}\xrightarrow{P}\boldsymbol{\theta}^*$, and the 2nd assumption, we have $$\frac{\partial U (\tau', \boldsymbol{\theta}')}{\partial \tau}\xrightarrow{P}\frac{\partial U (\tau^*, \boldsymbol{\theta}^*)}{\partial \tau}$$ and $$\frac{\partial U (\tau', \boldsymbol{\theta}')}{\partial \boldsymbol{\theta}^\top}\xrightarrow{P}\frac{\partial U (\tau^*, \boldsymbol{\theta}^*)}{\partial \boldsymbol{\theta}^\top}.$$
    By using the limiting distribution of U-statistics, we have
    \begin{align*}
        \sqrt{n} U_n (\tau^*, \boldsymbol{\theta}^*)=\frac{1}{\sqrt{n}}\sum_{i=1}^n l^*(\boldsymbol{C}_i;\tau^*,\boldsymbol{\theta}^*)+op(1),
    \end{align*}
    where $l^*(\boldsymbol{C}_i;\tau,\boldsymbol{\theta}) =\mathbb{E}\left\{l(\boldsymbol{C}_i,\boldsymbol{C}_j;\tau,\boldsymbol{\theta})\mid \boldsymbol{C}_i\right\}+\mathbb{E}\left\{l(\boldsymbol{C}_j,\boldsymbol{C}_i;\tau,\boldsymbol{\theta})\mid \boldsymbol{C}_i\right\}$. Therefore, we have
    \begin{align*}
        &\sqrt{n} (\hat{\tau} - \tau^*)\\
        \quad&=\left[ \frac{\partial U (\tau^*, \boldsymbol{\theta}^*)}{\partial \tau} \right]^{-1}  \frac{1}{\sqrt{n}}\sum_{i=1}^n \Biggl\{ \frac{\partial U (\tau^*, \boldsymbol{\theta}^*)}{\partial \boldsymbol{\theta}^\top} \left[\mathbb{E}\left\{\frac{\partial h(\boldsymbol{C}_i;{\boldsymbol{\theta}}^*)}{\partial \boldsymbol{\theta}^\top}\right\}\right]^{-1}h(\boldsymbol{C}_i;{\boldsymbol{\theta}}^*) \\&\qquad\qquad- l^*(\boldsymbol{C}_i;\tau^*,\boldsymbol{\theta}^*) \Biggr\}+op(1).
    \end{align*}
    Thus, we complete the proof by defining
    \begin{align*}
    u(\tau,\boldsymbol{\theta})=\frac{\partial U(\tau,\boldsymbol{\theta})}{\partial\tau},
    \end{align*}
    \begin{align*}
    \phi(\boldsymbol{C}_i;\tau,\boldsymbol{\theta})=\frac{\partial U(\tau,\boldsymbol{\theta})}{\partial\boldsymbol{\theta}^\top}\left[\mathbb{E}\left\{\frac{\partial h(\boldsymbol{C}_i;\boldsymbol{\theta})}{\partial\boldsymbol{\theta}^\top}\right\}\right]^{-1}h(\boldsymbol{C}_i;\boldsymbol{\theta})-l^*(\boldsymbol{C}_i;\tau,\boldsymbol{\theta}),
    \end{align*}
    and 
    \begin{align*}
    l^*(\boldsymbol{C}_i;\tau,\boldsymbol{\theta})=\mathbb{E}\left\{l(\boldsymbol{C}_i,\boldsymbol{C}_j;\tau,\boldsymbol{\theta})\mid \boldsymbol{C}_i\right\}+\mathbb{E}\left\{l(\boldsymbol{C}_j,\boldsymbol{C}_i;\tau,\boldsymbol{\theta})\mid \boldsymbol{C}_i\right\}.
    \end{align*}
    Therefore, we have
    \begin{align*}
    \sqrt{n}(\hat{\tau}-\tau^*)\rightarrow N\left(0,\operatorname{Var}\left[\left(u(\tau^*,\boldsymbol{\theta}^*)\right)^{-1}\phi(\boldsymbol{C}_i;\tau^*,\boldsymbol{\theta}^*)\right]\right).
\end{align*}
\end{proof}

\subsubsection{IPSW Estimator $\hat{\tau}_\text{IPSW}$}\label{app.asyNorm_ipsw}
\begin{theorem}\label{thm:asymptotic_ipsw}
    (Asymptotic normality for IPSW estimator): Assume $\{\frac{1}{\sqrt{n}}\sum_{i=1}^n\pi(\boldsymbol{X}_i;\balp),\balp\in \mathcal{A}\}$ is stochastically equicontinuous, $\hat{\balp}\xrightarrow{P}\balp^*$, and $\hat{\tau}\xrightarrow{P}\tau^*$. Then, 
    \begin{align*}
    \sqrt{n}\left(\hat{\tau}_{\text{IPSW}}-\tau^*\right)\rightarrow N\left(0,\operatorname{Var}\left[\left(u(\tau^*,\balp^*)\right)^{-1}\phi(\boldsymbol{C}_i;\tau^*,\balp^*)\right]\right),
\end{align*}
where 
\begin{align*}
    u(\tau^*,\balp^*)=\left.\frac{\partial U(\tau,\balp)}{\partial\tau}\right|_{\tau^*,\balp^*},
\end{align*}
\begin{align*}
\phi(\boldsymbol{C}_i;\tau,\balp)=\frac{\partial U(\tau,\balp)}{\partial\balp^\top}\left[\mathbb{E}\left\{\frac{\partial h(\boldsymbol{C}_i;\balp)}{\partial\balp^\top}\right\}\right]^{-1}h(\boldsymbol{C}_i;\balp)-l^*_{\text{IPSW}}(\boldsymbol{C}_i;\tau,\balp),
\end{align*}
and
\begin{align*}
    l^*_{\text{IPSW}}(\boldsymbol{C}_i;\tau,\balp)=\mathbb{E}\left\{l_\text{IPSW}(\boldsymbol{C}_i,\boldsymbol{C}_j;\tau,\balp)\mid\boldsymbol{C}_i\right\}+\mathbb{E}\left\{l_\text{IPSW}(\boldsymbol{C}_j,\boldsymbol{C}_i;\tau,\balp)\mid\boldsymbol{C}_i\right\}.
\end{align*}
\end{theorem}

\begin{proof}
Let $\hat{\tau}_{\text{IPSW}}$ be the solution of the estimating equation
\begin{align*}
    U_n(\tau,\boldsymbol{\hat{\balp}})=\frac{1}{n(n-1)}\sum_{i\neq j}l_\text{IPSW}(\boldsymbol{C}_i,\boldsymbol{C}_j;\tau,\hat{\balp})=0,
\end{align*}
where $$l_\text{IPSW}(\boldsymbol{C}_i,\boldsymbol{C}_j;\tau,\hat{\balp})=\tau  d^{\text{ipsw}}_0(\boldsymbol{C}_i,\boldsymbol{C}_j;\hat{\balp})-d^{\text{ipsw}}_1(\boldsymbol{C}_i,\boldsymbol{C}_j;\hat{\balp}).$$ As previously defined,  $d^{\text{ipsw}}_1(\boldsymbol{C}_i,\boldsymbol{C}_j;\hat{\balp})=w^{\text{ipsw}}(\boldsymbol{X}_i,\boldsymbol{X}_j;\hat{\balp}) \mathbb{I}(Y_i > Y_j, D_i = 1, D_j = 0, S_i = 1, S_j = 1)$ and $d^{\text{ipsw}}_0(\boldsymbol{C}_i,\boldsymbol{C}_j;\hat{\balp})=w^{\text{ipsw}}(\boldsymbol{X}_i,\boldsymbol{X}_j;\hat{\balp}) \mathbb{I}(D_i = 1, D_j = 0, S_i = 1, S_j = 1)$. Define 
\begin{align*}
    U(\tau,\balp) = \mathbb{E}\left\{U_n(\tau,\balp)\right\}=\mathbb{E}\left\{l_\text{IPSW}(\boldsymbol{C}_i, \boldsymbol{C}_j; \tau,\balp)\right\}
\end{align*} and assume $\{\frac{1}{\sqrt{n}}\sum_{i=1}^n\pi(\boldsymbol{X}_i;\balp),\balp\in \mathcal{A}\}$ is stochastically equicontinuous. Therefore, the sequence $\left\{\sqrt{n}\{U_n(\tau,\boldsymbol{\balp})-U(\tau,\boldsymbol{\balp})\},(\tau,\boldsymbol{\balp})\in \mathcal{A}\right\}$ is stochastically equicontinuous. Assume both the second and the third assumptions are held. Then, by Lemma \ref{lem3}, when $\hat{\balp}\xrightarrow{P}\balp^*$ and $\hat{\tau}\xrightarrow{P}\tau^*$, we have
\begin{align*}
    \sqrt{n}\left(\hat{\tau}_{\text{IPSW}}-\tau^*\right)\rightarrow N\left(0,\operatorname{Var}\left[\left(u(\tau^*,\balp^*)\right)^{-1}\phi(\boldsymbol{C}_i;\tau^*,\balp^*)\right]\right),
\end{align*}
where 
\begin{align*}
    u(\tau,\balp)=\frac{\partial U(\tau,\balp)}{\partial\tau}=\mathbb{E}\left\{d^{\text{ipsw}}_0(\boldsymbol{C}_i,\boldsymbol{C}_j;\balp)\right\}
\end{align*}
and
\begin{align*}
\phi(\boldsymbol{C}_i;\tau,\balp)=\frac{\partial U(\tau,\balp)}{\partial\balp^\top}\left[\mathbb{E}\left\{\frac{\partial h(\boldsymbol{C}_i;\balp)}{\partial\balp^\top}\right\}\right]^{-1}h(\boldsymbol{C}_i;\balp)-l^*_{\text{IPSW}}(\boldsymbol{C}_i;\tau,\balp).
\end{align*}
The $h(\boldsymbol{C}_i;\balp)$ constructs the estimating equation for $\balp$, i.e., $\frac{1}{n}\sum_{i=1}^nh(\boldsymbol{C}_i;\balp)=0$. As defined in the proof of Lemma \ref{lem3}, 
\begin{align*}
    l^*_{\text{IPSW}}(\boldsymbol{C}_i;\tau,\balp)&=\mathbb{E}\left\{l_\text{IPSW}(\boldsymbol{C}_i,\boldsymbol{C}_j;\tau,\balp)\mid\boldsymbol{C}_i\right\}\\&\qquad+\mathbb{E}\left\{l_\text{IPSW}(\boldsymbol{C}_j,\boldsymbol{C}_i;\tau,\balp)\mid\boldsymbol{C}_i\right\}.
\end{align*}
\end{proof}
Based on consistency of $\hat{\tau}_{\text{IPSW}}$ (Theorem \ref{thm:consistency_ipsw}), when $\hat{\balp}\rightarrow \balp^*$ and $\pi(\boldsymbol{X};\balp^*)=\text{Pr}(S=1\mid \boldsymbol{X})$, then
$\hat{\tau}_\text{IPSW}\xrightarrow{P}\tau_0$. The normal distribution of $\hat{\tau}_{\text{IPSW}}$ is around $\tau_0$ as:
\begin{align*}
    \sqrt{n}\left(\hat{\tau}_{\text{IPSW}}-\tau_0\right)\rightarrow N\left(0,\operatorname{Var}\left[\left(u(\tau_0,\balp^*)\right)^{-1}\phi(\boldsymbol{C}_i;\tau_0,\balp^*)\right]\right),
\end{align*}
where $u(\tau,\balp)$ and $\phi(\boldsymbol{C}_i;\tau,\balp)$ are defined as above. 

\subsubsection{CW Estimator $\hat{\tau}_\text{CW}$}\label{app.asyNorm_cw}
\begin{theorem}\label{thm:asymptotic_cw}
    (Asymptotic normality for CW estimator): Assume $\{\frac{1}{\sqrt{n}}\sum_{i=1}^nq(\boldsymbol{X}_i;\blam),\blam\in \mathcal{L}\}$ is stochastically equicontinuous, $\hat{\blam}\xrightarrow{P}\blam^*$, and $\hat{\tau}\xrightarrow{P}\tau^*$. Then, 
    \begin{align*}
    \sqrt{n}\left(\hat{\tau}_{\text{CW}}-\tau^*\right)\rightarrow N\left(0,\operatorname{Var}\left[\left(u(\tau^*,\blam^*)\right)^{-1}\phi(\boldsymbol{C}_i;\tau^*,\blam^*)\right]\right),
\end{align*}
where
\begin{align*}
    u(\tau^*,\blam^*)=\left.\frac{\partial U(\tau,\blam)}{\partial\tau}\right|_{\tau^*,\blam^*}=\left.\mathbb{E}\left\{d^{\text{cw}}_0(\boldsymbol{C}_i,\boldsymbol{C}_j;\blam)\right\}\right|_{\blam^*},
\end{align*}

\begin{align*}
\phi(\boldsymbol{C}_i;\tau,\blam)=\frac{\partial U(\tau,\blam)}{\partial\blam^\top}\left[\mathbb{E}\left\{\frac{\partial h(\boldsymbol{C}_i;\blam)}{\partial\blam^\top}\right\}\right]^{-1}h(\boldsymbol{C}_i;\blam)-l^*_{\text{CW}}(\boldsymbol{C}_i;\tau,\blam),
\end{align*}
and
\begin{align*}
    l^*_{\text{CW}}(\boldsymbol{C}_i;\tau,\blam)=\mathbb{E}\left\{l_\text{CW}(\boldsymbol{C}_i,\boldsymbol{C}_j;\tau,\blam)\mid\boldsymbol{C}_i\right\}+\mathbb{E}\left\{l_\text{CW}(\boldsymbol{C}_j,\boldsymbol{C}_i;\tau,\blam)\mid\boldsymbol{C}_i\right\}.
\end{align*}
\end{theorem}

\begin{proof}
    First, based on the identification of the CW method, we define 
    \begin{align*}
        &d_1^{\text{cw}}(\boldsymbol{C}_i,\boldsymbol{C}_j;\hat\blam)
        \\
        &\quad\qquad=w^{\text{cw}}(\boldsymbol{X}_i,\boldsymbol{X}_j;\hat\blam) \mathbb{I}(Y_i>Y_j, D_i=1, D_j=0, S_i=1,S_j=1)
    \end{align*} 
    and $$d_0^{\text{cw}}(\boldsymbol{C}_i,\boldsymbol{C}_j;\hat\blam)=w^{\text{cw}}(\boldsymbol{X}_i,\boldsymbol{X}_j;\hat\blam) \mathbb{I}( D_i=1, D_j=0, S_i=1,S_j=1),$$ where $w^{\text{cw}}(\boldsymbol{X}_i,\boldsymbol{X}_j;\hat\blam)=\hat{q}_i\hat{q}_j=\hat{q}_i(\boldsymbol{X}_i)\hat{q}_j(\boldsymbol{X}_j)=q(\boldsymbol{X}_i;\hat\blam)q(\boldsymbol{X}_j;\hat\blam)$. 
    
    Let $\hat{\tau}_{\text{CW}}$ be the solution of the estimating equation
    \begin{align*}
    U_n(\tau,\boldsymbol{\hat{\blam}})=\frac{1}{n(n-1)}\sum_{i\neq j}l_\text{CW}(\boldsymbol{C}_i,\boldsymbol{C}_j;\tau,\hat{\blam})=0,
    \end{align*}
    where $$l_\text{CW}(\boldsymbol{C}_i,\boldsymbol{C}_j;\tau,\hat{\blam})=\tau  d^{\text{cw}}_0(\boldsymbol{C}_i,\boldsymbol{C}_j;\hat{\blam})-d^{\text{cw}}_1(\boldsymbol{C}_i,\boldsymbol{C}_j;\hat{\blam}).$$
    Define 
    \begin{align*}
    U(\tau,\blam) = \mathbb{E}\left\{U_n(\tau,\blam)\right\}=\mathbb{E}\left\{l_\text{CW}(\boldsymbol{C}_i, \boldsymbol{C}_j; \tau,\blam)\right\}
    \end{align*} and assume $\{\frac{1}{\sqrt{n}}\sum_{i=1}^nq(\boldsymbol{X}_i;\blam),\blam\in \mathcal{L}\}$ is stochastically equicontinuous. Thus, the sequence $\left\{\sqrt{n}\{U_n(\tau,\boldsymbol{\blam})-U(\tau,\boldsymbol{\blam})\},(\tau,\boldsymbol{\blam})\in \mathcal{L}\right\}$ is stochastically equicontinuous. Assume the second and the third assumptions are satisfied. By Lemma \ref{lem3}, when $\hat{\blam}\xrightarrow{P}\blam^*$ and $\hat{\tau}\xrightarrow{P}\tau^*$, we have
    \begin{align*}
    \sqrt{n}\left(\hat{\tau}_{\text{CW}}-\tau^*\right)\rightarrow N\left(0,\operatorname{Var}\left[\left(u(\tau^*,\blam^*)\right)^{-1}\phi(\boldsymbol{C}_i;\tau^*,\blam^*)\right]\right),
\end{align*}
where 
\begin{align*}
    u(\tau,\blam)&=\frac{\partial U(\tau,\blam)}{\partial\tau}=\mathbb{E}\left\{d^{\text{cw}}_0(\boldsymbol{C}_i,\boldsymbol{C}_j;\blam)\right\}\\&=\mathbb{E}\left\{w^{\text{cw}}(\boldsymbol{X}_i,\boldsymbol{X}_j;\blam) \mathbb{I}( D_i=1, D_j=0)\right\}
\end{align*}
and
\begin{align*}
\phi(\boldsymbol{C}_i;\tau,\blam)=\frac{\partial U(\tau,\blam)}{\partial\blam^\top}\left[\mathbb{E}\left\{\frac{\partial h(\boldsymbol{C}_i;\blam)}{\partial\blam^\top}\right\}\right]^{-1}h(\boldsymbol{C}_i;\blam)-l^*_{\text{CW}}(\boldsymbol{C}_i;\tau,\blam).
\end{align*}
The estimating equation for $\blam$ is defined as $\frac{1}{n}\sum_{i=1}^nh(\boldsymbol{C}_i;\blam)=0$. The $l^*_{\text{CW}}(\boldsymbol{C}_i;\tau,\blam)$ is defined in the proof of Lemma \ref{lem3} as
\begin{align*}
    l^*_{\text{CW}}(\boldsymbol{C}_i;\tau,\blam)=\mathbb{E}\left\{l_\text{CW}(\boldsymbol{C}_i,\boldsymbol{C}_j;\tau,\blam)\mid\boldsymbol{C}_i\right\}+\mathbb{E}\left\{l_\text{CW}(\boldsymbol{C}_j,\boldsymbol{C}_i;\tau,\blam)\mid\boldsymbol{C}_i\right\}.
\end{align*}
\end{proof}
When $\hat{\blam}\xrightarrow{P}\blam^*$ and $q(\boldsymbol{X}_i;\blam^*)$ such that $\sum_{i=1}^nq(\boldsymbol{X}_i;\blam^*)\bg(\boldsymbol{X}_i)=\tilg$, then
\begin{align*}
    \sqrt{n}\left(\hat{\tau}_{\text{CW}}-\tau_0\right)\rightarrow N\left(0,\operatorname{Var}\left[\left(u(\tau_0,\blam^*)\right)^{-1}\phi(\boldsymbol{C}_i;\tau_0,\blam^*)\right]\right),
\end{align*}
where $u(\tau,\blam)$ and $\phi(\boldsymbol{C}_i;\tau,\blam)$ are defined as above. 

\subsubsection{OM+RWD Estimator $\hat{\tau}_\text{OM+RWD}$}\label{app.asyNorm_omrwd}
\begin{theorem}\label{thm:asymptotic_omrwd}
    (Asymptotic normality for OM+RWD estimator): Assume $\hat{\bbeta}\xrightarrow{P}\bbeta^*$ and $\hat{\tau}\xrightarrow{P}\tau^*$. Then, 
    \begin{align*}
    \sqrt{n}\left(\hat{\tau}_{\text{OM+RWD}}-\tau^*\right)\rightarrow N\left(0,\operatorname{Var}\left[\left(u(\tau^*,\bbeta^*)\right)^{-1}\phi(\boldsymbol{C}_i;\tau^*,\bbeta^*)\right]\right),
    \end{align*}
where 
$$ u(\tau^*,\bbeta^*)=\left.\frac{\partial U(\tau,\bbeta)}{\partial\tau}\right|_{\tau^*,\bbeta^*},$$
\begin{align*}
\phi(\boldsymbol{C}_i;\tau,\bbeta)&=\frac{\partial U(\tau,\bbeta)}{\partial\bbeta^\top}\left[\mathbb{E}\left\{\frac{\partial h(\boldsymbol{C}_i;\bbeta)}{\partial\bbeta^\top}\right\}\right]^{-1}h(\boldsymbol{C}_i;\bbeta)-l^*_{\text{OM+RWD}}(\boldsymbol{C}_i;\tau,\bbeta),
\end{align*}
and
\begin{align*}
    l^*_{\text{OM+RWD}}(\boldsymbol{C}_i;\tau,\bbeta)&=\mathbb{E}\left\{l_\text{OM+RWD}(\boldsymbol{C}_i,\boldsymbol{C}_j;\tau,\bbeta)\mid\boldsymbol{C}_i\right\}+\mathbb{E}\left\{l_\text{OM+RWD}(\boldsymbol{C}_j,\boldsymbol{C}_i;\tau,\bbeta)\mid\boldsymbol{C}_i\right\}.
\end{align*}
\end{theorem}

\begin{proof}
    First, define 
    \begin{align*}
        &d_1^{\text{om+rwd}}(\boldsymbol{C}_i,\boldsymbol{C}_j;\tau,\bbeta)\\&=\text{Pr}\left(Y_i>Y_j\mid D_i=1, D_j=0, S_i=0, S_j=0, \boldsymbol{X}_i,\boldsymbol{X}_j\right) \mathbb{I}(D_i=1, D_j=0, S_i=0, S_j=0)
    \end{align*}
    and
    $$d_0^{\text{om+rwd}}(\boldsymbol{C}_i,\boldsymbol{C}_j;\tau,\bbeta)=\mathbb{I}(D_i=1, D_j=0, S_i=0, S_j=0).$$
    Following the proof of Theorem \ref{thm:consistency_omrwd}, when assuming the outcome follows a normal distribution for simplicity, i.e., $Y_i\mid \boldsymbol{X}_i,D_i \sim N(M(\boldsymbol{X}_i,D_i;\bbeta), \sigma_1^2  D_i + \sigma_0^2  (1-D_i))$, where $M(\boldsymbol{X}_i,D_i=d;\bbeta)=\mathbb{E}\{Y_i\mid \boldsymbol{X}_i, D_i=d\}$($d\in\{0,1\}$), we have 
    \begin{align*}
         &d_1^{\text{om+rwd}}(\boldsymbol{C}_i,\boldsymbol{C}_j;\tau,\bbeta)\\&=\Phi\left(\frac{M(\boldsymbol{X}_i,D_i=1,S_i=0;\bbeta)-M(\boldsymbol{X}_j,D_j=0,S_j=0;\bbeta)}{\sqrt{{\sigma}_1^2+{\sigma}_0^2}}\right) \mathbb{I}(D_i=1, D_j=0, S_i=0, S_j=0),
    \end{align*}
    where the ${\sigma}_0^2$ and ${\sigma}_1^2$ are not of interest and depend on the distribution assumption. To be noted, the same as the proof of consistency for the OM+RWD estimator, the proof for asymptotic normality can also be generalized to other distributions as long as the CDF of the distribution exists and is differentiable. 

    Let $\hat{\tau}_{\text{OM+RWD}}$ be the solution of the estimating equation
    \begin{align*}
        U_m(\tau,\hat\bbeta)=\frac{1}{m(m-1)}\sum_{i\neq j}l_\text{OM+RWD}(\boldsymbol{C}_i,\boldsymbol{C}_j;\tau,\hat\bbeta)=0,
    \end{align*}
    where $$l_\text{OM+RWD}(\boldsymbol{C}_i,\boldsymbol{C}_j;\tau,\bbeta)=\tau d_0^{\text{om+rwd}}(\boldsymbol{C}_i,\boldsymbol{C}_j;\tau,\bbeta)-d_1^{\text{om+rwd}}(\boldsymbol{C}_i,\boldsymbol{C}_j;\tau,\bbeta).$$
    Define the expectation of $U_m(\tau,\bbeta)$ as
    \begin{align*}
        U(\tau,\bbeta)=\mathbb{E}\{U_m(\tau,\bbeta)\}=\mathbb{E}\{l_\text{OM+RWD}(\boldsymbol{C}_i,\boldsymbol{C}_j;\tau,\bbeta)\}.
    \end{align*}
    As shown in the proof of Theorem \ref{thm:consistency_omrwd}, 
    \begin{align*}
    \left\{\frac{1}{m(m-1)}\sum_{i\neq j}d^{\text{om+rwd}}_k(\boldsymbol{C}_i,\boldsymbol{C}_j;\bbeta)\right\}
    \end{align*}
    are stochastically equicontinuous. Assume both the second and the third assumptions in Lemma \ref{lem3} are held. Then, by Lemma \ref{lem3}, when $\hat\bbeta\xrightarrow{P}\bbeta^*$ and $\hat\tau\xrightarrow{P}\tau^*$, we have
    \begin{align*}
    \sqrt{n}\left(\hat{\tau}_{\text{OM+RWD}}-\tau^*\right)\rightarrow N\left(0,\operatorname{Var}\left[\left(u(\tau^*,\bbeta^*)\right)^{-1}\phi(\boldsymbol{C}_i;\tau^*,\bbeta^*)\right]\right),
    \end{align*}
    where 
\begin{align*}
    u(\tau,\bbeta)=\frac{\partial U(\tau,\bbeta)}{\partial\tau}&=\mathbb{E}\left\{d^{\text{om+rwd}}_0(\boldsymbol{C}_i,\boldsymbol{C}_j;\tau,\bbeta)\right\}\\
    &=\mathbb{E}\left\{\mathbb{I}(D_i=1, D_j=0, S_i=0, S_j=0)\right\}
\end{align*}
and
\begin{align*}
\phi(\boldsymbol{C}_i;\tau,\bbeta)&=\frac{\partial U(\tau,\bbeta)}{\partial\bbeta^\top}\left[\mathbb{E}\left\{\frac{\partial h(\boldsymbol{C}_i;\bbeta)}{\partial\bbeta^\top}\right\}\right]^{-1}h(\boldsymbol{C}_i;\bbeta)-l^*_{\text{OM+RWD}}(\boldsymbol{C}_i;\tau,\bbeta).
\end{align*}
The estimating equation for $\bbeta$ is $\frac{1}{m}\sum_{i=1}^mh(\boldsymbol{C}_i;\bbeta)=0$, which is constructed by $h(\boldsymbol{C}_i;\bbeta)$. Based on the construction in the proof of Lemma \ref{lem3}, 
\begin{align*}
    l^*_{\text{OM+RWD}}(\boldsymbol{C}_i;\tau,\bbeta)&=\mathbb{E}\left\{l_\text{OM+RWD}(\boldsymbol{C}_i,\boldsymbol{C}_j;\tau,\bbeta)\mid\boldsymbol{C}_i\right\}+\mathbb{E}\left\{l_\text{OM+RWD}(\boldsymbol{C}_j,\boldsymbol{C}_i;\tau,\bbeta)\mid\boldsymbol{C}_i\right\}.
\end{align*}
\end{proof}
When $\hat\bbeta\xrightarrow{P}\bbeta^*$ and $M(\boldsymbol{X},D;\bbeta^*)=\mathbb{E}\{Y\mid \boldsymbol{X},D\}$, then 
\begin{align*}
    \sqrt{n}\left(\hat{\tau}_{\text{OM+RWD}}-\tau_0\right)\rightarrow N\left(0,\operatorname{Var}\left[\left(u(\tau_0,\bbeta^*)\right)^{-1}\phi(\boldsymbol{C}_i;\tau_0,\bbeta^*)\right]\right),
\end{align*}
where $u(\tau,\bbeta)$ and $\phi(\boldsymbol{C}_i;\tau,\bbeta)$ are defined as above. 

\subsubsection{OM Estimator $\hat{\tau}_\text{OM}$}\label{app.asyNorm_om}
\begin{theorem}\label{thm:asymptotic_om}
    (Asymptotic normality for OM estimator): Assume $\{\frac{1}{\sqrt{n}}\sum_{i=1}^nq(\boldsymbol{X}_i;\blam),\blam\in \mathcal{L}\}$ is stochastically equicontinuous, $\hat{\blam}\xrightarrow{P}\blam^*$, $\hat{\boldsymbol{\bbeta}}\xrightarrow{P}\boldsymbol{\bbeta}^*$, and $\hat{\tau}\xrightarrow{P}\tau^*$. Then, 
    \begin{align*}
    \sqrt{n}\left(\hat{\tau}_{\text{OM}}-\tau^*\right)\rightarrow N\left(0,\operatorname{Var}\left[\left(u(\tau^*,\blam^*, \bbeta^*)\right)^{-1}\phi(\boldsymbol{C}_i;\tau^*,\blam^*, \bbeta^*)\right]\right),
\end{align*}
where 
\begin{align*}
    u(\tau^*,\blam^*,\bbeta^*)=\left.\frac{\partial U(\tau,\blam,\bbeta)}{\partial\tau}\right|_{\tau^*,\blam^*,\bbeta^*}=\left.\mathbb{E}\left\{d^{\text{om}}_0(\boldsymbol{C}_i,\boldsymbol{C}_j;\blam)\right\}\right|_{\blam^*},
\end{align*}

\begin{align*}
\phi(\boldsymbol{C}_i;\tau,\blam,\bbeta)&=\frac{\partial U(\tau,\blam,\bbeta)}{\partial\blam^\top}\left[\mathbb{E}\left\{\frac{\partial h_1(\boldsymbol{C}_i;\blam)}{\partial\blam^\top}\right\}\right]^{-1}h_1(\boldsymbol{C}_i;\blam)\\
&\quad+\frac{\partial U(\tau,\blam,\bbeta)}{\partial\bbeta^\top}\left[\mathbb{E}\left\{\frac{\partial h_2(\boldsymbol{C}_i;\bbeta)}{\partial\bbeta^\top}\right\}\right]^{-1}h_2(\boldsymbol{C}_i;\bbeta)-l^*_{\text{OM}}(\boldsymbol{C}_i;\tau,\blam,\bbeta),
\end{align*}
and
\begin{align*}
    l^*_{\text{OM}}(\boldsymbol{C}_i;\tau,\blam,\bbeta)=\mathbb{E}\left\{l_\text{OM}(\boldsymbol{C}_i,\boldsymbol{C}_j;\tau,\blam,\bbeta)\mid\boldsymbol{C}_i\right\}+\mathbb{E}\left\{l_\text{OM}(\boldsymbol{C}_j,\boldsymbol{C}_i;\tau,\blam,\bbeta)\mid\boldsymbol{C}_i\right\}.
\end{align*}
\end{theorem}
\begin{proof}
    First, same as the previous proof, we define 
    $$d^{\text{om}}_0(\boldsymbol{C}_i,\boldsymbol{C}_j;\blam)= w(\boldsymbol{X}_i,\boldsymbol{X}_j) \mathbb{I}(D_i = 1, D_j = 0, S_i = 1, S_j = 1).$$
    and
    \begin{align*}
     d^{\text{om}}_1(\boldsymbol{C}_i,\boldsymbol{C}_j;\blam,\bbeta)&=d^{\text{om}}_0(\boldsymbol{C}_i,\boldsymbol{C}_j;\blam) \mathcal{P}_{ij}(\boldsymbol{C}_i, \boldsymbol{C}_j; \bbeta, {\sigma_0},{\sigma_1})\\
     &=w(\boldsymbol{X}_i,\boldsymbol{X}_j) \mathcal{P}_{ij}(\boldsymbol{C}_i, \boldsymbol{C}_j; \bbeta, {\sigma_0},{\sigma_1})  \mathbb{I}(D_i = 1, D_j = 0, S_i = 1, S_j = 1)
     \end{align*} and 
    The weight is defined as $w(\boldsymbol{X}_i,\boldsymbol{X}_j)=w^{\text{cw}}(\boldsymbol{X}_i,\boldsymbol{X}_j;\blam)=q(\boldsymbol{X}_i;\lambda)q(\boldsymbol{X}_j;\lambda)$. Let $\hat{\tau}_{\text{OM}}$ be the solution of the estimating equation
    \begin{align*}
    U_n(\tau,\hat{\blam},\hat{\bbeta})=\frac{1}{n(n-1)}\sum_{i\neq j}l_\text{OM}(\boldsymbol{C}_i,\boldsymbol{C}_j;\tau,\hat{\blam},\hat{\bbeta})=0,
    \end{align*}
    where $$l_\text{OM}(\boldsymbol{C}_i,\boldsymbol{C}_j;\tau,\hat{\blam},\hat{\bbeta})=\tau  d^{\text{om}}_0(\boldsymbol{C}_i,\boldsymbol{C}_j;\hat{\blam})-d^{\text{om}}_1(\boldsymbol{C}_i,\boldsymbol{C}_j;\hat{\blam},\hat{\bbeta}).$$
    Define 
    $$U(\tau,\blam) = \mathbb{E}\left\{U_n(\tau,\blam)\right\}=\mathbb{E}\left\{l_\text{OM}(\boldsymbol{C}_i, \boldsymbol{C}_j; \tau,\blam,\bbeta)\right\}$$ and assume $\{\frac{1}{\sqrt{n}}\sum_{i=1}^nq(\boldsymbol{X}_i;\blam),\blam\in \mathcal{L}\}$ is stochastically equicontinuous.  Assume the second and the third assumptions are satisfied. By Lemma \ref{lem3}, when $\hat{\blam}\xrightarrow{P}\blam^*$, $\hat{\bbeta}\xrightarrow{P}\bbeta^*$, and $\hat{\tau}\xrightarrow{P}\tau^*$, we have
    \begin{align*}
    \sqrt{n}\left(\hat{\tau}_{\text{OM}}-\tau^*\right)\rightarrow N\left(0,\operatorname{Var}\left[\left(u(\tau^*,\blam^*,\bbeta^*)\right)^{-1}\phi(\boldsymbol{C}_i;\tau^*,\blam^*,\bbeta^*)\right]\right),
    \end{align*}
where 
$$u(\tau,\blam,\bbeta)=\frac{\partial U(\tau,\blam,\bbeta)}{\partial\tau}=\mathbb{E}\left\{d^{\text{om}}_0(\boldsymbol{C}_i,\boldsymbol{C}_j;\blam)\right\}$$
and
\begin{align*}
\phi(\boldsymbol{C}_i;\tau,\blam,\bbeta)&=\frac{\partial U(\tau,\blam,\bbeta)}{\partial\blam^\top}\left[\mathbb{E}\left\{\frac{\partial h_1(\boldsymbol{C}_i;\blam)}{\partial\blam^\top}\right\}\right]^{-1}h_1(\boldsymbol{C}_i;\blam)\\
&\quad+\frac{\partial U(\tau,\blam,\bbeta)}{\partial\bbeta^\top}\left[\mathbb{E}\left\{\frac{\partial h_2(\boldsymbol{C}_i;\bbeta)}{\partial\bbeta^\top}\right\}\right]^{-1}h_2(\boldsymbol{C}_i;\bbeta)-l^*_{\text{OM}}(\boldsymbol{C}_i;\tau,\blam,\bbeta).
\end{align*}
The $h_1(\boldsymbol{C}_i;\blam)$ constructs the estimating equation for $\blam$, i.e., ${n}^{-1}\sum_{i=1}^nh_1(\boldsymbol{C}_i;\blam)=0$. The estimating equation for $\bbeta$ is ${n}^{-1}\sum_{i=1}^nh_2(\boldsymbol{C}_i;\bbeta)=0$, which is based on $h_2(\boldsymbol{C}_i;\bbeta)$. Based on the construction in the proof of Lemma \ref{lem3}, 
$$l^*_{\text{OM}}(\boldsymbol{C}_i;\tau,\blam,\bbeta)=\mathbb{E}\left\{l_\text{OM}(\boldsymbol{C}_i,\boldsymbol{C}_j;\tau,\blam,\bbeta)\mid\boldsymbol{C}_i\right\}+\mathbb{E}\left\{l_\text{OM}(\boldsymbol{C}_j,\boldsymbol{C}_i;\tau,\blam,\bbeta)\mid\boldsymbol{C}_i\right\}.$$
\end{proof}

Under Assumption \ref{as1} - \ref{as:cw_loglinear}, we have $q(\boldsymbol{X}_i;\blam^*)$ be equivalent to $1/\pi(\boldsymbol{X};\alpha^*)$ as $n\rightarrow \infty$, where $\pi(\boldsymbol{X};\alpha^*)=Pr(S=1\mid \boldsymbol{X})$. Additionally, if we also have $M(\boldsymbol{X},D;\bbeta^*)=\mathbb{E}\{Y\mid \boldsymbol{X},D\}$. Then, based on Theorem \ref{thm:consistency_om}, $\hat{\tau}_{\text{OM}}\xrightarrow{P}\tau_0$, and the asymptotic distribution of $\hat{\tau}_{\text{OM}}$ becomes around $\tau_0$:
\begin{align*}
    \sqrt{n}\left(\hat{\tau}_{\text{OM}}-\tau_0\right)\rightarrow N\left(0,\operatorname{Var}\left[\left(u(\tau_0,\blam^*,\bbeta^*)\right)^{-1}\phi(\boldsymbol{C}_i;\tau_0,\blam^*,\bbeta^*)\right]\right),
    \end{align*}
where $u(\tau,\blam,\bbeta)$ and $\phi(\boldsymbol{C}_i;\tau,\blam,\bbeta)$ are similarly defined as above. 

\subsubsection{ACW Estimator $\hat{\tau}_\text{ACW}$}\label{app.asyNorm_acw}
\begin{proof}[Proof of Theorem~\upshape\ref{thm:asymptotic_acw}]
    We divide the $\hat{\tau}_{\text{ACW}}$ into two parts, \begin{align*}
    \hat{\tau}_{\text{ACW}}=\hat{\tau}_{\text{ACW1}}+\hat{\tau}_{\text{ACW2}}=(\hat{\tau}_{\text{CW}}-\hat{\tau}_{\text{OM}})+\hat{\tau}_{\text{OM+RWD}}
    \end{align*}
    First, we have
    \begin{align*}
    &\hat{\tau}_{\text{CW}}-\hat{\tau}_{\text{OM}}\\
    &= \frac{\sum_{i \neq j}^{n} \hat{w}^{\text{cw}}(\boldsymbol{X}_i, \boldsymbol{X}_j) \mathbb{I}(Y_i > Y_j, D_i = 1, D_j = 0, S_i = 1, S_j = 1)}{\sum_{i \neq j}^{n} \hat{w}^{\text{cw}}(\boldsymbol{X}_i, \boldsymbol{X}_j) \mathbb{I}(D_i = 1, D_j = 0, S_i = 1, S_j = 1)} \\ &\quad- 
    \frac{\sum_{i\neq j}^{n}\hat{w}^{\text{cw}}(\boldsymbol{X}_i, \boldsymbol{X}_j) \mathcal{P}_{ij}(\boldsymbol{V}_i, \boldsymbol{V}_j; \hat\bbeta, \hat{\sigma}_0,\hat{\sigma}_1) \mathbb{I}(D_i = 1, D_j = 0, S_i = 1, S_j = 1)}{\sum_{i\neq j}^{n}\hat{w}^{\text{cw}}(\boldsymbol{X}_i, \boldsymbol{X}_j) \mathbb{I}(D_i=1,D_j=0,S_i=1,S_j=1)}\\
    &=\frac{\sum_{i \neq j}^{n} \hat{w}^{\text{cw}}(\boldsymbol{X}_i, \boldsymbol{X}_j)\left[\left(\mathbb{I}(Y_i > Y_j)-\mathcal{P}_{ij}(\boldsymbol{V}_i, \boldsymbol{V}_j; \hat\bbeta, \hat{\sigma}_0,\hat{\sigma}_1)\right) \mathbb{I}(D_i = 1, D_j = 0, S_i = 1, S_j = 1)\right] }{\sum_{i \neq j}^{n} \hat{w}^{\text{cw}}(\boldsymbol{X}_i, \boldsymbol{X}_j) \mathbb{I}(D_i = 1, D_j = 0, S_i = 1, S_j = 1)},
    \end{align*}
    and thus
    \begin{align*}
    &d^{\text{acw1}}_1(\boldsymbol{V}_i,\boldsymbol{V}_j;{\blam},\bbeta)\\&=q(\boldsymbol{X}_i;\blam)q(\boldsymbol{X}_j;\blam)\Bigl[\left(\mathbb{I}(Y_i>Y_j)-\mathcal{P}_{ij}(\boldsymbol{V}_i, \boldsymbol{V}_j; \bbeta, {\sigma_0},{\sigma_1})\right)\mathbb{I}(D_i = 1, D_j = 0, S_i = 1, S_j = 1)\Bigr]
    \end{align*}
    and 
    $$d^{\text{acw1}}_0(\boldsymbol{V}_i,\boldsymbol{V}_j;{\blam})=q(\boldsymbol{X}_i;\blam)q(\boldsymbol{X}_j;\blam)\mathbb{I}(D_i = 1, D_j = 0, S_i = 1, S_j = 1).$$
    Assume $\tau_1$ is the true limit of $\hat{\tau}_{\text{ACW1}}=\hat{\tau}_{\text{CW}}-\hat{\tau}_{\text{OM}}$, which is around 0. Then, based on the definition in the previous proof, let 
    $$l_\text{ACW1}(\boldsymbol{V}_i,\boldsymbol{V}_j;\tau_1,\hat{\blam},\hat\bbeta)=
        \tau_1  d^{\text{acw1}}_0(\boldsymbol{V}_i,\boldsymbol{V}_j;\hat{\blam})-d^{\text{acw1}}_1(\boldsymbol{V}_i,\boldsymbol{V}_j;\hat{\blam},\hat\bbeta)$$
    and
    \begin{align*}
        l_\text{ACW2}(\boldsymbol{C}_i,\boldsymbol{C}_j;\tau,\hat{\bbeta})&=l_\text{OM+RWD}(\boldsymbol{C}_i,\boldsymbol{C}_j;\tau,\hat{\bbeta})\\
        &=\tau d_0^{\text{om+rwd}}(\boldsymbol{C}_i,\boldsymbol{C}_j;\tau,\hat{\bbeta})-d_1^{\text{om+rwd}}(\boldsymbol{C}_i,\boldsymbol{C}_j;\tau,\hat{\bbeta}).
    \end{align*}
    For the first term, let $\hat{\tau}_{\text{ACW1}}$ be the solution of the estimating equation
    \begin{align*}
        U_{n}^{\text{acw1}}(\tau_1,\hat{\blam},\hat\bbeta)=\frac{1}{n(n-1)}\sum_{i\neq j}l_\text{ACW1}(\boldsymbol{V}_i,\boldsymbol{V}_j;\tau_1,\hat{\blam},\hat\bbeta)=0,
    \end{align*}
    and define
    \begin{align*}
        U^{\text{acw1}}(\tau_1,\blam,\bbeta)=\mathbb{E}\left\{U_{n}^{\text{acw1}}(\tau_1,\blam,\bbeta)\right\}=\mathbb{E}\left\{l_\text{ACW1}(\boldsymbol{V}_i,\boldsymbol{V}_j;\tau_1,{\blam},\bbeta)\right\}.
    \end{align*}
    Assume $\{\frac{1}{\sqrt{n}}\sum_{i=1}^nq(\boldsymbol{X}_i;\blam),\blam\in \mathcal{L}\}$ is stochastically equicontinuous, and the second and third assumptions are held. Let $\tau_1^*$ be the limit of the difference of the first two terms, i.e., $\hat{\tau}_\text{CW}-\hat{\tau}_\text{OM}\xrightarrow{P}\tau_1^*$, which is not targeting $\tau$, but the true difference between the CW and OM estimators. By Lemma \ref{lem3}, when $\hat{\blam}\xrightarrow{P}\blam^*$, $\hat{\bbeta}\xrightarrow{P}\bbeta^*$, and $\hat{\tau}_\text{ACW1}\xrightarrow{P}\tau_1^*$, we have
    \begin{align*}
    \sqrt{n}\left(\hat{\tau}_{\text{ACW1}}-\tau_1^*\right)\rightarrow N\left(0,\operatorname{Var}\left[\left(u_1(\tau_1^*,\blam^*,\bbeta^*)\right)^{-1}\phi_1(\boldsymbol{V}_i;\tau_1^*,\blam^*,\bbeta^*)\right]\right),
    \end{align*}
    where 
    \begin{align*}
    u_1(\tau_1,\blam,\bbeta)&=\frac{\partial U^{\text{acw1}}(\tau_1,\blam,\bbeta)}{\partial\tau_1}=\mathbb{E}\left\{d^{\text{acw1}}_0(\boldsymbol{V}_i,\boldsymbol{V}_j;{\blam})\right\}\\
    &=\mathbb{E}\left\{q(\boldsymbol{X}_i;\blam)q(\boldsymbol{X}_j;\blam)\mathbb{I}(D_i=1,D_j=0,S_i=1,S_j=1)\right\}
    \end{align*}
    and
    \begin{align*}
\phi_1(\boldsymbol{V}_i;\tau_1,\blam,\bbeta)&=\frac{\partial U^{\text{acw1}}(\tau_1,\blam,\bbeta)}{\partial\blam^\top}\left[\mathbb{E}\left\{\frac{\partial h_1(\boldsymbol{V}_i;\blam)}{\partial\blam^\top}\right\}\right]^{-1}h_1(\boldsymbol{V}_i;\blam)\\&\quad +\frac{\partial U^{\text{acw1}}(\tau_1,\blam,\bbeta)}{\partial\bbeta^\top}\left[\mathbb{E}\left\{\frac{\partial h_2(\boldsymbol{V}_i;\bbeta)}{\partial\bbeta^\top}\right\}\right]^{-1}h_2(\boldsymbol{V}_i;\bbeta)-l^*_{\text{ACW1}}(\boldsymbol{V}_i;\tau_1,\blam,\bbeta).
\end{align*}

   The $h_1(\boldsymbol{V}_i;\blam)$ constructs the estimating equation for $\blam$, i.e., $\frac{1}{n}\sum_{i=1}^nh_1(\boldsymbol{V}_i;\blam)=0$. The $\frac{1}{n}\sum_{i=1}^nh_2(\boldsymbol{V}_i;\bbeta)=0$ is the estimating equation for $\bbeta$. The $l^*_{\text{ACW1}}(\boldsymbol{V}_i;\tau_1,\blam,\bbeta)$ is defined as
\begin{align*}
    l^*_{\text{ACW1}}(\boldsymbol{V}_i;\tau_1,\blam,\bbeta)=\mathbb{E}\left\{l_\text{ACW1}(\boldsymbol{V}_i,\boldsymbol{V}_j;\tau_1,\blam,\bbeta)\mid\boldsymbol{V}_i\right\}+\mathbb{E}\left\{l_\text{ACW1}(\boldsymbol{V}_j,\boldsymbol{V}_i;\tau_1,\blam,\bbeta)\mid\boldsymbol{V}_i\right\}.
\end{align*}

    Furthermore, the second term is the OM+RWD estimator, i.e., $\hat{\tau}_{\text{ACW2}}=\hat{\tau}_{\text{OM+RWD}}$. For simplicity, we omit the proof for $\hat{\tau}_{\text{ACW2}}$ here, which is the same as $\hat{\tau}_{\text{OM+RWD}}$. Let $\hat{\tau}_{\text{ACW2}}$ be the solution of the estimating equation
    $$U_{m}^{\text{acw2}}(\tau,\hat\bbeta)=\frac{1}{m(m-1)}\sum_{i\neq j}l_\text{ACW2}(\boldsymbol{C}_i,\boldsymbol{C}_j;\tau,\hat\bbeta)=0,$$
    and define
    $$U^{\text{acw2}}(\tau,\bbeta)=\mathbb{E}\left\{U_{m}^{\text{acw2}}(\tau,\bbeta)\right\}=\mathbb{E}\left\{l_\text{ACW2}(\boldsymbol{C}_i,\boldsymbol{C}_j;\tau,\bbeta)\right\}.$$
    By Lemma \ref{lem3}, assume $\hat{\tau}_{\text{ACW2}}\xrightarrow{P}\tau^*$, we have
    \begin{align*}
    \sqrt{n}\left(\hat{\tau}_{\text{ACW2}}-\tau^*\right)\rightarrow N\left(0,\operatorname{Var}\left[\left(u_2(\tau^*,\bbeta^*)\right)^{-1}\phi_2(\boldsymbol{C}_i;\tau^*,\bbeta^*)\right]\right),    
    \end{align*}
    where
    \begin{align*}
    u_2(\tau,\bbeta)=\frac{\partial U^{\text{acw2}}(\tau,\bbeta)}{\partial\tau}=\mathbb{E}\left\{d^{\text{om+rwd}}_0(\boldsymbol{C}_i,\boldsymbol{C}_j;\tau,\bbeta)\right\}=\mathbb{E}\left\{\mathbb{I}(D_i=1, D_j=0, S_i=0, S_j=0)\right\}
    \end{align*}
    and
    \begin{align*}
    \phi_2(\boldsymbol{C}_i;\tau,\bbeta)=\frac{\partial U^{\text{acw2}}(\tau,\bbeta)}{\partial\bbeta^\top}\left[\mathbb{E}\left\{\frac{\partial h_2(\boldsymbol{C}_i;\bbeta)}{\partial\bbeta^\top}\right\}\right]^{-1}h_2(\boldsymbol{C}_i;\bbeta)-l^*_{\text{OM+RWD}}(\boldsymbol{C}_i;\tau,\bbeta).
    \end{align*}
    The $l^*_{\text{OM+RWD}}(\boldsymbol{C}_i;\tau,\bbeta)$ and $h_2(\boldsymbol{C}_i;\bbeta)$ are the same as defined before in Section~\ref{app.asyNorm_omrwd}.

    Finally, by $\hat{\tau}_{\text{ACW1}}$ and $\hat{\tau}_{\text{ACW2}}$ are independent, we have
    \begin{align*}
    \sqrt{n}\left(\hat{\tau}_{\mathrm{ACW}}-(\tau_1^*+\tau^*)\right)
    &\rightarrow N\Bigl(0,\;\operatorname{Var}\!\left[\left(u_1(\tau_1^*,\blam^*,\bbeta^*)\right)^{-1}\phi_1(\boldsymbol{V}_i;\tau_1^*,\blam^*,\bbeta^*)\right]\\
    &\qquad\qquad\quad+\operatorname{Var}\!\left[\left(u_2(\tau^*,\bbeta^*)\right)^{-1}\phi_2(\boldsymbol{C}_i;\tau^*,\bbeta^*)\right]\Bigr).
    \end{align*}
    
    When either $M(\boldsymbol{X},D;\bbeta^*)=\mathbb{E}\{Y\mid \boldsymbol{X},D\}$ or $q(\boldsymbol{X}_i;\blam^*)=1/\pi(\boldsymbol{X};\alpha^*)=1/Pr(S=1\mid \boldsymbol{X})$ as $n\rightarrow \infty$, we have $\tau_1^*+\tau^*\xrightarrow{P}\tau_0$. Then, 
    \begin{align*}
    \sqrt{n}\left(\hat{\tau}_{\mathrm{ACW}}-\tau_0\right)
    &\rightarrow N\Bigl(0,\;\operatorname{Var}\!\left[\left(u_1(\tau_1,\blam^*,\bbeta^*)\right)^{-1}\phi_1(\boldsymbol{V}_i;\tau_1,\blam^*,\bbeta^*)\right]\\
    &\qquad\qquad\quad+\operatorname{Var}\!\left[\left(u_2(\tau_0,\bbeta^*)\right)^{-1}\phi_2(\boldsymbol{C}_i;\tau_0,\bbeta^*)\right]\Bigr).
    \end{align*}
\end{proof}

\subsubsection{AIPSW Estimator $\hat{\tau}_\text{AIPSW}$}\label{app.asyNorm_aipsw}
\begin{theorem}\label{thm:asymptotic_aipsw}
    (Asymptotic normality for AIPSW estimator): Assume $\{\frac{1}{\sqrt{n}}\sum_{i=1}^n\pi(\boldsymbol{X}_i;\balp),\balp\in \mathcal{A}\}$ is stochastically equicontinuous, and the second and third conditions in Lemma \ref{lem3} are satisfied. Let $\hat{\balp}\xrightarrow{P}\balp^*$ and $\hat{\bbeta}\xrightarrow{P}\bbeta^*$. If either $\pi(\boldsymbol{X};\balp^*)=\text{Pr}(S=1\mid \boldsymbol{X})$, or $M(\boldsymbol{X},D;\bbeta^*)=\mathbb{E}\{Y\mid \boldsymbol{X},D\}$, then
    \begin{align*}
        \sqrt{n}\left(\hat{\tau}_{\text{AIPSW}}-\tau_0\right)&\rightarrow N\Bigl(0,\operatorname{Var}\left[\left(u_1(\tau_1,\balp^*,\bbeta^*)\right)^{-1}\phi_1(\boldsymbol{V}_i;\tau_1,\balp^*,\bbeta^*)\right]\\
        &\qquad\qquad\quad
        +\operatorname{Var}\left[\left(u_2(\tau_0,\bbeta^*)\right)^{-1}\phi_2(\boldsymbol{C}_i;\tau_0,\bbeta^*)\right]\Bigr). 
    \end{align*}
    where $\tau_1$ is the true limit of $\hat{\tau}_{\text{AIPSW1}}=\hat{\tau}_{\text{IPSW}}-\hat{\tau}_{\text{OM}}$, which is around 0 and can be seen as fixed. For here,
    \begin{align*}
    u_1(\tau_1,\balp,\bbeta)=\mathbb{E}\left\{\frac{1}{\pi(\boldsymbol{X}_i;\balp)\pi(\boldsymbol{X}_j;\balp)}\mathbb{I}(D_i=1,D_j=0,S_i=1,S_j=1)\right\}
    \end{align*}
    and
   \begin{align*}
\phi_1(\boldsymbol{V}_i;\tau_1,\balp,\bbeta)&=\frac{\partial U^{\text{aipsw1}}(\tau_1,\balp,\bbeta)}{\partial\balp^\top}\left[\mathbb{E}\left\{\frac{\partial h_1(\boldsymbol{V}_i;\balp)}{\partial\balp^\top}\right\}\right]^{-1}h_1(\boldsymbol{V}_i;\balp)\\
&\quad+\frac{\partial U^{\text{aipsw1}}(\tau_1,\balp,\bbeta)}{\partial\bbeta^\top}\left[\mathbb{E}\left\{\frac{\partial h_2(\boldsymbol{V}_i;\bbeta)}{\partial\bbeta^\top}\right\}\right]^{-1}h_2(\boldsymbol{V}_i;\bbeta)-l^*_{\text{AIPSW1}}(\boldsymbol{V}_i;\tau_1,\balp,\bbeta)
\end{align*}
with
$$l^*_{\text{AIPSW1}}(\boldsymbol{V}_i;\tau_1,\blam,\bbeta)= \mathbb{E}\left\{l_\text{AIPSW1}(\boldsymbol{V}_i,\boldsymbol{V}_j;\tau_1,\blam,\bbeta)\mid\boldsymbol{V}_i\right\}+\mathbb{E}\left\{l_\text{AIPSW1}(\boldsymbol{V}_j,\boldsymbol{V}_i;\tau_1,\blam,\bbeta)\mid\boldsymbol{V}_i\right\}.$$
Additionally, 
\begin{align*}
u_2(\tau,\bbeta)&=\mathbb{E}\left\{d^{\text{om+rwd}}_0(\boldsymbol{C}_i,\boldsymbol{C}_j;\tau,\bbeta)\right\}=\mathbb{E}\left\{\mathbb{I}(D_i=1, D_j=0, S_i=0, S_j=0)\right\},
\end{align*}
and
\begin{align*}
\phi_2(\boldsymbol{C}_i;\tau,\bbeta)=\frac{\partial U^{\text{aipsw2}}(\tau,\bbeta)}{\partial\bbeta^\top}\left[\mathbb{E}\left\{\frac{\partial h_2(\boldsymbol{C}_i;\bbeta)}{\partial\bbeta^\top}\right\}\right]^{-1}h_2(\boldsymbol{C}_i;\bbeta)-l^*_{\text{OM+RWD}}(\boldsymbol{C}_i;\tau,\bbeta).
\end{align*}
with
\begin{align*}
    l^*_{\text{OM+RWD}}(\boldsymbol{C}_i;\tau,\bbeta)&=\mathbb{E}\left\{l_\text{OM+RWD}(\boldsymbol{C}_i,\boldsymbol{C}_j;\tau,\bbeta)\mid\boldsymbol{C}_i\right\}+\mathbb{E}\left\{l_\text{OM+RWD}(\boldsymbol{C}_j,\boldsymbol{C}_i;\tau,\bbeta)\mid\boldsymbol{C}_i\right\}.
\end{align*}
\end{theorem}
\begin{proof}
    The proof of $\hat{\tau}_{\text{AIPSW}}$ is similar to that of $\hat{\tau}_{\text{ACW}}$. The $\hat{\tau}_{\text{AIPSW}}$ is divided into two parts, 
    \begin{align*}
    \hat{\tau}_{\text{AIPSW}}&=\hat{\tau}_{\text{AIPSW1}}+\hat{\tau}_{\text{AIPSW2}}\\
    &=(\hat{\tau}_{\text{IPSW}}-\hat{\tau}_{\text{OM}})+\hat{\tau}_{\text{OM+RWD}}
    \end{align*}
    First, in $\hat{\tau}_{\text{OM}}$, we have $\hat{w}_{ij}=\hat{w}^{\text{ipsw}}(\boldsymbol{X}_i, \boldsymbol{X}_j)=\frac{1}{\pi(\boldsymbol{X}_i;\hat\balp)\pi(\boldsymbol{X}_j;\hat\balp)}$. Therefore,  
    \begin{align*}
    &\hat{\tau}_{\text{IPSW}}-\hat{\tau}_{\text{OM}}\\&= \frac{\sum_{i \neq j}^{n} \hat{w}^{\text{ipsw}}(\boldsymbol{X}_i, \boldsymbol{X}_j) \mathbb{I}(Y_i > Y_j, D_i = 1, D_j = 0, S_i = 1, S_j = 1)}{\sum_{i \neq j}^{n} \hat{w}^{\text{ipsw}}(\boldsymbol{X}_i, \boldsymbol{X}_j) \mathbb{I}(D_i = 1, D_j = 0, S_i = 1, S_j = 1)}\\&\qquad - 
    \frac{\sum_{i\neq j}^{n}\hat{w}_{ij} \mathcal{P}_{ij}(\boldsymbol{V}_i, \boldsymbol{V}_j; \hat\bbeta, \hat{\sigma}_0,\hat{\sigma}_1) \mathbb{I}(D_i = 1, D_j = 0, S_i = 1, S_j = 1)}{\sum_{i\neq j}^{n}\hat{w}_{ij} \mathbb{I}(D_i=1,D_j=0,S_i=1,S_j=1)}\\
    &=\frac{\sum_{i \neq j}^{n} \hat{w}^{\text{ipsw}}(\boldsymbol{X}_i, \boldsymbol{X}_j)\left[\left(\mathbb{I}(Y_i > Y_j)-\mathcal{P}_{ij}(\boldsymbol{V}_i, \boldsymbol{V}_j; \hat\bbeta, \hat{\sigma}_0,\hat{\sigma}_1)\right) \mathbb{I}(D_i = 1, D_j = 0, S_i = 1, S_j = 1)\right] }{\sum_{i \neq j}^{n} \hat{w}^{\text{ipsw}}(\boldsymbol{X}_i, \boldsymbol{X}_j) \mathbb{I}(D_i = 1, D_j = 0, S_i = 1, S_j = 1)},
    \end{align*}
    and thus
    \begin{align*}
    &d_1^{\text{aipsw1}}(\boldsymbol{V}_i, \boldsymbol{V}_j;\balp,\bbeta)\\&=\frac{1}{\pi(\boldsymbol{X}_i;\balp)\pi(\boldsymbol{X}_j;\balp)}\left[\left(\mathbb{I}(Y_i>Y_j)-\mathcal{P}_{ij}(\boldsymbol{V}_i, \boldsymbol{V}_j; \bbeta, {\sigma_0},{\sigma_1})\right)\mathbb{I}(D_i = 1, D_j = 0, S_i = 1, S_j = 1)\right]
    \end{align*}
    and 
    \begin{align*}
    d_0^{\text{aipsw1}}(\boldsymbol{V}_i, \boldsymbol{V}_j;\balp)=\frac{1}{\pi(\boldsymbol{X}_i;\balp)\pi(\boldsymbol{X}_j;\balp)}\mathbb{I}(D_i = 1, D_j = 0, S_i = 1, S_j = 1)
    \end{align*}
    Let $\tau_1$ be the true limit of $\hat{\tau}_{\text{AIPSW1}}=\hat{\tau}_{\text{IPSW}}-\hat{\tau}_{\text{OM}}$, which is around 0. Then, let 
    \begin{align*}
        l_\text{AIPSW1}(\boldsymbol{V}_i,\boldsymbol{V}_j;\tau_1,\hat{\balp},\hat\bbeta)=
        \tau_1  d^{\text{acw1}}_0(\boldsymbol{V}_i,\boldsymbol{V}_j;\hat{\balp})-d^{\text{acw1}}_1(\boldsymbol{V}_i,\boldsymbol{V}_j;\hat{\balp},\hat\bbeta)
    \end{align*}
    and
    \begin{align*}
        l_\text{AIPSW2}(\boldsymbol{C}_i,\boldsymbol{C}_j;\tau,\hat{\bbeta})&=l_\text{OM+RWD}(\boldsymbol{C}_i,\boldsymbol{C}_j;\tau,\hat{\bbeta})\\
        &=\tau d_0^{\text{om+rwd}}(\boldsymbol{C}_i,\boldsymbol{C}_j;\tau,\hat{\bbeta})-d_1^{\text{om+rwd}}(\boldsymbol{C}_i,\boldsymbol{C}_j;\tau,\hat{\bbeta}).
    \end{align*}
    For the first term, let $\hat{\tau}_{\text{AIPSW1}}$ be the solution of the estimating equation
    \begin{align*}
        U_{n}^{\text{aipsw1}}(\tau_1,\hat{\balp},\hat\bbeta)=\frac{1}{n(n-1)}\sum_{i\neq j}l_\text{AIPSW1}(\boldsymbol{V}_i,\boldsymbol{V}_j;\tau_1,\hat{\balp},\hat\bbeta)=0,
    \end{align*}
    and define
    \begin{align*}
        U^{\text{aipsw1}}(\tau_1,\balp,\bbeta)=\mathbb{E}\left\{U_{n}^{\text{aipsw1}}(\tau_1,\balp,\bbeta)\right\}=\mathbb{E}\left\{l_\text{AIPSW1}(\boldsymbol{V}_i,\boldsymbol{V}_j;\tau_1,{\balp},\bbeta)\right\}.
    \end{align*}
    Assume $\{\frac{1}{\sqrt{n}}\sum_{i=1}^n\pi(\boldsymbol{X}_i;\balp),\balp\in \mathcal{A}\}$ is stochastically equicontinuous, and the second and the third assumptions are held. Let $\tau_1^*$ be the limit of the difference of the first two terms, i.e., $\hat{\tau}_\text{IPSW}-\hat{\tau}_\text{OM}\xrightarrow{P}\tau_1^*$. By Lemma \ref{lem3}, when $\hat{\balp}\xrightarrow{P}\balp^*$, $\hat{\bbeta}\xrightarrow{P}\bbeta^*$, and $\hat{\tau}_\text{AIPSW1}\xrightarrow{P}\tau_1^*$, we have
    \begin{align*}
    \sqrt{n}\left(\hat{\tau}_{\text{AIPSW1}}-\tau_1^*\right)\rightarrow N\left(0,\operatorname{Var}\left[\left(u_1(\tau_1^*,\balp^*,\bbeta^*)\right)^{-1}\phi_1(\boldsymbol{V}_i;\tau_1^*,\balp^*,\bbeta^*)\right]\right),
    \end{align*}
    where 
    \begin{align*}
    u_1(\tau_1,\balp,\bbeta)&=\frac{\partial U^{\text{aipsw1}}(\tau_1,\balp,\bbeta)}{\partial\tau_1}\\&=\mathbb{E}\left\{d^{\text{aipsw1}}_0(\boldsymbol{V}_i,\boldsymbol{V}_j;{\balp})\right\}\\
    &=\mathbb{E}\left\{\frac{1}{\pi(\boldsymbol{X}_i;\balp)\pi(\boldsymbol{X}_j;\balp)}\mathbb{I}(D_i=1,D_j=0,S_i=1,S_j=1)\right\}
    \end{align*}
    and
    \begin{align*}
\phi_1(\boldsymbol{V}_i;\tau_1,\balp,\bbeta)=&\frac{\partial U^{\text{aipsw1}}(\tau_1,\balp,\bbeta)}{\partial\balp^\top}\left[\mathbb{E}\left\{\frac{\partial h_1(\boldsymbol{V}_i;\balp)}{\partial\balp^\top}\right\}\right]^{-1}h_1(\boldsymbol{V}_i;\balp)\\&+\frac{\partial U^{\text{aipsw1}}(\tau_1,\balp,\bbeta)}{\partial\bbeta^\top}\left[\mathbb{E}\left\{\frac{\partial h_2(\boldsymbol{V}_i;\bbeta)}{\partial\bbeta^\top}\right\}\right]^{-1}h_2(\boldsymbol{V}_i;\bbeta)-l^*_{\text{AIPSW1}}(\boldsymbol{V}_i;\tau_1,\balp,\bbeta).
\end{align*}

   The $n^{-1}\sum_{i=1}^nh_1(\boldsymbol{V}_i;\balp)=0$ is the estimating equation for $\balp$, and $n^{-1}\sum_{i=1}^nh_2(\boldsymbol{V}_i;\bbeta)=0$ is the estimating equation for $\bbeta$. The $l^*_{\text{AIPSW1}}(\boldsymbol{V}_i;\tau_1,\balp,\bbeta)$ is defined as
\begin{align*}
    l^*_{\text{AIPSW1}}(\boldsymbol{V}_i;\tau_1,\blam,\bbeta)= \mathbb{E}\left\{l_\text{AIPSW1}(\boldsymbol{V}_i,\boldsymbol{V}_j;\tau_1,\blam,\bbeta)\mid\boldsymbol{V}_i\right\}+\mathbb{E}\left\{l_\text{AIPSW1}(\boldsymbol{V}_j,\boldsymbol{V}_i;\tau_1,\blam,\bbeta)\mid\boldsymbol{V}_i\right\}.
\end{align*}

    Furthermore, the second term is the OM+RWD estimator, i.e., $\hat{\tau}_{\text{AIPSW2}}=\hat{\tau}_{\text{OM+RWD}}$. For simplicity, we omit the proof for $\hat{\tau}_{\text{AIPSW2}}$ here, which is the same as $\hat{\tau}_{\text{OM+RWD}}$. Let $\hat{\tau}_{\text{AIPSW2}}$ be the solution of the estimating equation
    \begin{align*}
        U_{m}^{\text{aipsw2}}(\tau,\hat\bbeta)=\frac{1}{m(m-1)}\sum_{i\neq j}l_\text{AIPSW2}(\boldsymbol{C}_i,\boldsymbol{C}_j;\tau,\hat\bbeta)=0,
    \end{align*}
    and define
    \begin{align*}
        U^{\text{aipsw2}}(\tau,\bbeta)=\mathbb{E}\left\{U_{m}^{\text{aipsw2}}(\tau,\bbeta)\right\}=\mathbb{E}\left\{l_\text{AIPSW2}(\boldsymbol{C}_i,\boldsymbol{C}_j;\tau,\bbeta)\right\}.
    \end{align*}
    By Lemma \ref{lem3}, assume $\hat{\tau}_{\text{AIPSW2}}\xrightarrow{P}\tau^*$, we have
    \begin{align*}
    \sqrt{n}\left(\hat{\tau}_{\text{AIPSW2}}-\tau^*\right)\rightarrow N\left(0,\operatorname{Var}\left[\left(u_2(\tau^*,\bbeta^*)\right)^{-1}\phi_2(\boldsymbol{C}_i;\tau^*,\bbeta^*)\right]\right),    
    \end{align*}
    where
    \begin{align*}
    u_2(\tau,\bbeta)&=\frac{\partial U^{\text{aipsw2}}(\tau,\bbeta)}{\partial\tau}\\&=\mathbb{E}\left\{d^{\text{om+rwd}}_0(\boldsymbol{C}_i,\boldsymbol{C}_j;\tau,\bbeta)\right\}\\
    &=\mathbb{E}\left\{\mathbb{I}(D_i=1, D_j=0, S_i=0, S_j=0)\right\}
    \end{align*}
    and
    \begin{align*}
    \phi_2(\boldsymbol{C}_i;\tau,\bbeta)=&\frac{\partial U^{\text{aipsw2}}(\tau,\bbeta)}{\partial\bbeta^\top}\left[\mathbb{E}\left\{\frac{\partial h_2(\boldsymbol{C}_i;\bbeta)}{\partial\bbeta^\top}\right\}\right]^{-1}h_2(\boldsymbol{C}_i;\bbeta)-l^*_{\text{OM+RWD}}(\boldsymbol{C}_i;\tau,\bbeta).
    \end{align*}
    The $l^*_{\text{OM+RWD}}(\boldsymbol{C}_i;\tau,\bbeta)$ and $h_2(\boldsymbol{C}_i;\bbeta)$ are the same as defined before. Finally, by $\hat{\tau}_{\text{AIPSW1}}$ and $\hat{\tau}_{\text{AIPSW2}}$ are independent, we have

    \begin{align*}
&\sqrt{n}\left(\hat{\tau}_{\mathrm{AIPSW}}-(\tau_1^*+\tau^*)\right)
\rightarrow  N\Bigl(0,\;\operatorname{Var}\!\left[\left(u_1(\tau_1^*,\balp^*,\bbeta^*)\right)^{-1}\phi_1(\boldsymbol{V}_i;\tau_1^*,\balp^*,\bbeta^*)\right]
\\&\qquad \qquad\qquad \qquad \qquad\qquad\qquad\quad+\operatorname{Var}\!\left[\left(u_2(\tau^*,\bbeta^*)\right)^{-1}\phi_2(\boldsymbol{C}_i;\tau^*,\bbeta^*)\right]\Bigr).
\end{align*}

    When either $\pi(\boldsymbol{X};\balp^*)=\text{Pr}(S=1\mid \boldsymbol{X})$, or $M(\boldsymbol{X},D;\bbeta^*)=\mathbb{E}\{Y\mid \boldsymbol{X},D\}$, we have $\tau_1^*+\tau^*\xrightarrow{P}\tau_0$. Thus,
    \begin{align*}
    \sqrt{n}\left(\hat{\tau}_{\mathrm{AIPSW}}-\tau_0\right)
    &\rightarrow N\Bigl(0,\;\operatorname{Var}\!\left[\left(u_1(\tau_1,\balp^*,\bbeta^*)\right)^{-1}\phi_1(\boldsymbol{V}_i;\tau_1,\balp^*,\bbeta^*)\right]\\
    &\qquad\qquad\quad+\operatorname{Var}\!\left[\left(u_2(\tau_0,\bbeta^*)\right)^{-1}\phi_2(\boldsymbol{C}_i;\tau_0,\bbeta^*)\right]\Bigr).
    \end{align*}
\end{proof}

\section{Additional Discussions about Methods}\label{sec:OtherMethodsDiscussions}

\subsection{Comparison of CW and IPSW}\label{sec:CWvsIPSW}
Table~\ref{tab:CWvsIPSW} compares the CW and IPSW estimators in terms of data requirements, model specification, double robustness, and computational burden. In summary, IPSW requires individual-level data, which may be restricted when only summary statistics are available from disease registries or population-based observational studies. By contrast, CW remains feasible as long as covariate summary statistics are available, providing valid inference for the target population AUC without access to patient-level data. The double robustness property of CW has been formally established \citep{Zhao2017}. IPSW relies on the correct specification of the sampling model, whereas CW does not; its performance instead depends on the choice of calibration functions $\bg(\boldsymbol{X})$. 

\begin{table*}[htbp]%
\centering %
\renewcommand{\arraystretch}{0.7}
\caption{Comparison of CW and IPSW \label{tab:CWvsIPSW}}%
\begin{tabular*}{\textwidth}{@{\extracolsep\fill}lll@{\extracolsep\fill}}
\toprule
\textbf{} & \textbf{CW}  & \textbf{IPSW}  \\
\midrule
Information required & Covariate summary statistics  & Individual-level data \\
Model specification & No sampling model required  & Correct sampling model  \\
Double robustness & $\checkmark$  & $\times$   \\
Computing load & Moderate & Low \\
\bottomrule
\end{tabular*}
\end{table*}

\section{Additional Simulation Information}\label{sec:app_Simulation}
\subsection{Details of the Simulation Setup}\label{sec:SimulationDetails_setup}
The data for both the RWD $\mathcal{R}$ and the finite validation population $\mathcal{V}$ are generated as follows. For each subject $i=1$, let $\boldsymbol{X}_i = (X_{i1}, X_{i2}, X_{i3})^\top$ denote the $p=3$ dimensional covariate vector, where
\[
X_{i1} \sim N(1,0.5^2), \quad 
X_{i2} \sim N(-1,0.5^2), \quad 
X_{i3} \sim U(0,1),
\]
independently across $j$ and $i$. The binary outcome indicator $D_i$ is generated from a logistic model
\[
\text{logit}\{\Pr(D_i=1 \mid \boldsymbol{X}_i)\}
= 0.2 - 0.25 X_{i1} - 0.15 X_{i2} + 0.25 X_{i2}X_{i3} + 0.3 X_{i3}^2.
\]
The continuous biomarker $Y_i$ is generated according to
\begin{align*}
    Y_i = &0.2 - 0.15 X_{i1} + 0.2 X_{i3} + 0.1 X_{i2}X_{i3} - 0.1 X_{i2}^2 + 0.15 D_i + 0.4 D_i X_{i1}^2 + 0.2 D_i X_{i1}X_{i3} + \epsilon_i,
\end{align*}

where $\epsilon_i \sim N(0,0.5^2)$. This outcome model also serves as the oracle outcome model, representing the correctly specified specification. In addition, a sampling model is introduced to reflect the probability of an individual being selected into the validation cohort from the finite validation population. The sampling indicator $S_i$ is generated from the logistic regression model
\[
\text{logit}\{\Pr(S_i=1 \mid \boldsymbol{X}_i)\} 
= \alpha_0 + \alpha_1 X_{i1}^2 + \alpha_2 X_{i2}^2 + \alpha_3 X_{i1}X_{i3},
\]
where $\balp=(\alpha_0,\alpha_1,\alpha_2,\alpha_3)$ controls the degree of covariate shift between the validation cohort and the target population. A detailed setup is provided in Table~\ref{tab:scena_covshift}.

To better reflect real-world practice in our simulation, we do not use the unselected part of the validation cohort as the RWD for estimating weights in the IPSW method. Instead, we treat the generated RWD as the sample with $S=1(\tilS=1)$ and the validation cohort as the sample with $S=1 (\tilS=0)$. Accordingly, the IPSW weights are derived by fitting a sampling model for $S$, using the combined individual-level data from both the RWD and the validation cohort. To implement the IPSW method, the sampling model can be estimated from the combined data, a strategy applicable to both simulation studies and real-world applications. However, in practice, the actual non-selected sample is unobserved, and RWD serves only as a proxy for the portion of the population left after selecting the validation cohort ($S=1$).

For calibration weight (CW) estimation, we consider two ways of covariate calibration: (i) Moment balance: calibrating the first and second moments of the covariates, i.e., $\bg_1(\boldsymbol{X}_i)=\left[X_{i1}, X_{i2}, X_{i3}, X_{i1}^2, X_{i2}^2, X_{i3}^2\right]$, and (ii) Moment + interaction balance: calibrating the first and second moments along with their interactions, i.e., $\bg_2(\boldsymbol{X}_i)=\left[X_{i1}, X_{i2}, X_{i3}, X_{i1}^2, X_{i2}^2, X_{i3}^2, 
X_{i1}X_{i2}, X_{i1}X_{i3}, X_{i2}X_{i3}\right]$. For the implementation of the CW methods, summary statistics from the RWD are used to estimate individual-level calibration weights. These weights can be further updated using summary statistics from the target population, if such information is available. 
\begin{table*}[htbp]
\centering
\begin{threeparttable}
\caption{Basic summary statistics under different degrees of covariate shift}
\label{tab:scena_covshift}
\small 
\setlength{\tabcolsep}{0pt} 
\renewcommand{\arraystretch}{0.7}
\begin{tabular*}{\textwidth}{@{\extracolsep{\fill}}llccccc@{}}
\toprule
 & & & \multicolumn{4}{c}{\textbf{Na\"ive method\tnote{1}}} \\
\cmidrule(lr){4-7}
\textbf{Covariate shift} & $\alpha$ & $\tau_0$ & \textbf{Relative bias\tnote{2} (\%)} & \textbf{Bias/SE} & \textbf{RMSE\tnote{3}} & \textbf{CP\tnote{4}} \\
\midrule
No shift       & (0.15, 0, 0, 0)           & 0.81 & 0.018 & 0.007 & 0.015 & 0.946 \\
Moderate shift & (0.15, 0.30, -0.10, 0.10) & 0.81 & 1.914 & 1.107 & 0.022 & 0.780 \\
Severe shift   & (0.15, 0.45, -0.25, 0.20) & 0.81 & 2.860 & 1.657 & 0.028 & 0.602 \\
\bottomrule
\end{tabular*}

\begin{tablenotes}[flushleft]
\footnotesize
\item[1] Without handling the covariate shift.
\item[2] Percentage bias relative to the true AUC in the target population.
\item[3] Square root of MSE.
\item[4] Coverage probability.
\end{tablenotes}
\end{threeparttable}
\end{table*}

\section{Additional Case Study Results}\label{sec:app_CaseStudy}
\subsection{Baseline Covariate Characteristics in the Case Studies}\label{sec:app_CaseStudy_baseCov}
Table \ref{tab:case1_table1} summarizes the baseline covariate characteristics for Case Study 1. The validation cohort from the POWER trials differs noticeably from the target population represented by the NSCLC dataset, particularly in sex, performance status, smoking status, baseline weight, and histology. These differences suggest substantial covariate shift between the validation cohort and the target population, implying that AUC estimates based solely on the validation cohort may not accurately reflect biomarker performance in the target population.

Figure \ref{tab:case2_table1} presents the baseline covariate characteristics for the two POWER trials in Case Study 2. Although the two trial populations are broadly similar, some imbalance remains, especially in smoking status and histology. Therefore, even in this setting, adjustment for covariate shift is important to ensure that AUC comparisons are made with respect to the same target population and are therefore interpretable and fair.

\begin{table}[!ht]
\centering
\caption{Baseline characteristics of the representative dataset (NCDB) and validation cohort.}
\label{tab:case1_table1}
\small
\renewcommand{\arraystretch}{0.7}
\begin{tabularx}{\textwidth}{lXXX}
\toprule
 & \textbf{NCDB} & \textbf{Validation Cohort} & \textbf{Overall} \\
 & \textbf{(N=7119)} & \textbf{(N=116)} & \textbf{(N=7235)} \\
\midrule
\multicolumn{4}{l}{\textbf{Age}} \\
\quad Mean (SD) & 62.2 (9.93) & 65.0 (9.78) & 62.2 (9.94) \\
\quad Median [Min, Max] & 63.0 [20.0, 89.5] & 64.5 [45.0, 88.0] & 63.0 [20.0, 89.5] \\

\multicolumn{4}{l}{\textbf{Gender}} \\
\quad Male & 4468 (62.8\%) & 60 (51.7\%) & 4528 (62.6\%) \\
\quad Female & 2651 (37.2\%) & 56 (48.3\%) & 2707 (37.4\%) \\

\multicolumn{4}{l}{\textbf{Race}} \\
\quad White & 6092 (85.6\%) & 100 (86.2\%) & 6192 (85.6\%) \\
\quad Non-White & 1027 (14.4\%) & 16 (13.8\%) & 1043 (14.4\%) \\

\multicolumn{4}{l}{\textbf{Performance status}} \\
\quad 0 (Normal activity) & 6702 (94.1\%) & 48 (41.4\%) & 6750 (93.3\%) \\
\quad 1 (Limited activity) & 417 (5.9\%) & 68 (58.6\%) & 485 (6.7\%) \\

\multicolumn{4}{l}{\textbf{Smoking status}} \\
\quad Current smoker & 1988 (27.9\%) & 53 (45.7\%) & 2041 (28.2\%) \\
\quad Former smoker & 4611 (64.8\%) & 44 (37.9\%) & 4655 (64.3\%) \\
\quad Non-smoker & 520 (7.3\%) & 19 (16.4\%) & 539 (7.4\%) \\

\multicolumn{4}{l}{\textbf{Baseline weight}} \\
\quad Mean (SD) & 76.1 (15.6) & 72.1 (13.5) & 76.1 (15.6) \\
\quad Median [Min, Max] & 74.0 [30.0, 195] & 71.0 [42.0, 111] & 74.0 [30.0, 195] \\

\multicolumn{4}{l}{\textbf{Histology}} \\
\quad Adenocarcinoma & 1916 (26.9\%) & 73 (62.9\%) & 1989 (27.5\%) \\
\quad Squamous & 1143 (16.1\%) & 34 (29.3\%) & 1177 (16.3\%) \\
\quad Other & 4060 (57.0\%) & 9 (7.8\%) & 4069 (56.2\%) \\
\bottomrule
\end{tabularx}
\end{table}

\begin{table}[!ht]
\centering
\caption{Baseline characteristics of POWER 1 and POWER 2 trials.}
\label{tab:case2_table1}
\small
\renewcommand{\arraystretch}{0.7}
\begin{tabularx}{\textwidth}{lXXX}
\toprule
 & \textbf{POWER 1} & \textbf{POWER 2} & \textbf{Overall} \\
 & \textbf{(N=321)} & \textbf{(N=304)} & \textbf{(N=625)} \\
\midrule
\multicolumn{4}{l}{\textbf{Age}} \\
\quad Mean (SD) & 61.5 (8.77) & 61.4 (7.79) & 61.5 (8.31) \\
\quad Median [Min, Max] & 61.0 [34.0, 88.0] & 61.0 [40.0, 81.0] & 61.0 [34.0, 88.0] \\

\multicolumn{4}{l}{\textbf{Gender}} \\
\quad Male & 232 (72.3\%) & 214 (70.4\%) & 446 (71.4\%) \\
\quad Female & 89 (27.7\%) & 90 (29.6\%) & 179 (28.6\%) \\

\multicolumn{4}{l}{\textbf{Race}} \\
\quad White & 311 (96.9\%) & 297 (97.7\%) & 608 (97.3\%) \\
\quad Non-White & 10 (3.1\%) & 7 (2.3\%) & 17 (2.7\%) \\

\multicolumn{4}{l}{\textbf{Performance status}} \\
\quad 0 (Normal activity) & 101 (31.5\%) & 91 (29.9\%) & 192 (30.7\%) \\
\quad 1 (Limited activity) & 220 (68.5\%) & 213 (70.1\%) & 433 (69.3\%) \\

\multicolumn{4}{l}{\textbf{Smoking status}} \\
\quad Current smoker & 181 (56.4\%) & 187 (61.5\%) & 368 (58.9\%) \\
\quad Former smoker & 79 (24.6\%) & 69 (22.7\%) & 148 (23.7\%) \\
\quad Non-smoker & 61 (19.0\%) & 48 (15.8\%) & 109 (17.4\%) \\

\multicolumn{4}{l}{\textbf{Baseline BMI}} \\
\quad Mean (SD) & 24.9 (4.21) & 24.4 (3.99) & 24.7 (4.11) \\
\quad Median [Min, Max] & 25.0 [14.0, 36.0] & 24.0 [15.0, 42.0] & 25.0 [14.0, 42.0] \\

\multicolumn{4}{l}{\textbf{Histology}} \\
\quad Adenocarcinoma & 145 (45.2\%) & 124 (40.8\%) & 269 (43.0\%) \\
\quad Squamous & 144 (44.9\%) & 147 (48.4\%) & 291 (46.6\%) \\
\quad Other & 32 (10.0\%) & 33 (10.9\%) & 65 (10.4\%) \\
\bottomrule
\end{tabularx}
\end{table}

\subsection{AUC Estimation and Generalization from Study to Population}\label{sec:CaseStudy_c1}
In this case study, we estimate the AUC of baseline SCP for predicting 6-month survival in a broader target population defined as U.S. trial-eligible patients with advanced NSCLC receiving chemotherapy. To extrapolate from the POWER trials to this target population, we use a large dataset that aggregates 210 adult lung-cancer trials (1990-2012) conducted by NCI cooperative groups \citep{pang2016}. After restricting to NSCLC patients without prior chemotherapy, we identified 8,232 individuals across four sponsoring groups. This cohort serves as a proxy for the national trial-eligible population and is used to calibrate the POWER trials and estimate the target-population AUC.

Previous analyses of the placebo arms from both trials indicated that a higher baseline SCP was associated with a greater relative SCP loss by Day 84 \citep{kinsey2018}. Given the established link between muscle mass and survival in cancer patients \citep{Rodrigues2013}, we aim to evaluate the prognostic utility of baseline SCP to discriminate which NSCLC patients are likely to survive beyond six months. 
As a descriptive check before pooling, we tested for interactions between baseline SCP and trial ($p=0.726$) and between baseline SCP and treatment arm ($p=0.984$). These non-significant tests did not indicate strong departures from a common prognostic association in this dataset. Therefore, we pool patients from both treatment arms and both POWER trials to assess the discriminatory performance (AUC) of baseline SCP as a prognostic marker for 6-month survival in this case study to enhance power. 

Because our target population is U.S. trial-eligible NSCLC patients, we exclude non-U.S. participants and keep 116 participants from the POWER trials. In the representative dataset, we further restricted the analysis to individuals with complete baseline covariate information and observed 6-month response outcomes, leaving 7,119 participants for analysis. We adopted this complete-case approach to ensure that the covariates required for calibration and outcome modeling were consistently available across all estimators. 
As shown in Table \ref{tab:case1_table1}, notable imbalances exist in gender, performance status, smoking history, and histology. Thus, covariate calibration is required before generalizing the baseline SCP AUC to the broader target population.

To implement CW and OM, following the simulation results, we calibrate covariates using only the first and second marginal moments from the representative dataset. To obtain sampling weights, we fit a logistic model for the indicator $S$ (POWER $S= 1$; representative data $S= 0$) and computing inverse probability weights based on $\hat{\pi}(\boldsymbol{X}) = \Pr(S = 1 \mid \boldsymbol{X})$. 
Here, the binary outcome is survival beyond 6 months, and baseline SCP is the biomarker of interest. Baseline covariates, including age, gender, race, performance status, smoking history, weight, and histology, are included as main effects in both the sampling and biomarker-outcome model. CW matches first moments for all covariates and second moments for continuous only, as squaring dichotomous variables does not yield additional information. To limit the influence of a few extreme weights, we truncate the weights at the 0.1\% and 99.9\% quantiles and then normalize them \citep{cole2008trunc}.

\begin{figure}[ht]
    \centering
    \includegraphics[width=0.6\linewidth]{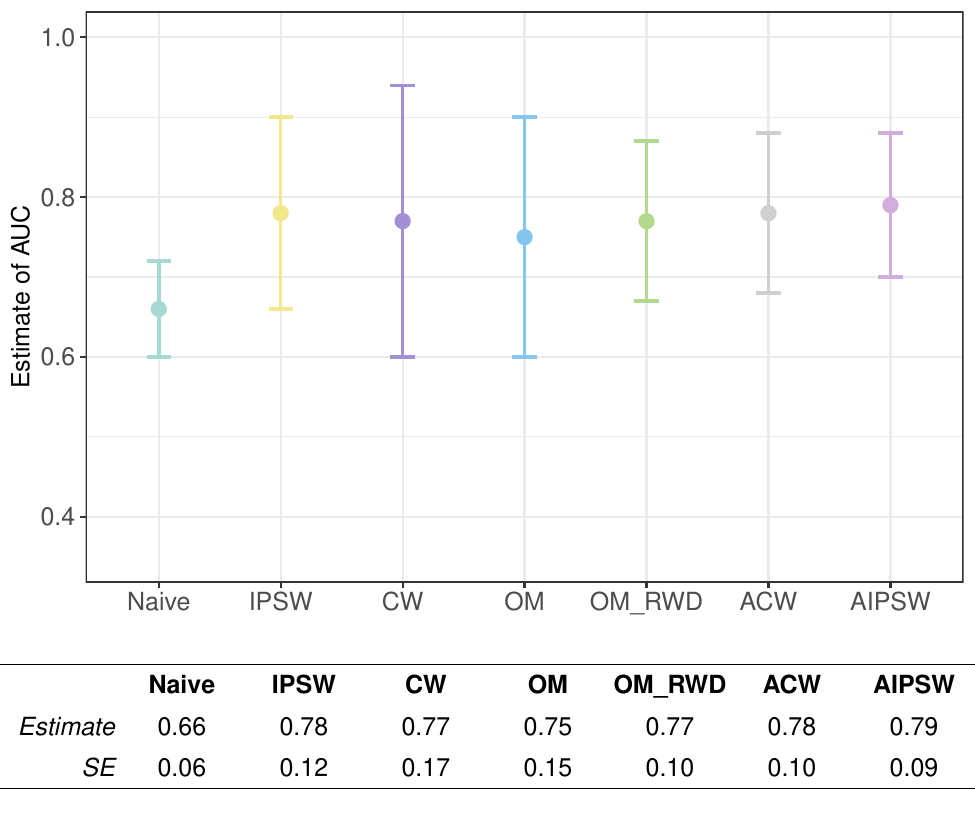}
    \caption{Estimation results for AUC in the broader target population}
    \label{fig:res_Case1}
\end{figure}

As shown in Figure~\ref{fig:res_Case1}, proposed covariate‐shift–adjusted estimators yield similar AUCs of about $0.77$, all higher than the Na\"ive estimate of $0.66$.
This suggests that SCP's prognostic performance among all trial-eligible patients with NSCLC would be underestimated if covariate differences between POWER trials and the target population were ignored. Aligned with simulation results, the augmented estimators achieve higher efficiency in AUC generalization. The estimated lower confidence bounds are above $0.5$, providing some evidence that baseline SCP has discriminatory ability for predicting 6-month survival, though the evidence should be viewed cautiously given that the lower bounds are close to the null value. The augmented estimators, ACW and AIPSW, yield standard errors of approximately $0.10$, showing an efficiency gain compared with CW alone (SE = $0.17$) and OM alone (SE = $0.15$). This pattern is consistent with the simulation findings in Section~\ref{sec:simulation}.

\end{document}